\documentclass{aa}

\usepackage{placeins}

\usepackage{amsmath}
\usepackage{graphicx}
\usepackage{txfonts}
\usepackage{lipsum}
\usepackage{multirow}

\usepackage{hyperref}
\hypersetup{
    colorlinks=true,
    linkcolor=black,
    filecolor=magenta,      
    urlcolor=cyan,
    citecolor=black,
}

\begin{document}

\newcommand{\ha}{\ensuremath{{\rm H \alpha}} }
\newcommand{\hb}{\ensuremath{{\rm H \beta} } }

\title{Recently launched AGN-driven outflow in a lensed quiescent galaxy at Cosmic Noon}

   \author{L. Bugiani\inst{1,2}\corrauth{letizia.bugiani@unibo.it} \and A. B. Newman\inst{3,4}
        \and S. Belli\inst{1}
        \and C. Marconcini\inst{11}
        \and P. Bergamini\inst{2,5}
        \and S. Gangula\inst{3,4}
        \and C. Grillo\inst{5,6}
        \and A. Bolamperti\inst{8, 6}
        \and K. Whitaker\inst{9,10}
        \and A. Beverage\inst{3,12}
        \and R. Ellis\inst{7}
        \and A. H. Khoram\inst{1,2}
        \and A. Matthews\inst{3}
        \and M. Lombardi\inst{5}}

   \institute{Dipartimento di Fisica e Astronomia, Universit\`a di Bologna, Bologna, Italy.
   \and INAF, Osservatorio di Astrofisica e Scienza dello Spazio, Via Piero Gobetti 93/3, I-40129, Bologna, Italy.
   \and The Observatories of the Carnegie Institution for Science, Pasadena, CA, USA.
   \and Department of Physics and Astronomy, University of Southern California, Los Angeles, CA, USA.
   \and Dipartimento di Fisica, Universit{\`a} degli Studi di Milano, Milano, Italy.
   \and Instituto di Astrofisica Spaziale e Fisica cosmica Milano, Istituto Nazionale di Astrofisica, Milano, Italy.
   \and Department of Physics \& Astronomy, University College London, London, UK.
   \and Max-Planck-Institut f{\"u}r Astrophysik, Karl-Schwarzschild-Str. 1, D-85748 Garching, Germany
   \and Department of Astronomy, University of Massachusetts, Amherst, Massachusetts, USA
   \and Cosmic Dawn Center (DAWN), University of Copenhagen, Denmark
   \and INAF, Osservatorio Astrofisico di Arcetri, Largo E. Fermi 5, I-50125, Firenze, Italy
   \and Department of Astrophysical Sciences, Princeton University, 4 Ivy Lane, Princeton, NJ 08544, USA}

   \date{Received September 30, 20XX}

 
  \abstract
  {We analyze JWST/NIRSpec IFU spectroscopy of MRG-M0150, a gravitationally lensed, massive quiescent galaxy at $z=2.6$ from the INQUEST-JWST survey. Thanks to the combination of lensing magnification with JWST's angular resolution, we are able to build spatially resolved maps of the ionized gas spatial distribution, kinematics, and ionization properties in this system. We detect ionized gas emission throughout the galaxy: the lack of residual star formation inferred from line-ratio diagnostic diagrams, together with the decoupling between the gas and stellar kinematics, suggest that the ionized gas is in a wind. We use the 3D gas kinematics modeling code \texttt{MOKA3D} to constrain the geometry and physical properties of the ionized gas wind. We find that the observations are consistent with a biconical wind driven by the low-luminosity AGN at the center of the galaxy, which is detected in X-rays. The wind reaches high velocities of  $\approx1000-1400$ km s${}^{-1}$, implying that part of the gas may escape the halo. The ionized gas mass outflow rate, $(0.6 \pm 0.2)$~${\rm M}_{\odot}$~yr${}^{-1}$, is modest but exceeds the upper limit on the residual star formation rate, and it is likely that the wind is multiphase and carries more mass. We show that typical unresolved observations would not detect the wind, suggesting that the incidence of such outflows may be higher than previously recognized. Simple geometric arguments show that the observed outflow must have been launched $\approx 0.45$~Myr ago from the time of observation. This evidence suggests that low-level AGN activity is an important mechanism for maintaining quiescence in massive galaxies at cosmic noon, potentially over cosmological timescales.}
   \maketitle
   \nolinenumbers

\section{Introduction} 
The physical processes responsible for the quenching of star-formation in massive galaxies remain one of the key open problems in galaxy formation and evolution. Numerous mechanisms have been proposed to explain the quenching of massive galaxies, including active galactic nuclei (AGN), stellar feedback, and virial shock heating \citep[see][for a recent review]{withaker_quenchingreview_2026}, but no single mechanism seems to be universally applicable. Semi-analytical models and cosmological hydrodynamical simulations typically invoke AGN feedback as a mechanism capable of quenching massive galaxies. Feedback from the AGN can drive galaxy-scale outflows in multiple gas phases, quenching the galaxy both by expelling gas from the system and by heating the circumgalactic medium, thereby preventing the cooling and accretion of gas that could otherwise fuel subsequent star formation \citep[e.g.][]{dimatteo_2005,hopkins_2006,bower_2006,somerville_2008,dave_2019}. 
However, direct observational evidence linking AGN feedback to the quenching of star formation remains elusive.
At cosmic noon ($z\sim2-3$), a significant fraction of massive galaxies stopped forming new stars \citep[e.g.][]{muzzin_2013-cosmosultravista,Straatman2014,Tomczak2014,Weaver2023}. Observing recently quenched systems at this epoch is crucial for investigating the main  mechanisms responsible for quenching star formation in the most massive galaxies.

\begin{figure*}[t]
    \centering
    \includegraphics[width=\linewidth]{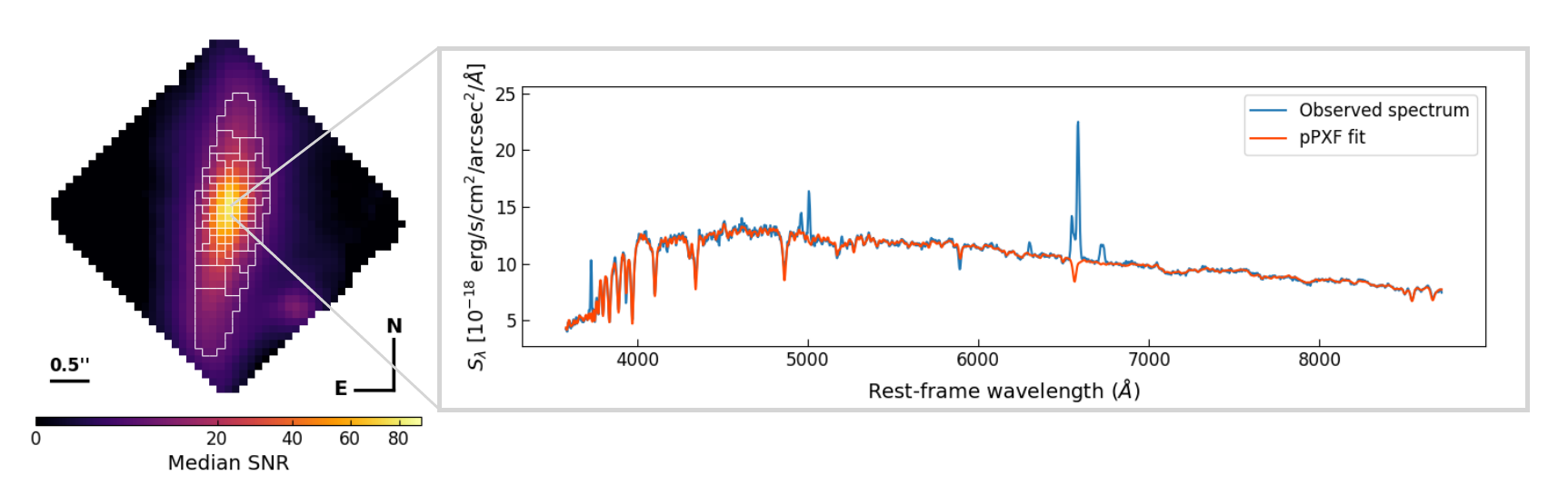}
    \caption{
    Left: Median signal-to-noise ratio along the wavelength axis of the IFU data cube. Overplotted in white are the boundaries of the Voronoi bins analyzed.
    Right: One example spectrum extracted from the central spaxel (blue) with the pPXF stellar continuum fit overlaid in red. 
    }
    \label{fig:general image}
\end{figure*}

Galaxy-scale gas outflows have been extensively studied at $z \sim 1 - 3$ through large galaxy surveys. It has been found that outflows are ubiquitous in massive star-forming galaxies at cosmic noon \citep{Steidel2010,weiner_2009,forster_schreiber_kmos3d_2019} and in those hosting the most luminous AGN \citep{canodiaz_2012,cresci_blowin_2015,kakkad_tracing_2016,davies_nuclear_2020}, where they are expected to be powerful enough to be able to suppress star formation. However, the study of gas outflows and nuclear activity in the first quiescent galaxies at high redshift is challenging from the ground, relying either on deep exposures of individual galaxies \citep[e.g.][]{heavymetalsurvey,belli_2017} or stacked spectra \citep[e.g.][]{belli2019}.

The launch of the James Webb Space Telescope (JWST) has opened a new window on the high-redshift Universe, revealing an abundant population of massive quiescent galaxies already in place at $z>2$ \citep{carnall_surprising_2023,weibel_rubies_2025,nanayakkara_population_2024,de_graaff_efficient_2025}. 
Early JWST spectra of individual galaxies have revealed the presence of AGN activity in these systems \citep{carnall_massive_2023,bellinat_2024,deugenio_fast-rotator_2024,wupofeng_2025} and subsequent studies with larger samples have confirmed a high AGN incidence in massive quiescent galaxies at $z>2$ \citep{mypaper25,skarbinkimaya26}. These AGN, for the most part, have been classified as low-luminosity AGN (LL-AGN), based on their faint inferred bolometric luminosities and the lack of more overt tracers of powerful AGN activity such as X-rays detection or IR excess \citep{almaini2025,hewitt_monsters_2026}. 

At the same time, JWST observations have revealed a widespread presence of gas outflows in massive quiescent galaxies at high redshift, detected both in the neutral and ionized phases \citep{rebecca_neutral24,taylor26,valentino_recently_quenched,matteos_article_sodium_2026,Sun_2026_neutral}. These outflows can be remarkably powerful, often carrying large amounts of cold gas.
In particular, observations of neutral-phase outflows in $z>2$ galaxies through the Na D absorption line have shown that powerful winds are prevalent among massive galaxies at cosmic noon and may represent the dominant pathway for quenching at this epoch \citep{rebecca_neutral24,matteos_article_sodium_2026}. In some systems, the inferred neutral-phase mass outflow rates exceed the current star-formation rates by more than an order of magnitude, suggesting that efficient removal of cold gas may contribute directly to the rapid quenching observed in these galaxies \citep{bellinat_2024,deugenio_fast-rotator_2024}. 

However, the origin of these galactic winds and their connection to star-formation quenching remain unclear, as the low-level AGN activity observed in many massive systems struggles to explain the high energy and momentum rates characterizing the observed outflows. While some studies attribute the origin of the observed winds to ongoing LL-AGN activity \citep[e.g.][]{bellinat_2024,wupofeng_2025,mypaper25}, others interpret them as relics of past starburst episodes or previous periods of more powerful black hole activity, with the central AGN having since faded into a low-accretion state \citep{valentino_recently_quenched,Sun_2026_neutral,taylor26}. Distinguishing between these scenarios is therefore crucial for establishing whether the outflows observed in massive quiescent galaxies represent ongoing AGN feedback or the remnants of the events that originally quenched the galaxy.


With only a few exceptions \citep{gangula_resolved_2026,deugenio_fast-rotator_2024}, most studies rely on spatially unresolved data and therefore are not able to directly constrain the geometry, kinematics, and timescales of the outflows.
While the ionized phase may represent a small contribution to the total outflowing mass, as observed in high-$z$ quiescent galaxies \citep{bellinat_2024} and predicted by simulations of galactic winds \citep{kim_smaug_2020}, it is the best suited to measuring these properties. Ionized gas can be detected directly in emission without requiring a bright background continuum. Its emission traces the full spatial extent of the wind, and, provided that it can be spatially resolved, yields detailed information about the kinematics and 3D structure of the outflow. This can reduce some of the uncertainties associated with assumptions about the geometry and kinematics when deriving mass outflow rates and energetics. Moreover, reconstructing the extent, geometry, and velocity of the winds provides a way to constrain their characteristic timescales, which can be used to distinguish between winds powered by ongoing AGN activity and fossil outflows launched by a past quasar or starburst event. Resolved observations of the ionized phase can therefore offer a unique view on the connection between AGN activity, galactic outflows, and the maintenance of quiescence in massive galaxies at high redshift.

In this work, we investigate ionized gas emission in MRG-M0150, a strongly-lensed, massive quiescent galaxy at $z=2.64$ \citep{newman_discovery_2015} using spatially resolved observations obtained by JWST/NIRSpec in IFU mode. We make use of the combined sensitivity of JWST and gravitational lensing magnification to characterize the geometry, velocity, and energetics of the ionized gas outflow observed in this system, and use it to investigate the connection between the outflow and the quenching of star-formation in this galaxy.
In Section \ref{sec:target}, we present the observations used in this paper. In Section \ref{sec:emissionlinefit}, we describe the emission-line fitting and discuss the resulting maps and resolved emission-line diagnostics, as well as the AGN activity and residual star formation. In Section \ref{sec:ionizeedgasoutflow}, we construct a 3D kinematic model of the observed ionized gas outflow and derive its main properties. In Section \ref{sec:sedfitting}, we perform SED fitting of the integrated light from the galaxy to reconstruct its star-formation history (SFH). We discuss our results in Section \ref{sec:discussion} and summarize our main conclusions in Section \ref{sec:conclusions}. Throughout this paper, we assume the WMAP9 cosmology \citep{hinshaw13} and a \citet{CHABRIER_IMF} initial mass function.


\section{Target and observations} \label{sec:target}
MRG-M0150 is a massive quiescent galaxy at $z=2.64$, multiply imaged by the galaxy cluster MACS J0150.3$-$1005 at $z=0.365$ \citep{newman_discovery_2015}. A JWST/NIRCam image of the cluster, with the multiple images of our target labeled, can be found in Fig.~\ref{fig:cluster_and_lensmodel} in Appendix~\ref{appendixlens}. The galaxy has a stellar mass of $\log (M/M_{\odot}) = 11.30 \pm 0.04$ -- derived by the integrated SED fitting, see Sect.~\ref{sec:sedfitting} -- and is classified as a ``fast rotator,'' exhibiting a rotationally supported stellar disk \citep{newman_resolving_2018,newman_resolving2_2018}. It has a circularized half-light radius of $R_e \sim 0.2$ arcsec (corresponding to 1.6~kpc on the source plane), as found by the lens modeling (Appendix~\ref{appendixlens}). Dynamical models indicate that the inclination of the stellar disk is $\sim 30^{\circ}$ \citep{newman_resolving2_2018}.\footnote{These models were constrained by ground-based slit spectra. Dynamical modeling of the JWST IFU data indicates a similar inclination of $\sim40^{\circ}$ (Newman et al., in prep.).}

The target belongs to the INQUEST-JWST galaxy sample, which includes 5 gravitationally lensed, massive and quiescent galaxies at redshift $z=2.1 - 2.6$. 
Data used in this paper comes from the Cycle 3 JWST GO-4903 observational program. The program observed one image of MRG-M0150, labeled Image 1 in the cluster image visible in Fig.\ref{fig:cluster_and_lensmodel}, using the JWST/NIRSpec IFU with the G140M/F100LP and G235M/F170LP grating/filter pairs. Exposure times were 4.4 hours for each setting. Additionally, the same program acquired NIRCam imaging with total exposure times of 20-30 minutes in the F115W, F200W, F277W and F444W filters. We also make use of archival imaging data from the Hubble Space Telescope (HST) Advanced Camera for Surveys (ACS) through the F814W filter \citep{newman_discovery_2015} for the full SED fitting (Sect.\ref{subsec:prospector}).

The NIRSpec data cubes were reduced using the JWST Science Data Processing system version 2025\_3a and the JWST Science Calibration Pipeline version 2.0.0. Spectral ``wiggles'' -- artifacts affecting NIRSpec data due to undersampling of the PSF -- were mitigated using the adaptive trace model step in the pipeline. We detected some residual wiggles, which we modeled and removed using the procedure described in \cite{newman_stellar_2025}. 
We created output data cubes with a $0.1''$ spaxel scale for the G140M and G235M observations. The fluxes and variances are logarithmically rebinned onto a wavelength grid with 132 km s$^{-1}$ per pixel and shifted to the systemic rest frame ($z_{\rm sys}=2.6355$). Then they are combined by splicing them together at $\lambda_{rest} = 0.505 \: \mu$m. Noise spectra coming out of the pipeline are rescaled based on residuals to the stellar continuum model fits (Sect. \ref{sec:emissionline_fit}).
The resulting data cube is spatially binned to a signal-to-noise target of 40 using the \texttt{vorbin} algorithm \citep{cappellari_vorbin_2012,cappellari_vorbin_2003}. A visualization of the resulting Voronoi bins can be seen in the left panel of Figure \ref{fig:general image}. We also measure PSF-matched photometry in each spatial bin for every photometric filter.

The final step is to reconstruct the galaxy onto the source plane by applying the lens model described in Appendix \ref{appendixlens}. The gravitational magnification of a factor of $\mu \sim 5$ in the central parts of the galaxy delivers a resolution for the IFU data of $\sigma_{psf} = 29$ mas $= 235$ pc, calculated by casting a PSF placed at the galaxy's centroid to the source plane (see Appendix \ref{appendixmoka}).

\section{Spatially resolved gas emission} \label{sec:emissionlinefit}

\subsection{Emission lines fit} \label{sec:emissionline_fit}
In order to isolate emission lines due to ionized gas, we first have to model the underlying emission from the stellar continuum. We fit the stellar continuum contribution in each spaxel using the \texttt{pPXF} code. The model used for the stellar population synthesis with \texttt{pPXF} and the E-MILES library \citep{emiles_library} is analogous to the one used by \cite{newman_stellar_2025}. The best-fit stellar continuum model is subtracted from the data to isolate the emission lines. An example spectrum and stellar continuum fit is given in Figure \ref{fig:general image}. To measure the stellar kinematics, we perform an additional \texttt{ppxf} run in which the strongest emission lines are masked, since tests showed that fitting them could bias the stellar velocity dispersions (Newman et al, in prep.).

All rest-frame optical emission lines -- consisting of \ha, \hb, [OII]$\lambda\lambda 3727,3729$, [OIII]$\lambda\lambda 5007,4959$, [NII]$\lambda\lambda 6548,6581$, [SII]$\lambda\lambda 6717,6732$, [NeIII]$\lambda 3968$ and [OI]$\lambda 6300$ -- are fitted together for each spatial bin. The fitting is performed using \texttt{EMCEE}, an affine invariant Markov Chain Monte Carlo (MCMC) ensemble sampler \citep{emcee}. We employ single Gaussian profiles fitting the same velocity shift and dispersion for all lines, and taking into account the NIRSpec wavelength-dependent spectral broadening. The flux ratios [OIII]5007/[OIII]4959 and [NII]6581/[NII]6548 are fixed at the theoretical value of 3:1 \citep{osterbrockfernland}. Although our spectral resolution is not sufficiently high to resolve the [OII]$\lambda\lambda3727,3729$ doublet, we fit the lines separately and impose the $1.5 <\mathrm{[OII]3729/[OII]3727}< 2.5$ theoretical limits on their flux ratio \citep{osterbrockfernland}. 

We then repeat the fit using double Gaussian components for each line profile, in which the kinematics of both components are tied together across all lines during the fit. We do not assign any physical meaning to each Gaussian component, but rather use them to better reproduce the observed line profiles. Finally, for each spatial bin, we select which model better reproduces the data based on residuals and the Bayesian Information Criterion (BIC) score of the fits. The line surface brightness values provided throughout this paper correspond to the median values from the \texttt{EMCEE} parameters' posterior distributions, with uncertainties represented by the $\pm1\sigma$ percentiles around the median values. We consider a line above the detection threshold when it is detected with a signal-to-noise ratio $>3$.

We find that ionized gas emission is detected across the entire galaxy, with many bins requiring double Gaussian profile to reproduce the observed data. The high spatial resolution highlights interesting features in the ionized gas lines, that make it possible to thoroughly map the kinematics and physical properties of the ionized gas. 
In Fig. \ref{fig:panels_kinematics}, we show the integrated surface brightness maps for the \ha and [OIII]$\lambda5007$ emission lines, and we compare the kinematics of the gas emission and the stellar component. \ha is well detected in all bins, while [OIII] is mostly detected in the central regions and along the minor axis (upper left and upper right panels, respectively). Both lines' surface brightness distribution peak towards the center, with the brightest bin overlapping the centroid of the galaxy. 

\subsection{Gas kinematics}
\begin{figure*}
    \centering
    \includegraphics[width=0.7\linewidth]{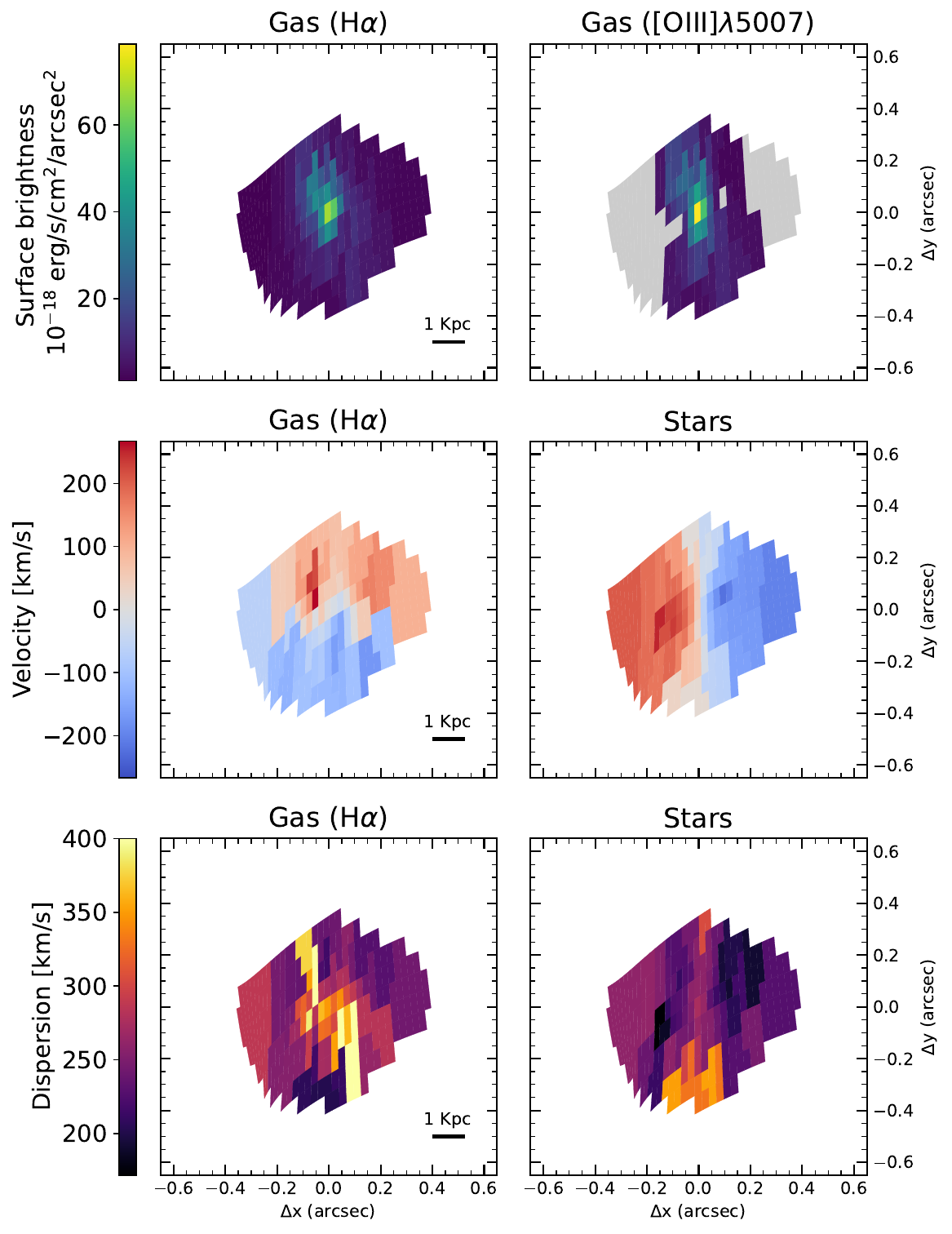}
    \caption{Source plane maps of MRG-M0150. First row: integrated surface brightness of the fitted \ha (left) and [OIII]$\lambda5007$ emission lines (right); light gray areas mark non-detections. Second row: velocity field of the gas (left) from fitted \ha line profile, and of the stars (right) from pPXF. Third row: gas (left) and stellar (right) velocity dispersions, obtained in the same way. Maps are rotated to align the x and y axis to the major and minor axis, respectively. }
    \label{fig:panels_kinematics}
\end{figure*}
In the second and third rows of panels in Fig.~\ref{fig:panels_kinematics} we compare the velocity and dispersion of the gas (which are taken as the moment 1 and moment 2 of the best-fit profile of the \ha emission line) to the stellar kinematics. It is immediately visible that the kinematics of stars and ionized gas are not well coupled in this system. The stellar velocity field traces a well defined rotating disk, as already observed by \cite{newman_resolving2_2018} (right panel), whereas the gas does not follow at all the stellar rotational pattern, with the red- and blue-shifted sides of the gas velocity field positioned almost perpendicular to the stellar rotational axis. 
The velocity dispersion field (last row of panels) is also quite different between stars and gas. The gas map shows a symmetric, high-velocity dispersion feature almost aligned with the minor axis of the galaxy that is absent from the stellar map, as well as exhibiting generally higher values of $\sigma$ in the central regions.

The decoupling between stellar and gas velocity field suggests that the gas emission we are detecting is not produced by star-forming regions rotating in the disk, but may instead be coming from a galaxy-wide ionized gas outflow, as commonly observed in AGN host galaxies with the same gas kinematics signature \citep{fabian2012,fiore_agnwinds_2017}. The outflowing gas scenario could also explain the widespread emission from ionized gas observed in this quenched galaxy, in the absence of the star-formation needed to sustain the observed emission. In this case, an outflow centrally driven by an active nucleus would explain both the observed line fluxes and gas kinematics.

\subsection{Resolved line ratios} \label{sec:lineratios}

\begin{figure*}
    \centering
    \includegraphics[width=0.8\linewidth]{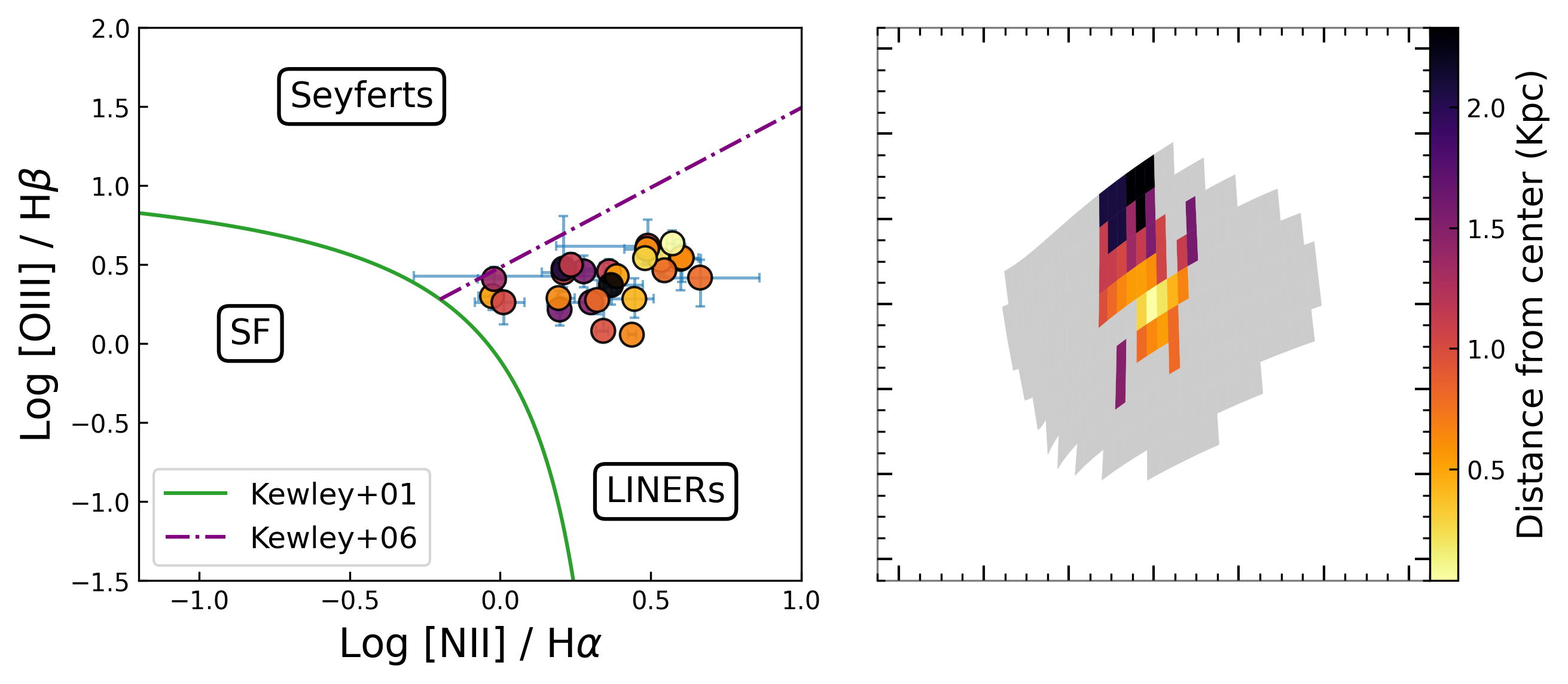}
    \includegraphics[width=0.8\linewidth]{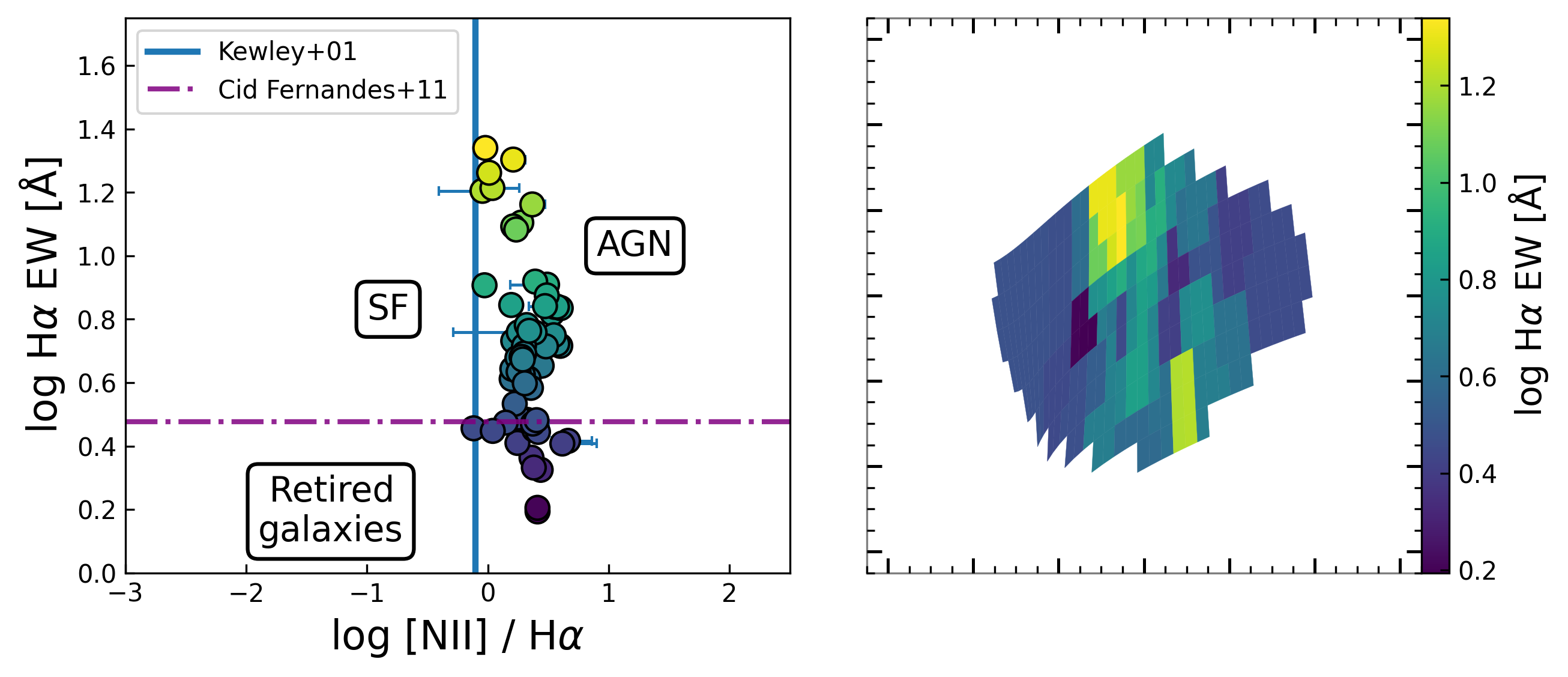}

    \caption{Resolved emission line ratios in MRG-M0150. Top: resolved BPT \citep{baldwin_classification_1981} diagram (left), in which bins have been colored according to their median distance from the center of the galaxy in kpc, mapped onto the right panel. Bottom: wHaN diagnostic diagram \citep{cid_fernandes_comprehensive_2011} (left) colored by EW of \ha (right): spaxels below the ``retired galaxies'' purple line show emission that might be produced by an evolved stellar population; all the rest show definite evidence of AGN or shock contribution.}
    \label{fig:bpts}
\end{figure*}

We investigate the possibility that MRG-M0150 hosts an AGN by using resolved rest-frame optical line diagnostics, which are widely used to discern between the contributions of star formation (SF) and AGN emission to the ionization of the interstellar medium (ISM). In the top panel of Fig.~\ref{fig:bpts} we show the BPT diagram \citep{baldwin_classification_1981}, i.e., [OIII]/H$\beta$ as a function of [NII]/\ha, for each spatial bin in which all lines are detected. The resolved BPT diagram of MRG-M0150 indicates the presence of ionizing radiation that is not due to star formation: as noticeable in the upper panel of Fig.~\ref{fig:bpts}, all spatial bins for which all 4 emission lines are detected -- colored in the map on the right -- are clustered in the same region of the BPT diagram. Line ratios typical of Low Ionization Nuclear Emission-line Regions (LINERs) are detected from the center to the edges of the galaxy. LINER-like ratios can arise from AGN photoionization \citep[e.g.][]{ho_nuclear_2008}, shocked gas \citep[e.g.][]{dopita_2015b} or ionization by hot evolved stellar populations \citep[e.g.][]{belfiore_2016} and the BPT diagram alone cannot definitively differentiate between excitation mechanisms.

We construct the resolved WHaN diagram \citep{cid_fernandes_comprehensive_2011} by showing the equivalent width (EW) of \ha as a function of [NII]/\ha,in Fig.~\ref{fig:bpts}, lower left panel. The WHaN diagram is useful to identify regions of gas ionized by hot evolved stellar populations, such as post-AGB stars, which can heavily contribute to the total ionizing flux budget particularly in quiescent galaxies \citep{cidfernandes_2010}. We find that most of the spatial bins in MRG-M0150 are located above the ``retired galaxy'' line at 3 \r{A} in \ha EW \citep{cid_fernandes_comprehensive_2011}: this implies that most of the observed emission is not excited by ionization due to old stars, but instead a contribution from AGN radiation or shocks is needed to explain the observations. Furthermore,~\cite{Byler2019} found that the EW of H$\alpha$ line emission due to post-AGB stars is weaker for younger stellar populations (stellar ages $<1$ Gyr), shifting down the retired galaxies threshold for high-z quiescent galaxies. 
Estimating the contribution of shocks from AGN or star-formation feedback to the total ionizing budget of MRG-M0150 is a more difficult endeavor. The negligible level of current star formation in this galaxy, as inferred from \ha luminosity (Sect.~\ref{subsec:AGN_activity}), makes it unlikely that shocks from star formation feedback are the main source of ionization. Instead, the observed LINER-like line ratios are more likely to be due to AGN photoionization or shocks from a central AGN-driven outflow.

We map the \ha EW distribution across the galaxy in the lower-right panel of Fig.~\ref{fig:bpts}: unlike the \ha surface brightness map in Fig.~\ref{fig:panels_kinematics}, the \ha EW is higher in the outskirts of the galaxy than in the center. This radial trend cannot be explained by gradients in star formation or dust attenuation, given that the line ratios show little indication of either (Sect.~\ref{subsec:AGN_activity}). It is also inconsistent with hot, evolved stars powering the line emission, since in that case the line emission should more closely follow the stellar light.
Instead, we propose that the \ha~EW pattern is analogous to that seen in 'red geysers', a category that represents $\sim 10\%$ of all massive quiescent galaxies at $z \sim 0$. Red geyser galaxies are characterized by the presence of a low-luminosity AGN (LL-AGN) in their center and centrally-driven ionized gas outflows \citep{cheung_suppressing_2016, roy_evidence_2021}. In these galaxies, the same \ha EW pattern is due to the fact that the \ha emission is tracing the outflowing gas, while the continuum is tracing the stellar population: going from the center to the outskirts of the galaxy, the stellar surface brightness profile decreases more rapidly then the \ha emissivity in the outflow, generating the observed inside-out EW distribution. The distribution of \ha EW in MRG-M0150 is thus further evidence in support of the presence of a centrally-driven, galaxy-wide ionized gas outflow.

\subsection{Residual star formation and AGN activity in MRG-M0150} \label{subsec:AGN_activity}
Although line ratio diagnostics and gas kinematics indicate that the line emission is not powered by star formation, we can still use the integrated \ha line luminosity to place an upper limit on the residual star formation rate. We obtain the total integrated \ha flux by first correcting the flux measured in each spaxel for magnification, and then correcting for dust attenuation. In a small number of bins (13 out of 61) where both \ha and \hb lines are above the detection limit of $3\sigma$, we are able to use the Balmer decrement \ha/\hb to estimate the dust attenuation A$_V$; for all other bins, we use the value estimated by the pPXF fits. We find generally low dust attenuation throughout the galaxy using both methods, with a median value of A$_V = 0.3_{-0.3}^{+0.4}$. We then apply the \cite{calzetti2000} dust attenuation curve and obtain the unabsorbed \ha luminosity. Assuming the conversion provided by \cite{kennicutt_sf98}, we derive a current SFR for MRG-M0150 of $0.30 \pm 0.08$ M$_{\odot}$ yr$^{-1}$, which has to be considered an \textit{upper limit} on the true SFR of the galaxy, since most of the \ha emission is excited by processes other than star formation, according to the diagnostics of Fig.~\ref{fig:bpts}. For a galaxy of stellar mass $\log (M/M_{\odot}) = 11.30$, this implies a specific star-formation rate of $\log (\text{sSFR/yr}^{-1}) = -11.8$. This is very strong evidence of the currently quenched nature of the galaxy, and confirms the presence of an AGN in the center of the galaxy as the main engine powering the observed ionized gas emission.

\renewcommand{\arraystretch}{1.2}
\begin{table*}[h]
    \centering
    \begin{tabular}{c|c|c|c|c|c|c|c|c}
        \hline \hline
         \textbf{Emission} & Flux & \multicolumn{2}{|c|}{$v_{\rm out}$} & \multicolumn{2}{|c|}{$\beta$} & $\mathrm{\dot{M}_{out}}$ & $\mathrm{\dot{E}_{kin}}$ & $\dot{p}_{\rm out}$\\
         \textbf{line} & $10^{-18}$ erg s$^{-1}$ cm$^{-2}$ &  \multicolumn{2}{|c|}{km s$^{-1}$} & \multicolumn{2}{|c|}{degrees} &  M$_{\odot}$ yr$^{-1}$ & $10^{41}$ erg s$^{-1}$ & $10^{33}$ dyn\\
         \cline{2-9}
          &  \textit{total} & \textit{(+) side }& \textit{($-$) side} &  \textit{(+) side} & \textit{($-$) side} & \textit{total}  & \textit{total} & \textit{total} \\
         \hline
         \ha & $3.0 \pm 0.8$ & $1200 \pm 57$& $1150 \pm 73$ & $80 \pm 5$ & $100 \pm 3$  & $0.6 \pm 0.2$ & $2.8 \pm 0.8$  & $5 \pm 2$ \\
         $\rm [OIII]\lambda 5007$ & $3 \pm 1$ & $1000 \pm 66$& $1050 \pm 49$ & $80 \pm 3$ & $100 \pm 5$ & $0.08 \pm 0.02$ & $0.3 \pm 0.1$ & $0.5 \pm 0.2$ \\
         \hline 
    \end{tabular}\
    \vspace{10pt}
    \caption{Results from the 3D kinematics fitting with a bi-conical wind model, for both the \ha and [OIII] line fits.}
    \label{tab:placeholder}
\end{table*}

\begin{figure*}[h]
    \centering
    \includegraphics[width=0.95\linewidth]{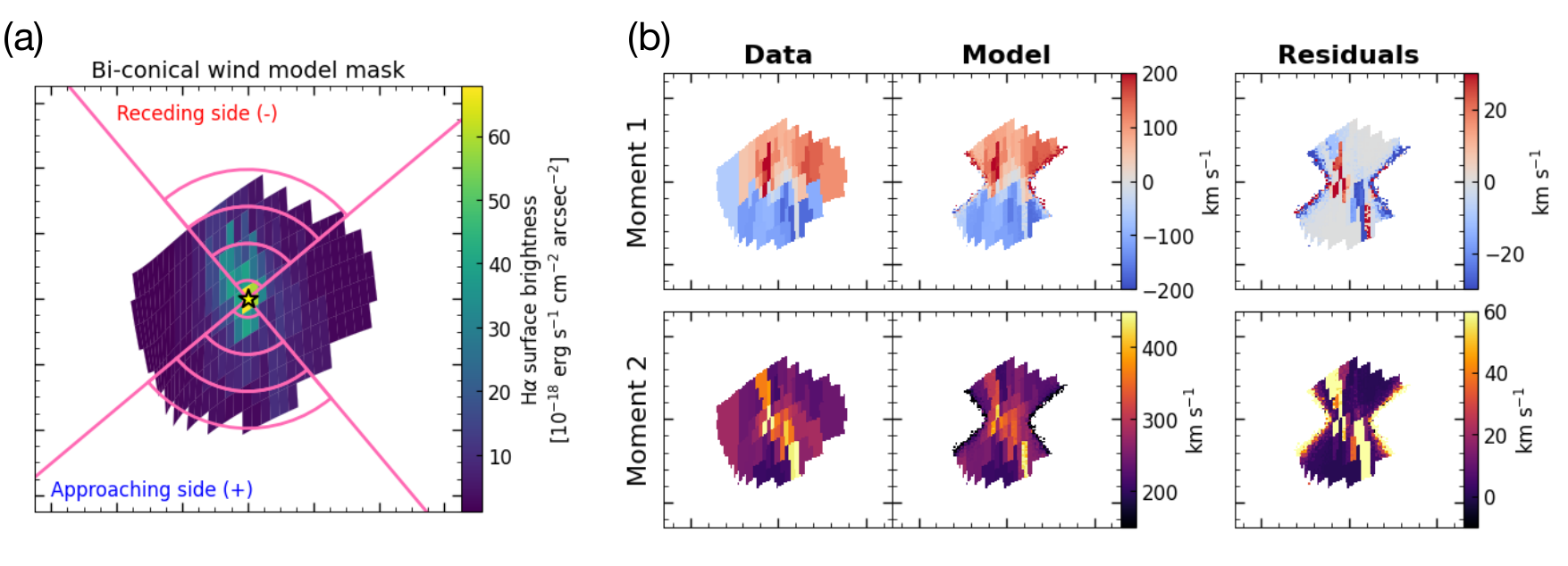}
    \caption{Best-fit results of the kinematic fit with MOKA3D. Panel (a): bi-conical wind model mask overlayed on the observed \ha surface brightness distribution; the model is made up of two cones delimited by the pink diagonal lines; each cone is further divided into 5 concentric shells. Panel (b): comparison between the observed and model gas velocity (moment 1) and dispersion (moment 2) maps from \ha line profiles.}
    \label{MOKA3D_bestfit}
\end{figure*}

We investigate the AGN properties by estimating its bolometric luminosity using the [O~III] 5007 line luminosity in the central spaxel, corrected for aperture losses assuming a point-like emission. We correct for dust attenuation using the Balmer decrement and compute a central [O~III] luminosity of $L_{\rm [O~III]} = (9 \pm 2)\cdot 10^{39}$ erg s$^{-1}$. Assuming the bolometric correction factor from \cite{heckman_growth_2004}, we derive the AGN bolometric luminosity $L_{\rm bol} = (3.2 \pm 0.9)\cdot 10^{43}$ erg s$^{-1}$. We estimate the Eddington luminosity of the SMBH from the black hole mass, which we infer from the M$_{\rm BH} - \sigma$ relation of \cite{mcconnell_blackholes_2011}, adopting the light-weighted second stellar velocity moment within the effective radius to define $\sigma = 282$~km~s${}^{-1}$. We find an Eddington ratio of $L_{\rm bol}/L_{\rm Edd} \approx 10^{-3.8}$. Using the [O~III] line luminosity integrated over the whole galaxy -- as might be done for unresolved studies -- yields a bolometric AGN luminosity almost $6\times$ higher. 


We further constrained $L_{\rm bol}$ using a 27~ks observation with the Advanced CCD Imaging Spectrometer (ACIS) onboard the Chandra X-ray Observatory (program 11800090, PI H. Ebeling). We used the same methodology as \cite{newman_stellar_2025} to measure $L_X$, which we define from $2-10$~keV in the rest frame, and to estimate $L_{\rm bol} = C_X L_X$ using $C_X = 15.8$ \citep{ho_2009_bolcorr}. For the multiple images 1, 2, and 3, we find $L_{\rm bol} / 10^{44}~{\rm erg}~{\rm s}^{-1} = 1.3 \pm 1.2$, $2.9 \pm 2.3$, and $5.1 \pm 1.9$, respectively. We thus marginally detect X-ray emission in each image. We combined the Poisson posteriors for $L_{\rm bol}$ from the three multiple images to estimate $L_{\rm bol} = 3.3^{+1.2}_{-1.0} \cdot 10^{44}~{\rm erg}~{\rm s}^{-1}$. This is an order of magnitude higher than [O~III]-based estimate, which could be an underestimate due to the LINER-type spectrum. Nonetheless, the implied Eddington ratio is still low, $L_{\rm bol}/L_{\rm Edd} \approx 10^{-2.8}$. 

The low bolometric luminosity and Eddington ratio classifies the SMBH as an LL-AGN, a class of AGN occupying the faintest end of the luminosity function and characterized by very low accretion rates, $L_{\rm bol}/L_{\rm Edd} \leq 10 ^{-3}$ \citep[see][for a review]{ho_nuclear_2008}. Recent studies of similar quenched systems at cosmic noon and even higher redshift have found a high incidence of LL-AGN activity \citep{mypaper25,skarbinkimaya26} based on rest-frame optical line diagnostics.

\section{Ionized gas outflows} \label{sec:ionizeedgasoutflow}
\begin{figure}
    \centering
    \includegraphics[width=\linewidth]{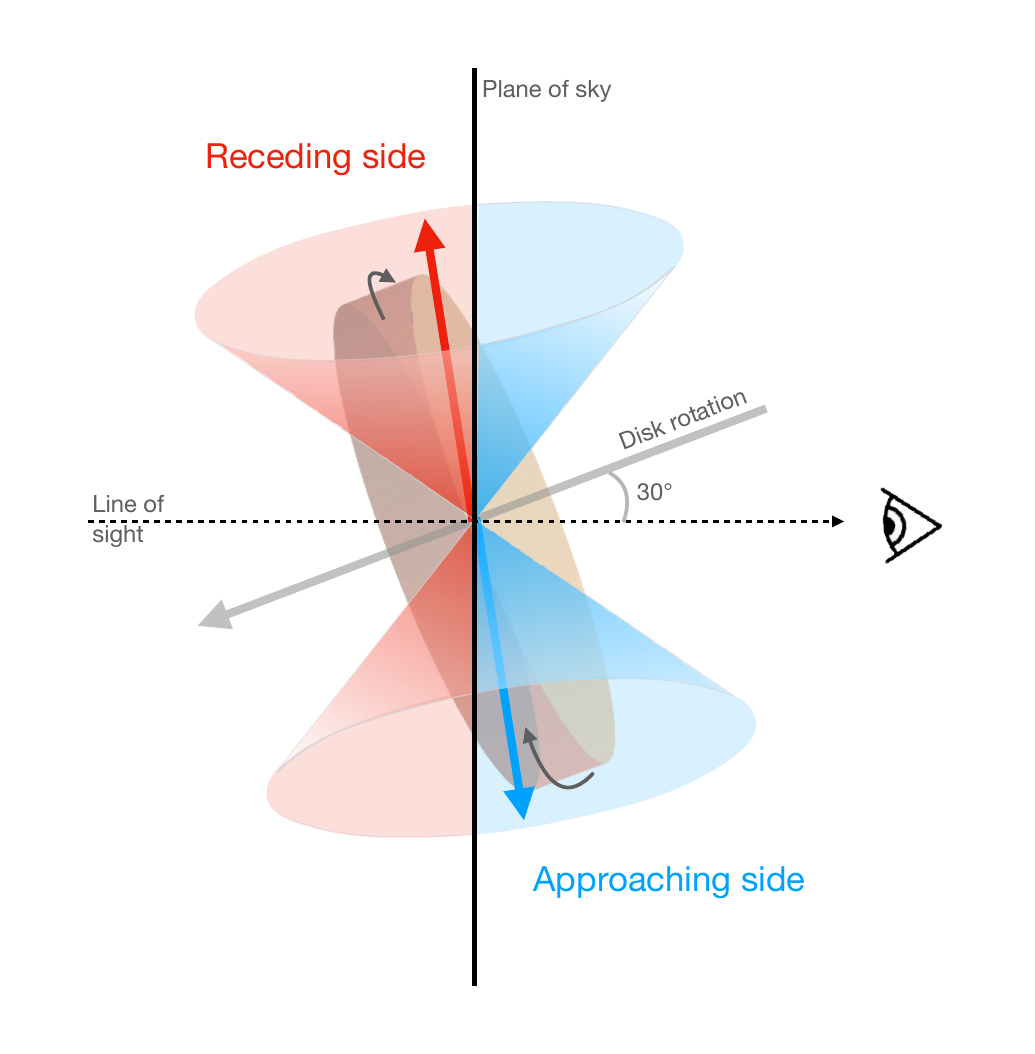}
    \caption{3D representation of the geometry of the bi-conical outflow fitted with MOKA3D with respect to the stellar disk and our line-of-sight. The galaxy is seen at an inclination of $i = 30^{\circ}$ (gray disk), while the outflow is inclined at $\beta = 80^{\circ}$ and $\beta = 100^{\circ}$ for the approaching and receding cones, indicated with a blue and red arrow respectively. Each cone has an aperture angle of $\alpha = 45^{\circ}$ by construction. We have color-coded the cones according to their apparent velocity along the line-of-sight, with blue and red corresponding to approaching and receding velocities, respectively.}
    \label{fig:wind_schematics}
\end{figure}

\subsection{Wind model}
All the evidence so far points to the presence of an AGN-driven galaxy-wide ionized gas outflow originating from the central part of the galaxy. To test whether our observations are compatible with such model we employ the kinematics fitting code ``Modeling Outflows and Kinematics of AGN in 3D'' \citep[MOKA3D;][]{marconcini_moka3d_2023} and apply it to the de-lensed maps of observed emission line profiles.
MOKA3D reconstructs the three-dimensional gas distribution in the outflow by weighting model emitting clouds according to the observed line emission profile and velocity in each spaxel (Sect.\ref{sec:emissionline_fit}). It then generates a model data cube with the same spatial and spectral resolution as the observations, including PSF convolution. This tomographic reconstruction enables the geometry and kinematics of the outflow to be constrained, and provides estimates of its mass outflow rate and energetics.

The model adopted is shown in Fig.~\ref{MOKA3D_bestfit}a. We assume a bi-conical geometry for the wind, with a semi-aperture angle of 45 degrees consistent with other AGN-driven outflows studied at lower or equal redshift. The origin is fixed in the brightest spaxel in the [OIII] emission line map, co-spatial with the geometric center of the galaxy, and the position angle of the bi-cone's axis is fixed to $-$5 degrees with respect to the minor axis after visual inspection of the observed gas velocity distribution. We adopt a constant radial velocity profile in both the receding and the approaching cones, which are fitted independently from each other. For each cone, we find best-fit values of radial velocity $v_{out}$ and inclination angle $\beta$, defined as the angle between the cone's axis and the observer's direction, of the wind in 5 shells of incremental radius, as shown in Fig.~\ref{MOKA3D_bestfit}a. Global values for the inclination and intrinsic velocity of the wind in the cone are given by averaging the values obtained in the shells. We repeat the fit twice -- once fitting the emission line profiles of the [OIII] line and once with \ha -- and find consistent results. Further details on the 3D kinematics modeling of the gas emission in MRG-M0150 are provided in Appendix \ref{appendixmoka}. 

\begin{figure*}[t]
    \centering 
    \includegraphics[width=\linewidth]{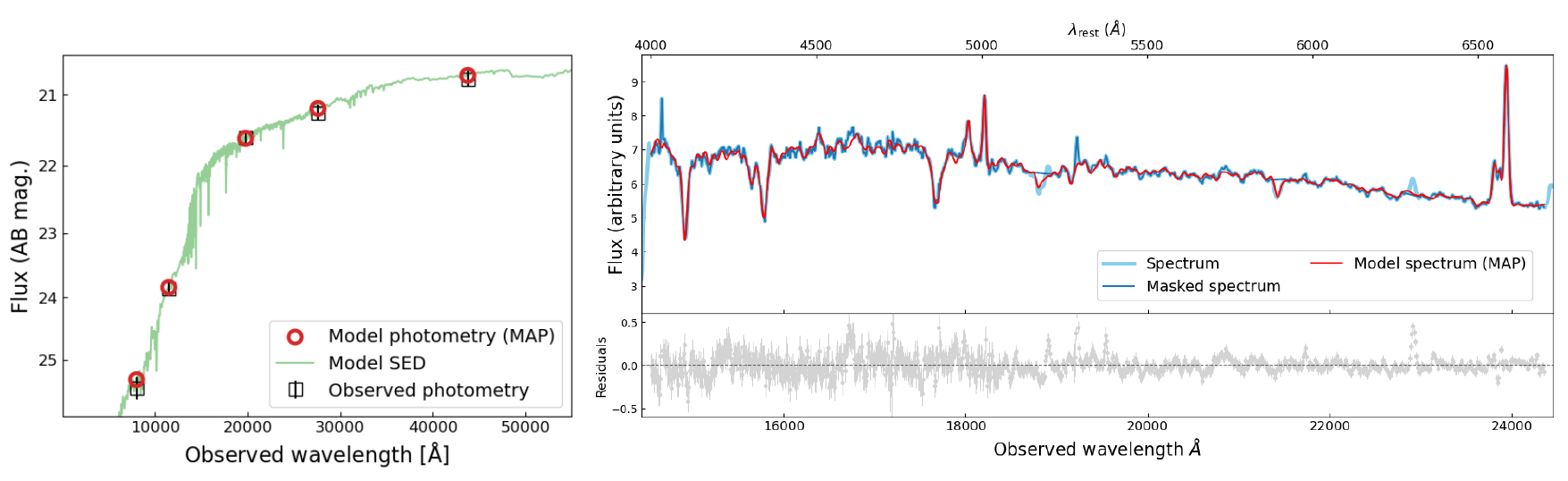}
    \caption{Results of the SED fitting procedure on the integrated photometry and spectrum. Left: observed and predicted photometric points, and galaxy SED from the best-fit \texttt{Prospector} model (green plot). Right: the observed integrated spectrum (blue) and  the best-fit model (red) in the $4000-6700$ \r{A} rest-frame wavelength range, with residuals in the lower panel.}
    \label{fig:prospector}
\end{figure*}

\begin{figure}
    \includegraphics[width=\linewidth]{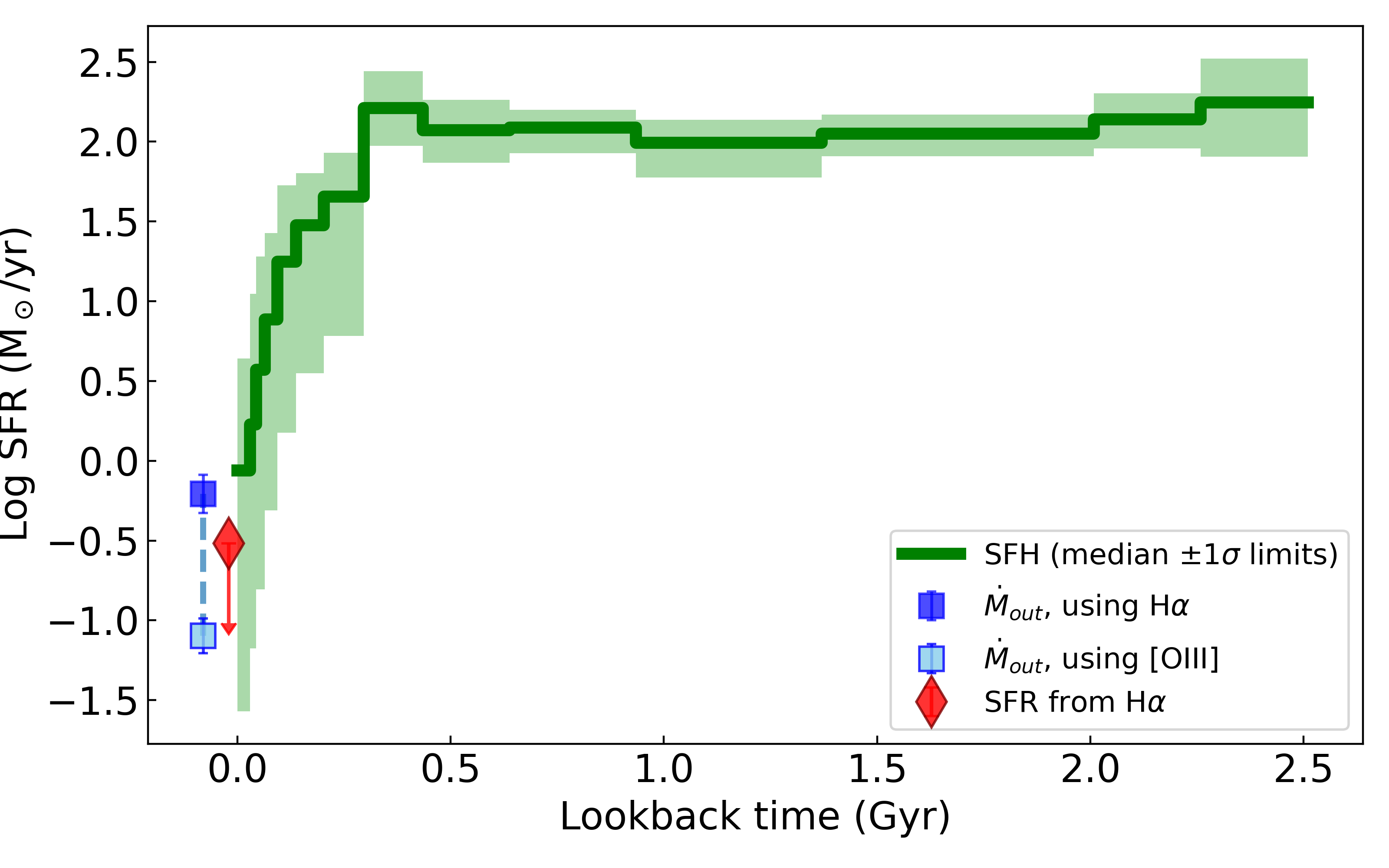}
    \caption{Reconstructed star-formation history of MRG-M0150 according to the \texttt{Prospector} fit. The solid green line indicates the median SFR per time bin, with the light green shaded areas indicating the 16th -- 84th confidence intervals from the posterior distributions. The current SFR and mass outflow rates are shown at a lookback time of approximately zero. }
    \label{fig:sfh}
\end{figure}

\subsection{Outflow properties and energetics} \label{subsec:outflow_properties}
The results of the MOKA3D modeling using the \ha line profiles and a bi-conical wind prescription are summarized in Fig.~\ref{MOKA3D_bestfit}. Analogous results and plots for the [OIII] line fitting results can be found in Appendix \ref{appendixmoka}. The gas kinematics are well reproduced by the model, as visible from the residuals map. The observations match a bi-conical wind whose axis is nearly within the plane of the sky: a schematic representation
of the geometry of the outflow with respect to the stellar disk and our line-of-sight is shown in Fig.~\ref{fig:wind_schematics}. 
For the approaching $(+)$ side, we find an inclination angle $\beta = 80$ degrees with respect to the line-of-sight and an intrinsic de-projected velocity of $v_+ = 1200 \pm 57$ km s$^{-1}$; for the receding $(-)$ side 
an inclination angle $\beta = 100$ degrees and similar velocity $v_+ = 1150 \pm 49$ km s$^{-1}$.  Consistent results are obtained from the model fitted to the [OIII] line emission, although we measure slightly lower values of wind velocity (see Table \ref{tab:placeholder}).
The two cones extend out to the edge of the fitted spaxels, corresponding to an extension of at least 3.3 kpc ($2.3$ times the effective radius $R_e$ in F444W).

As visible in Fig.~\ref{fig:wind_schematics}, the bi-conical outflow's inclination implies strong interactions between the galaxy's disk and the outflow. This is not often seen in classical QSOs and starburst systems hosting such large-scale ionized phase winds \citep{veilleux_galactic_winds_2005}, as usually the outflow chooses the path of least resistance out of the galaxy and is seen positioned perpendicular to the disk plane. 
The quiescent nature of MRG-M0150 may be a contributor to the relative orientation of the outflow and the disk: indeed, a centrally-driven wind expanding inside a quiescent system with a negligible amount of gas in its disk does not encounter much resistance from the ISM along its path and is free to propagate in any direction. This is the case for MRG-M0150, for which ALMA observations of the dust continuum \citep[REQUIEM survey,][]{whitaker_quenching_2021} give an upper limit of 1\% on its molecular gas fraction. Interestingly, recent semi-analytical models of dust evolution in quiescent galaxies find that AGN feedback can rapidly destroy dust in after quenching \citep{Donevski2026}, in line with both the ALMA results and the AGN activity detected in MRG-M0150. On the other hand, the close alignment between the outflow axis and the stellar disk could favor strong interactions between the wind and the ISM, contributing to a negative feedback mechanism. The higher residuals between model and observations in the moment 2 (fig.~\ref{MOKA3D_bestfit}, bottom right panel) reflects closely the feature of enhanced gas velocity dispersion of Fig.~\ref{fig:panels_kinematics} and may be tracing injected turbulence by the wind in the densest ISM regions as a consequence of wind-disk interactions.

\begin{figure*}[t]
    \centering
    \includegraphics[width=0.8\linewidth]{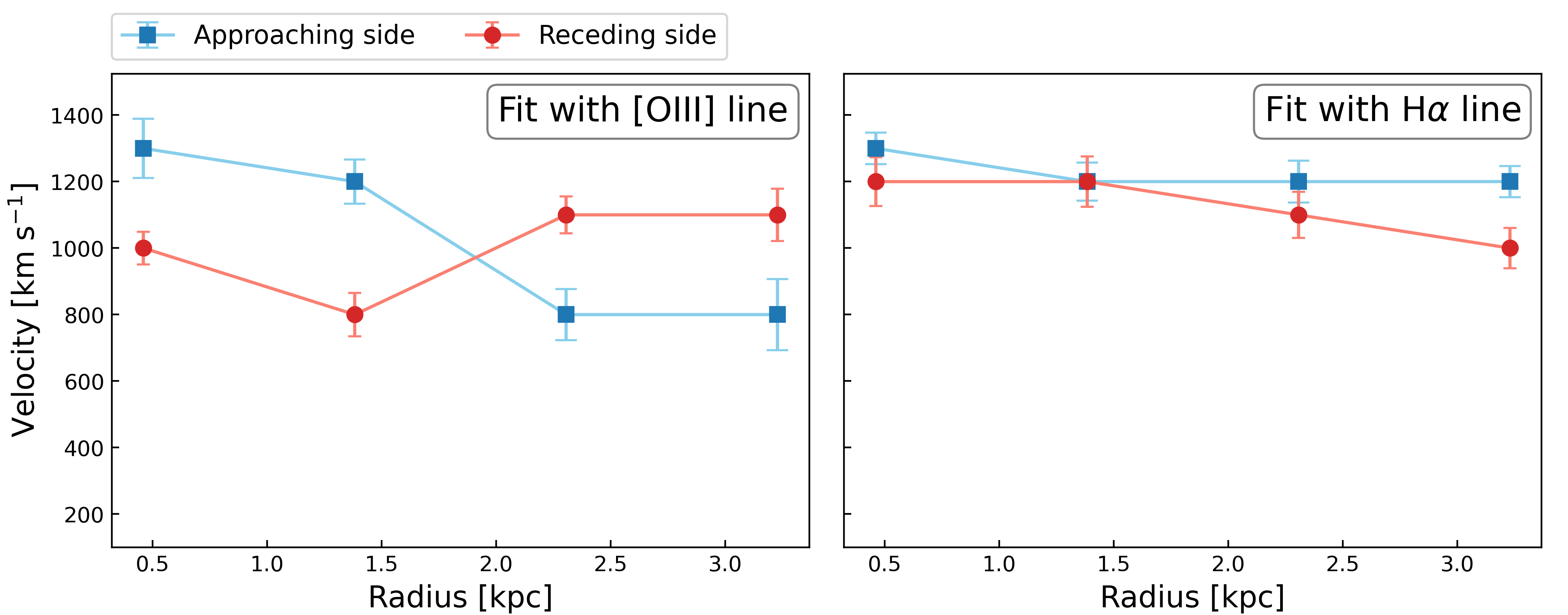}
    \caption{Radial velocity profile of the ionized gas outflows in MRG-M0150, according to the 3D kinematics fit. The left panel shows the profile for the approaching (blue) and receding (red) sides of the bi-conical outflow derived from [OIII]; the right one shows the values derived from \ha.}
    \label{fig:velocity_radial_profiles}
\end{figure*}

We use the model to constrain the wind's mass outflow rate and energetics, following \cite{marconcini_moka3d_2023}. The tomographical reconstruction of emission clouds in the outflow by the MOKA3D code enables us to accurately estimate the mass outflow rate of the ionized phase outflow and to better constrain wind energetics with respect to standard approaches used when the outflow physical properties are unknown \citep{carniani_outflows_2015,canodiaz_2012}. But first, we need to estimate an electron density $n_e$. This is usually derived using either the [OII]$\lambda\lambda 3727,3729$ or the [SII]$\lambda\lambda6717,6732$ doublet ratio \citep{osterbrockfernland}, or is assumed from a broad range of values -- from $1000$ cm$^{-3}$ \citep{venturi_ganifs_2025,bertola_mapping_outflows_2025} to more widely adopted choices of $n_e \sim 100 - 200$ cm$^{-3}$ \citep[e.g.]{kakkad_tracing_2016,brusa_xshooter_2015,cresci_blowing_2015,fiore_agnwinds_2017}. We infer the average electron density of the gas in the red- and blue-shifted sides of the outflow from the observed [SII]$\lambda\lambda6717,6732$ doublet ratio, since the [OII] doublet is not resolved at NIRSpec's medium spectral resolution. The most rigorous approach involves deriving a resolved map of the [SII] doublet ratio to infer $n_e$ in each spaxel: however, the [SII] doublet is below detection limit for many bins of the cube, on top of being difficult to de-blend in most of the rest. Thus, we choose to adopt a constant electron density value $n_e = 600$ cm$^{-3}$ for both sides of the wind, corresponding to the average value found in all spaxels with an accurate estimate of the [SII] doublet ratio, assuming a temperature of $10^4$ K. This choice constitutes one of the most important sources of uncertainties in the outflow's energetics estimation -- indeed, assuming a constant electron density can significantly impact the outflow energetics \citep{daviesbaron2020,veilleux2018} -- thus, we attach a 50\% systematic uncertainty to $n_e$ to account both for the averaging procedure and the uniform density assumptions. Moreover, different values adopted for the constant electron density only affect the (inverse) scaling of the mass outflow rate $\dot{M}_{out}$ and related quantities by a numerical factor; thus our inferred $n_e=600$ cm$^{-3}$ results in more conservative $\dot{M}_{out}$ values than the typical assumptions of $n_e \sim 100 -200$ cm$^{-3}$ would yield.

We finally obtain a total mass outflow rate from the ionized gas of $0.08 \pm 0.02$ M$_{\odot}$ yr$^{-1}$ from the [OIII] line flux, and $0.6 \pm 0.2$ M$_{\odot}$ yr$^{-1}$ from the \ha line flux. This discrepancy in mass outflow rate values between [OIII] and Balmer lines tracers is a consequence of the different volumes in which [OIII] and \ha are emitted: [OIII] is a higher-ionization tracer that usually misses a substantial amount of lower-ionized gas mass \citep{carniani_outflows_2015}, especially in LINERs such as MRG-M0150. Furthermore, no dust correction was applied to the line luminosities before calculations, thus [OIII] may be more affected by attenuation than the redder \ha line, as well as being under the detection limit in many spaxels embedded in the outflow (see Fig.\ref{fig:moka3d_results_o3} in Appendix \ref{appendixmoka}).
Thus, we adopt the mass outflow rates and derived properties given by \ha  as our fiducial values. Consequently, we can calculate the energy and momentum rates of the ionized wind as:
\begin{align}
    &\dot{E}_{kin} = \frac{1}{2} \dot{M}_{out} v_{out}^2  = (2.8 \pm 0.2) \cdot 10^{41} \text{ erg s}^{-1} \\
    &\dot{p}_{out} =  \dot{M}_{out} v_{out} = (5 \pm 2) \cdot 10^{33} \text{ dyne}
\end{align}
All the derived outflow energetics are summed up in Table \ref{tab:placeholder}.

\subsection{Alternative models}

We also consider a tilted gaseous disk geometry to explain the observed emission. To test this scenario, we use MOKA3D to fit a rotating disk model to the observed \ha line emission, allowing it to be freely inclined with respect to the stellar disk. We find that the rotating disk model requires extreme rotational velocities of $v_{rot} \sim 600$ km s$^{-1}$ and an inclination of the gas disk of 80 degrees. Thus, we can exclude this model from the discussion. 
We also experiment with an alternative model in which we constrain the inclination angle of the bi-cone to be $30^{\circ}$, in order to simulate a (more commonly observed) nearly-polar bi-cone with a 3D outflow velocity more comparable to the observed projected one. We find, however, that such model does a poor job in fitting the observed data (see Appendix \ref{appendixmoka}).

\section{SED fitting and star-formation history} \label{sec:sedfitting}

We carry out full SED fitting on the integrated photometry and spectroscopy of MRG-M0150 using \texttt{Prospector} \citep{Johnson_prospector_2021}, to reconstruct its star-formation history (SFH) and verify the quenched state of the galaxy.  
The integrated photometry and spectrum are extracted from all spatial bins included in the analysis, corresponding to an aperture of $2.3 \text{ R}_e$. Flux values extracted in each spaxel are corrected for magnification according to the lens model before being added together, and an additional $10 \%$ error is added in quadrature to all photometric points to account for systematic uncertainties.

We employ a model akin to the one used in \cite{park2024}, adopting MIST isochrones \citep{mist_isochrones}, the MILES stellar library \citep{miles_lib}, and a non-parametric star-formation history (SFH) defined on 14 logarithmically-spaced age bins. 
A ``continuity prior'' on the SFH is adopted, which discourages drastic variations of SFR between adjacent time bins \citep{Leja19_SFH}. We fit photometry from the ACS/WFC F814W filter and JWST/NIRCam's F115W, F200W, F277W, F444W filters, together with the integrated spectrum from NIRSpec/IFU in the $4000-6700$ \r{A} wavelength range. Given that nebular emission lines are tracing the AGN activity and/or the outflowing gas and not star formation, we choose to marginalize over them during the fit. We do not include any AGN emission in the model. 


The quality of the photometry + spectrum fitting is shown in Fig.~\ref{fig:prospector}. In Fig.~\ref{fig:sfh}, we plot MRG-M0150's non-parametric star formation history derived by \texttt{Prospector} (green line). This is compared to our upper limit on the current SFR measured from the integrated \ha luminosity (red diamond) and to the mass outflow rate of the ionized gas outflow (blue squares). The reconstructed SFH of MRG-M0150 is similar to other recently quenched systems observed at $z>2$ \citep{park2024,carnall_massive_2023,bellinat_2024}, showing a sharp decrease of the SFR after the peak at $\sim 500$ Myr ago in lookback time. We further discuss the SFH of MRG-M0150 and the impact of outflows in Sect. \ref{subsec:prospector}.

\section{Discussion} \label{sec:discussion}

\begin{figure*}[t]
    \centering 
    \includegraphics[width=\linewidth]{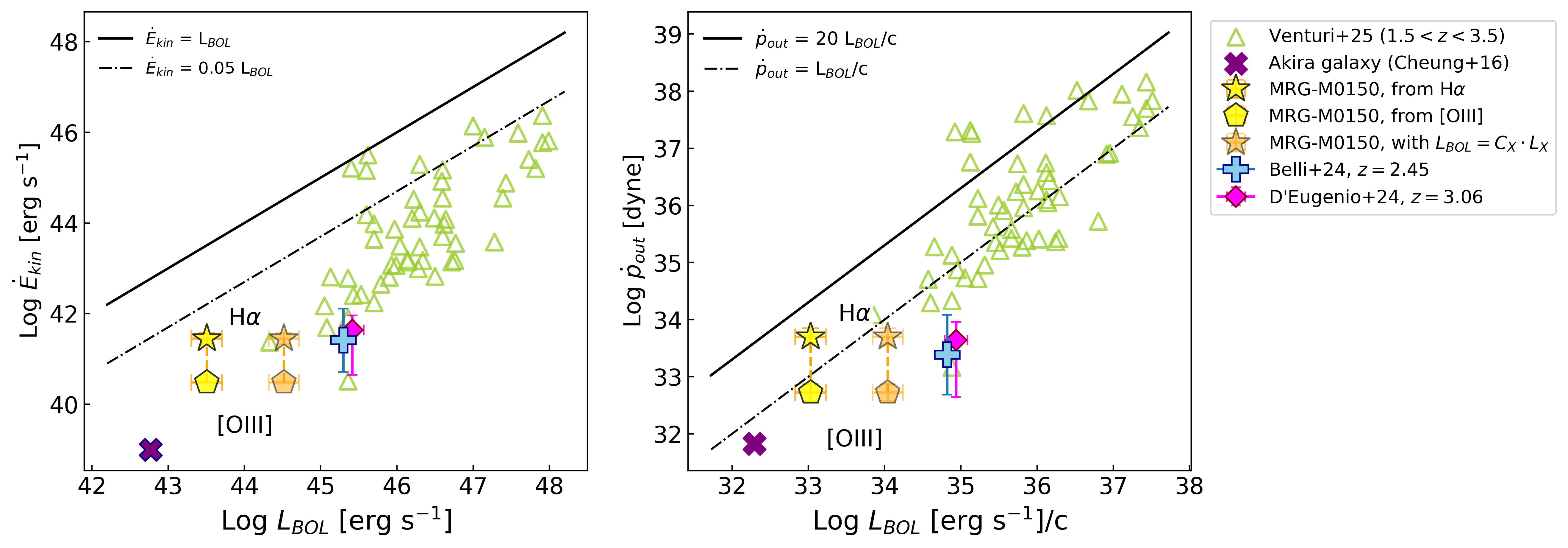}
    \caption{Energy (left) and momentum rate (right) of the ionized winds in MRG-M0150 estimated from the \ha line (star-shaped points) and from the [OIII] line (pentagons) compared to other ionized-phase outflows in the literature, as a function of the AGN bolometric luminosity. Yellow scatter points have been plotted using $\rm L_{bol}$ computed from [OIII] line luminosity, while orange points use $\rm L_{bol}$ computed from the X-rays. Ionized-phase outflows observed in AGN at \(1.5 < z < 3.5\) are marked with empty green triangles \citep{venturi_ganifs_2025}. The blue and magenta filled scatter points are taken from \cite{bellinat_2024} and \cite{deugenio_fast-rotator_2024}, respectively; the purple X represents the prototypical \textit{red geyser} galaxy ``Akira'' at $z\sim 0$ \citep{cheung_suppressing_2016}.}
    \label{fig:energetics}
\end{figure*}

We have detected widespread line emission throughout MRG-M0150. Given the upper limit on the residual star formation from H$\alpha$ and the observed emission line ratios, we concluded that the ionized gas emission is not powered by star formation but instead requires an AGN contribution. Moreover, the bi-conical, centrally-driven outflow morphology and its inclination with respect to the stellar disk (Fig.~\ref{fig:wind_schematics}) further argues against a stellar origin of the observed winds, which would produce a much different wind geometry \cite[see][for examples]{veilleux_galactic_winds_2005}. Finally, the velocities we measure are substantially higher than what observed in the most powerful star-formation driven outflows at $z\sim2$ \citep{Davies2019}. Thus, we conclude an AGN in MRG-M0150 is driving the observed outflow, for which we measure a mass outflow rate of $0.6 \pm 0.2$ $\rm M_\odot~yr^{-1}$.

\subsection{Recently launched ionized gas outflows from a LL-AGN}

Thanks to the high spatial resolution achieved by NIRSpec, the lensing magnification in MRG-M0150, and our ionized wind kinematics fitting, we can use this system to directly measure the timescales of an AGN-driven outflow in a quiescent galaxy at $z>2$.
In Fig.~\ref{fig:velocity_radial_profiles}, we plot the radial wind velocity along both the blue-shifted and red-shifted sides of the bi-conical outflow as fitted by MOKA3D using the [OIII] emission line and the H$\alpha$ line. We find ionized gas velocities  $v_{\rm out}\sim1200$~km~s$^{-1}$, similar to those of prototypical Seyfert galaxies \citep{fiore_agnwinds_2017} and much higher than the $v_{\rm out}\sim300-400$~km~s$^{-1}$ usually observed in the local ``red geyser'' galaxies, but consistent with ionized-phase outflow velocities found in similar massive, quenched galaxies at cosmic noon, such as \citet{bellinat_2024} and \citet{deugenio_fast-rotator_2024}. These results suggest a difference in the nature of feedback mechanisms and impact of ionized gas outflows in massive quiescent systems between $z>2$ and $z=0$.

Using the radial velocity profile of the wind in MRG-M0150, we can constrain the launching timescales of the observed wind. Considering the full extent of the wind, we infer that the ionized outflow in MRG-M0150 has been launched over, at least, the past $2.6$ Myr. 
Narrowing our focus on the innermost shell of the bicone and using the lowest measured velocity from the [OIII] line fit, we infer that the innermost resolved portion of the wind must have launched no more than $\sim 0.45$ Myr ago from the nuclear region: this shows that the central engine has been recently launching material. Moreover, the SFH inferred by the SED fitting (Fig.\ref{fig:sfh}) further rules out a fossil origin for the observed outflow due to a past starburst, given that the travel times that we measure for the bi-conical winds are much smaller than the time elapsed since the last burst in star formation ($\sim 500$ Myr). The biconical and centrally-driven outflow geometry in MRG-M0150 also differs from that predicted for fossil AGN outflows, which hydrodynamical simulations show to become detached from the nucleus, lopsided and slow once the driving AGN has switched off \citep{Zubovas2023}.

Following \cite{rebecca_neutral24} and \cite{taylor26}, in Figure \ref{fig:energetics} we compare the wind energetics to the AGN bolometric luminosity for MRG-M0150. We also show the ionized outflows reported in the literature for Cosmic-Noon AGN -- as compiled by \cite{venturi_ganifs_2025}, converted to $n_e \sim 600$ cm$^{-3}$  (green triangles) -- as well as those observed in two massive quiescent galaxies at $z>2$ by \cite{bellinat_2024} and \cite{deugenio_fast-rotator_2024}, and in the prototype red geyser ``Akira'' \citep{cheung_suppressing_2016}. Energy-conserving AGN-driven outflows carry kinetic energy rates to up to $5\%$ of the AGN bolometric luminosity \citep{king_pounds_2015}, while their momentum rate $\dot{p}_{out} = L_{BOL}/c $ can be boosted by a factor of 5-20 due to interactions with the ISM \citep{fachur_gigeurts_2012}: we mark these theoretical limits with black dashed lines in the plots. As expected, we find substantially different energy and momentum rates when using \ha rather then [OIII] (see Sect.\ref{subsec:outflow_properties}), however in both cases we find that the system is broadly consistent with an extrapolation of the  \cite{venturi_ganifs_2025} outflows to lower $L_{\rm bol}$, and within the theoretical limits expected given the power of the central engine. MRG-M0150's outflow energy and momentum rates (from \ha) are consistent with those found in the other two quiescent galaxies at high-z, even though they host AGN $\sim2$ dex more luminous. Since outflow properties for both of these objects have been estimated using [OIII] emission, a substantial fraction of mass outflow in the ionized phase may be missed by relying only on higher-ionization lines. Thus, especially in the case of LL-AGN hosted in quiescent galaxies, mass outflow rates and related properties estimated from [OIII] should be treated as lower limits on the true values. We also note that the X-ray-based estimate of $L_{\rm bol}$ is about an order of magnitude higher than the [OIII]-based one (Sect.~3.4), so that the ratio between the outflow energetics and the AGN luminosity decreases by about one dex when the former is adopted.
We conclude that the current LL-AGN in MRG-M0150 is powerful enough to sustain the observed winds, and that a previous, more luminous AGN episode is not required by the energetics.

\subsection{Incidence of ionized gas outflows}
 
\begin{figure}
    \centering
    \includegraphics[width=0.9\linewidth]{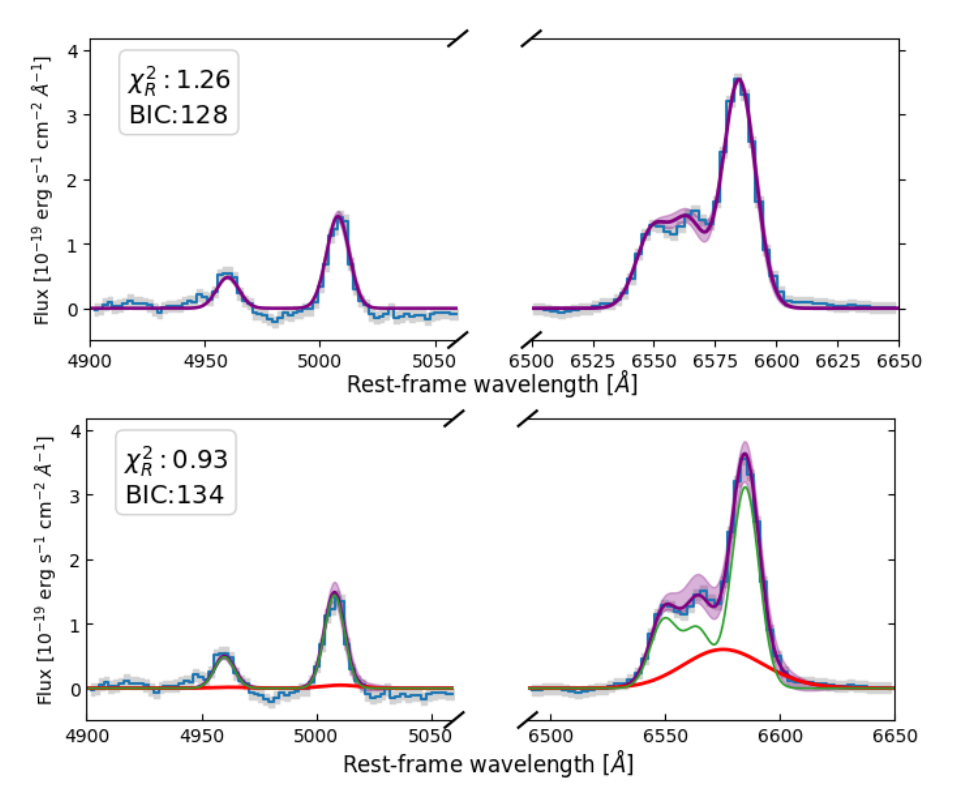}
    \caption{The difficult detection of outflows in unresolved spectra: [OIII] and [NII]/\ha emission lines observed (blue) and fitted (purple) in the simulated unresolved spectrum of MRG-M0150. Top: fit to the emission lines using single Gaussian components; bottom: fit using broad (red) + narrow (green) components. 
    }
    \label{fig:1d_simulatedspec}
\end{figure}

The incidence of AGN-driven ionized gas outflows in high-z massive quiescent galaxies appears to be relatively high. In particular, \cite{mypaper25} report an incidence of $\sim 20\%$, and comparable values are found by studies targeting outflows in the neutral phase \citep{matteos_article_sodium_2026,rebecca_neutral24}. This is higher  than  what is found in the local Universe -- in which the \textit{red geyser} galaxies constitute around $10\%$ of the entire quiescent population -- indicating that ionized gas outflows in high-z quiescent galaxies are not only more powerful but also more frequent than in their local counterparts.
Interestingly, our detection of the ionized outflows in MRG-M0150 is highly dependent on the availability of both \textit{lensed} and \textit{spatially resolved} data. To illustrate this, we simulate the unlensed and unresolved NIRSpec/MSA spectrum of the target with a slit loss of $\sim 50\%$ by extracting the integrated spectrum within 1 $R_e$ of the galaxy, corrected for magnification: when fitting the emission lines in this spectrum, the broad component is no longer statistically preferred over a single-component model (see the BIC values for both fits in Fig. \ref{fig:1d_simulatedspec}), implying that such outflows may remain undetected in unresolved observations of high-z galaxies. 
The same applies to the X-ray emission of the central LL-AGN, which is only marginally detected in each of the multiple images (Sect.~\ref{subsec:AGN_activity}).
This suggests that the relatively high incidence of outflows previously reported is a lower limit, supporting the idea that winds driven by a LL-AGN are a common phase in the evolution of massive quiescent galaxies at cosmic noon, while they become rarer in the present-day Universe.

\subsection{The fossil outflow scenario}

The relationship between the observed multiphase outflows in massive galaxies at $z>2$ and the low levels of AGN activity observed in the same sample remains unclear, as these AGN are rarely detected in X-rays and are primarily classified as active through rest-frame optical diagnostic diagrams \citep{mypaper25,skarbinkimaya26}. Studies of neutral-phase outflows in massive galaxies at high redshift have raised the idea that these constitute mostly \textit{relics} of a past starburst or fossils of past quasar events -- that is, short-lived bursts of black-hole activity that occurred at earlier epochs than the time of observation \citep{taylor26,almaini2025,Sun_2026_neutral}.
The remnant nature of outflows observed in high-z quenched galaxies hosting inactive or low-luminosity AGN is supported by hydrodynamical simulations \citep[e.g.][]{zubovas_outflows_2022} and energetics arguments \citep{almaini2025,taylor26}, which for many of the observed galaxies struggle to reconcile the lack of powerful AGN signatures with the energetics of the observed neutral-phase winds (while ionized-phase winds generally do not pose a problem, as shown in Fig.~\ref{fig:energetics}).
In this scenario, the low incidence of X-ray AGN among recently quenched galaxies at cosmic noon ($\sim6\%$; \citealt{almaini2025}) is reconciled with the high incidence of fast outflows if the luminous AGN phase is short-lived, while the outflow it launches remains visible for up to $\sim10$~Myr after the AGN has faded \citep{almaini2025}.
For example, \citet{taylor26} assume that the detection rate of neutral outflows/inflows in quenched galaxies from the EXCELS survey is set by the duty cycle timescales of the AGN hosted at their center, and conclude that the visibility window of both an X-ray active AGN and the outflow launched by the same event is extremely short, of the order of 1 Myr. 

It is useful to distinguish between two versions of the fossil outflow scenario. In its weaker form, it postulates that the outflows observed today were often launched by an AGN in a substantially different state from the one we observe, so that the present-day $L_{\rm bol}$ is poorly correlated with the wind energetics, even if a lower-luminosity AGN is still active. A single galaxy whose outflow does not exceed the energy budget of its current AGN, such as MRG-M0150, cannot test this version. In its stronger form, the AGN switches between ``on'' and ``off'' states, or at least undergoes drastic changes in $L_{\rm bol}$, and only the ``on'' phases are able to drive outflows. MRG-M0150 can be used as a test case for this second version.

Our kinematic modelling shows that the innermost portion of the wind was launched no more than $\sim0.45$~Myr ago, well within the 1~Myr time window in which an outflow and the X-ray bright AGN that launched it should be observable at the same time \citep{taylor26}. 
We do detect the AGN, but it is a LL-AGN with $\lambda_{Edd}\approx10^{-3}$ (Sect.~\ref{subsec:AGN_activity}) almost an order of magnitude below the Eddington ratio adopted by \citet{taylor26} as the threshold for X-ray detection and for AGN activity capable of driving outflows ($\lambda\gtrsim0.027$). This suggests that AGN in a lower accretion state can also launch and sustain galaxy-scale winds. Such systems are, however, difficult to identify: neither the AGN nor the ionized outflow in MRG-M0150 would have been detected without lensing and spatially resolved spectroscopy (Sect.~6.2), and in unlensed, unresolved samples, galaxies like MRG-M0150 would likely be counted as hosting neither. Accounting for this population may therefore lengthen the inferred overlap between the AGN duty cycle and outflow phases.

The energetics argument in favor of fossil outflows is also sensitive to how $L_{\rm bol}$ is estimated. In most quiescent galaxies at these redshifts, $L_{\rm bol}$ can only be derived from the [OIII] line, an estimate subject to significant systematic uncertainties related to dust obscuration and to the physical state of the AGN \citep[see the discussion in][]{taylor26}. In MRG-M0150, the X-ray-based $L_{\rm bol}$ is an order of magnitude higher than our [OIII]-based estimate (Sect.~3.4). If similar offsets apply to other LL-AGN in quiescent galaxies, their current AGN may account for a larger fraction of the observed outflow energetics than [OIII]-based estimates would suggest.

These considerations suggest an interpretation in which the AGN luminosity in quenched galaxies follows a continuous distribution that declines after quenching, rather than alternating between two distinct states. The AGN may flicker on short timescales, but its overall luminosity fades as the galaxy ages, as also illustrated by the scenario proposed by \citet{almaini2025}, in which AGN activity broadly traces the decline of the SFR after the starburst. Consistently, \citet{Valentino2026} find that the most luminous AGN in quenched galaxies at $z\sim1.5-4.5$ are preferentially found in the youngest post-starburst systems.
Observed shortly after quenching, galaxies such as COSMOS-11142 \citep{bellinat_2024} host massive outflows and luminous AGN. MRG-M0150 would represent a later stage, in which both the AGN and the outflow have faded below the sensitivity of typical unlensed observations. Finally, older quiescent galaxies show no outflows and, in some cases, evidence of gas accretion, as in the lensed galaxy MRG-M0138 \citep{gangula_resolved_2026}. Within this framework, MRG-M0150 shows that AGN just below the detection limits of current samples can still drive galaxy-scale outflows. Such weak activity may be common: stacking the X-ray emission of individually undetected post-starburst galaxies at cosmic noon reveals, on average, low-level AGN accretion ($L/L_{\rm Edd}\lesssim10^{-3}$; \citealt{almaini2025}).

\subsection{Quenching by AGN-driven outflows} \label{subsec:prospector}

A major open question regarding observations of gas outflows in high-redshift quiescent galaxies is whether the presence of these winds can be directly linked to the quenching of the galaxy itself. While it is not trivial to link the rapid quenching of MRG-M0150 in the last 0.5~Gyr with the observed ionized-phase outflows, the mass outflow rate of $0.6 \pm 0.2$~M$_\odot$~yr$^{-1}$ is higher than the current levels of SF, for which we have an upper limit of 0.30~M$_\odot$~yr$^{-1}$ from H$\alpha$, consistent with the SFR estimated by \texttt{Prospector} in the most recent time bin of [0--30]~Myr (SFR$=0.2^{+0.2}_{-0.8}$~M$_\odot$~yr$^{-1}$, see Fig.~7). Furthermore, the detection of outflowing neutral gas through the Na\,D absorption line in the central parts of the galaxy (Gangula et al., in prep.) implies a multi-phase nature of the observed winds. It has been observed that neutral-phase gas can reach mass outflow rates more than one order of magnitude higher than in the ionized phase, both at high redshift \citep{bellinat_2024} and in the local Universe \citep{roberts_multiphase_2020,avery_cooloutflows_2022}. Thus, the ionized-phase mass outflow rate that we measure is a strictly lower limit on the true rate of the multiphase wind in MRG-M0150.

The high velocities reached by the ionized outflow means that some of the gas may be capable of escaping the gravitational potential of the galaxy. The halo escape velocity is expected to be 2.5--3 times the rotational velocity of the galaxy \citep[e.g.][]{weiner_2009,swinbank_energetics_2019}. Using the rotational velocity of MRG-M0150 inferred from stellar kinematic modelling of $v_{\rm rot}\approx450$~km~s$^{-1}$ (Newman et al., in prep.), we estimate an escape velocity between $\sim 1100-1300 \rm~km~s^{-1}$. The highest velocities measured in the ionized outflow are comparable to these estimates, suggesting that a substantial amount of ionized outflowing gas may be able to escape the halo.
This type of ``ejective'' feedback from AGN-driven outflows has been invoked also in the post-starburst high-$z$ galaxies of \citet{bellinat_2024} and \citet{deugenio_fast-rotator_2024}, in which the ejection of cold gas at a faster rate than it can be converted into stars has been invoked as the main mechanism through which the galaxy has been quenched. While the fast-moving ionized phase does not transport as much mass as the cold phase, what we observe in MRG-M0150 and similar high-$z$ systems is qualitatively different than the ``maintenance mode'' feedback of local \textit{red geysers}, in which the ionized-phase gas is unable to escape the galaxy and instead keeps the galaxy quenched through the heating of gas in the halo \citep{cheung_suppressing_2016}.

Although the short launching timescales associated with the current wind prevents us from directly attributing the main quenching event to the observed outflow, the ongoing wind provides evidence that AGN feedback remains active well after star-formation has been quenched. 
Outflows such as the one observed here can be intermittent, and it is possible that similar, possibly more powerful episodes have occurred repeatedly since the last peak of star formation. The cumulative effect of recurrent AGN-driven winds, rather than a single event, may therefore have been ultimately responsible for quenching MRG-M0150.
The mass outflow rate we measure exceeds the residual SFR, suggesting that winds driven by a low-luminosity AGN can substantially affect the remaining gas reservoir and suppress subsequent star formation well after the initial ``blowout'' phase. Given that most of an active galaxy's lifetime is spent in a low-accretion state \citep{ho_nuclear_2008}, this type of long-term feedback may represent the key to maintenance of quiescence in massive systems over much longer timescales, potentially linking the population of recently quenched galaxies at cosmic noon to the massive quiescent population observed in the local Universe.

\section{Summary and conclusions} \label{sec:conclusions}

We have used JWST/NIRSpec IFU observations of the gravitationally lensed massive quiescent galaxy MRG-M0150 at $z=2.64$, from the INQUEST-JWST sample. Owing to the spatial resolution and sensitivity of NIRSpec's IFU and the magnification offered by the strong lensing, we are able for the first time to reconstruct in 3D the bi-conical ionized outflow of a massive quiescent galaxy at $z>2$, making it possible investigate the impact of AGN-driven outflows on star-formation quenching at high redshift. The main findings of this study are the following:

\begin{itemize}
\item \textbf{MRG-M0150 hosts a previously undetected low-luminosity AGN}, identified through LINER-like line ratios and marginal X-ray detections, with $L_{\rm bol}\approx3\times10^{43}-3\times10^{44}$~erg~s$^{-1}$ and $L/L_{\rm Edd}\sim10^{-3}$.
 
\item \textbf{The ionized gas traces a galaxy-wide, centrally driven bi-conical outflow} reaching $\sim1200$~km~s$^{-1}$, whose mass outflow rate is substantially underestimated when relying only on high-ionization lines such as [OIII] as opposed to low-ionization lines such as \ha.
 
\item \textbf{The outflow has been recently launched by the central LL-AGN}: its innermost portion was launched no more than 0.45~Myr ago, and the current AGN is powerful enough to sustain it without requiring a previous, more luminous episode.
 
\item \textbf{Neither the AGN nor the outflow would be detected without lensing and spatially resolved spectroscopy}, suggesting that similar systems are missed in unresolved samples, affecting incidence-based estimates of the visibility timescales of AGN and outflows.
 
\item \textbf{The ionized mass outflow rate ($0.6\pm0.2$~M$_\odot$~yr$^{-1}$), a lower limit on the multiphase wind, exceeds the residual SFR ($<0.30$~M$_\odot$~yr$^{-1}$)}, and part of the gas may escape the halo, implying ejective feedback is still active in the galaxy after the main quenching event. Combined with the relatively high incidence of such outflows, our results suggest that low-luminosity AGN activity may provide an important long-term mechanism for maintaining quiescence in massive galaxies at cosmic noon, potentially over cosmological timescales.
\end{itemize}

Overall, our work shows that AGN-driven galaxy-wide outflows can be sustained by low-luminosity AGN in massive quiescent galaxies at cosmic noon. These outflows may play a key role in the maintenance of quiescence over long timescales, and are sufficiently common to likely represent a major mechanism in preventing new star-formation in massive galaxies at this epoch. At the same time, our results highlight the importance of spatially resolved spectroscopy, combined with the magnification provided by gravitational lensing, for uncovering and constraining such outflows in the high-redshift Universe.

\begin{acknowledgements}
We acknowledge support by the ERC grant 101076080 ``Red Cardinal''. 
This work is based on observations made with the NASA/ESA/CSA James Webb Space Telescope. The data were obtained from the Mikulski Archive for Space Telescopes at the Space Telescope Science Institute, which is operated by the Association of Universities for Research in Astronomy, Inc., under NASA contract NAS5-03127 for JWST. These observations are associated with program GO-4903. Support for US investigators in program GO-4903 was provided by NASA through a grant from the Space Telescope Science Institute.
\end{acknowledgements}



\appendix

\section{Lens Model} \label{appendixlens}

The modeling was performed using imaging data of the foreground lensing cluster MACSJ0150.3-1005 at $z=0.365$ in the NIRCam F200W bands obtained as part of the same Cycle 3 JWST program that provided the spectroscopic data discussed in the text. Additionally, we identified the cluster members and measured their redshifts making use of the IFU spectroscopic data taken with the Multi Unit Spectroscopic Explorer (MUSE) at the Very Large Telescope (VLT), which covers a field of view of $1\arcmin \times 1\arcmin$.  MACSJ0150.3-1005 was observed in September 2019 for one OB of 48~min (program 0103.A-0777, PI:~A.~Edge). We reduced the MUSE data and measured the redshifts following the procedure described in \citet{Bolamperti2024}.

\begin{figure*}[h!]
    \centering
    \includegraphics[width=0.7\linewidth]{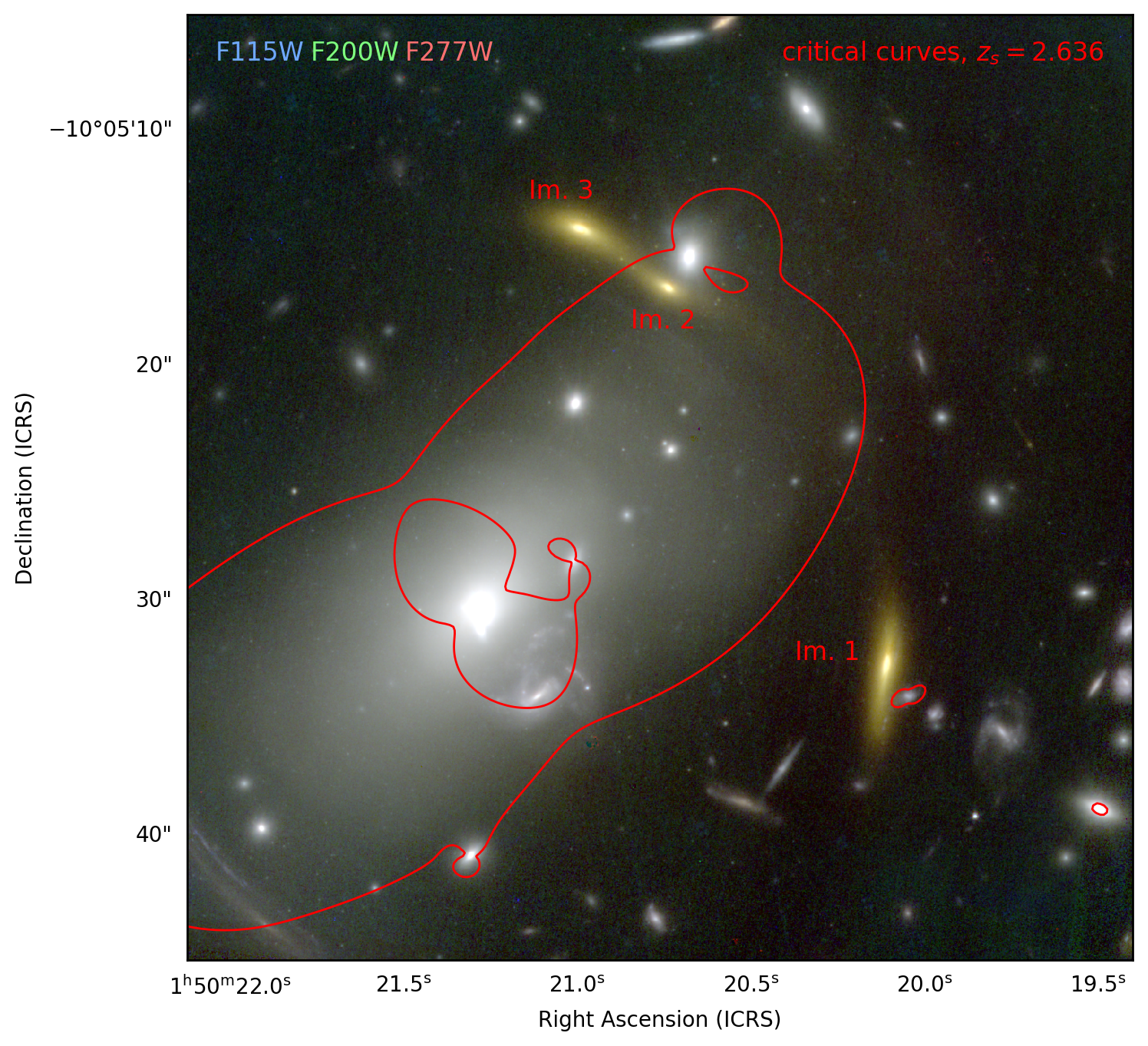} 
    \caption{Lens model of the MACS J0150 cluster. We show the JWST/NIRCam F115W/F200W/F277W images with an asinh stretch. The three images of MRG-M0150 are labeled; red curves trace the critical line.}
    \label{fig:cluster_and_lensmodel}
\end{figure*}

We construct the strong lensing model using the software \texttt{Gravity.jl} \citep{lombardi_gravity_2024}, which allows for the reconstruction of the total mass distribution of the cluster using a multiplane approach. The F200W image of the cluster and multiple images of MRG-M0150, together with the final lens model, can be seen in Figure~\ref{fig:cluster_and_lensmodel}.
The following mass components were included into the model. The main cluster-scale DM halo of the cluster, which is parametrized with a nearly isothermal ellipsoid (NIE) mass distribution characterized by six free parameters: the on-sky coordinates (two parameters), ellipticity, position angle, velocity dispersion, and core radius. An elliptical coreless Jaffe profile is used to describe one of the cluster galaxies lying close to Image 2 of MRG-M0150 along the line of sight. Contrary to the NIE profile, its core radius is substituted with a truncation radius, and the galaxy on-sky coordinates are fixed, for a total of 4 free parameters. A second elliptical coreless Jaffe profile with 4 free parameters is used to parameterize a galaxy located in the background at $z=0.4907$ (as determined from the NIRSpec IFU data) that is close in projection to Image 1 of MRG-M0150. This galaxy is included in the model employing the multiplane approach possible within \texttt{Gravity.jl}. Finally, 18 additional spherical Jaffe profiles are used to describe 18 spectroscopically confirmed cluster galaxies identified from the MUSE data, including the BCG. For these galaxies, the central velocity dispersion $\sigma$ and truncation radius $r_{cut}$ are scaled according to the galaxy luminosity in the HST/F160W band assuming the following scaling relations:
\begin{align}
     &\sigma^{gal} = \sigma^{ref}\left(\frac{L_{gal}}{L_{ref}} \right)^\alpha, \\
     &r^{gal}_{cut} = r^{ref}_{cut} \left(\frac{L_{gal}}{L_{ref}} \right)^\beta;
\end{align}
where $L_{gal}$ is the F200W luminosity of the cluster member. In the model we assume $\alpha=0.25$ and $\beta=0.50$, which is equivalent to assuming a constant mass-to-light ratio across the cluster population \citep[see][]{Bergamini2019, Oguri2010}, and thus our only free parameters for the cluster members' modeling are $\sigma^{ref}$ and $r^{ref}_{cut}$ . 

To constrain the lens model with the data we adopt a parametric extended modeling approach. In addition to modeling the lens, we also model the brightness distribution of the multiply-imaged source (that is, MRG-M0150) using a S\'{e}rsic profile: this adds as free parameters the coordinates, the effective radius, total flux, the ellipticity and the position angle of the source. 

In total the model contains 21 free parameters and is constrained by 22781 pixels. The model is optimized to reproduce the observed multiple images of the source; our goodness-of-fit estimation is given by the residuals between the model-predicted and observed shape of the multiple images. To sample the full parameter space of the models employed in this work, we carry out the \texttt{Gravity.jl} fits using a Markov Chain Monte Carlo (MCMC) algorithm to obtain posterior distributions.

\section{Kinematic modeling with MOKA3D} \label{appendixmoka}
We perform 3D kinematics fitting of the ionized gas emission in MRG-M0150 using the de-lensed maps of ionized gas surface brightness in the source plane, along with the line profiles in each spaxel, in order to to reconstruct the geometry and velocity structure of the outflow. 
The MOKA3D code requires the observations' grid to be regularly sampled and to convolve the model with the PSF of the instrument. Since we have to work on the source plane, we need to deal with the distortion of the NIRSpec/IFU spaxels and PSF caused by the application of the lens model, and to make them suitable to be fed to MOKA3D. 
We resample the source-plane cubes onto a regular grid of pixels with area much smaller than the minimum size of the original pixels in the following way: we first construct the regular grid, then for each of the original pixels we assign the surface signal and noise values to the closest pixel in the new grid, and put NaN values elsewhere. MOKA3D performs a convolution of the model with the PSF of the instrument; in this way we can ignore any correlated noise deriving from the resampling as long as the PSF convolution is applied, since the dimension of the PSF is much greater than the pixels in the new grid. In order to obtain the PSF on the source plane, we apply the same lens model to the instrumental PSF of NIRSpec/IFU. The 2D PSF of NIRSpec/IFU is first positioned to overlap the centroid of the galaxy in the image plane and then de-lensed onto the source plane.

\begin{figure}
    \centering
    \includegraphics[width=0.9\linewidth]{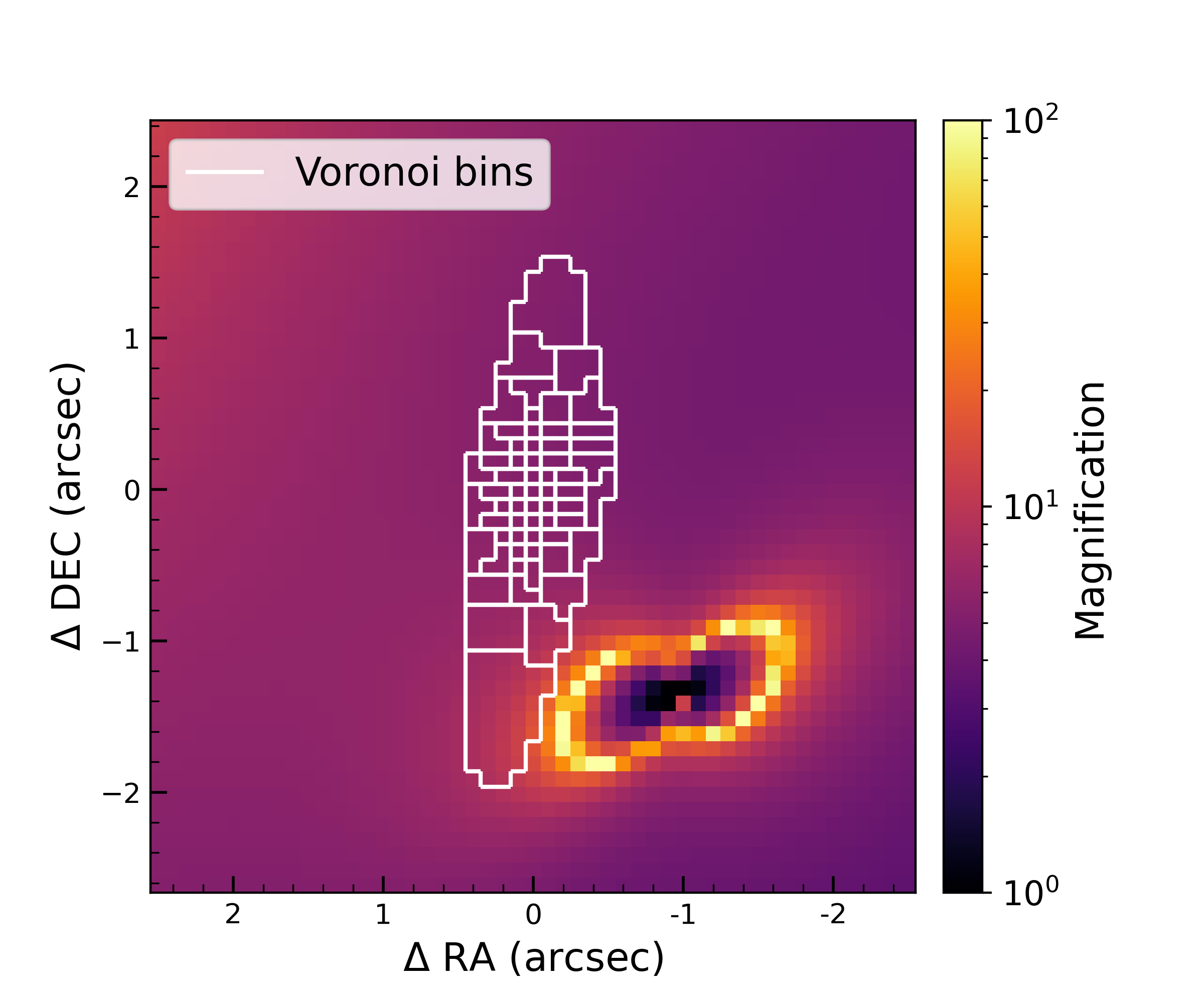}
    \caption{Magnification map near Image 1, with contours of the analyzed Voronoi bins of MRG-M0150 overlayed in white.}
    \label{fig:magnification_map}
\end{figure}

\begin{figure}
    \centering
    \includegraphics[width=\linewidth]{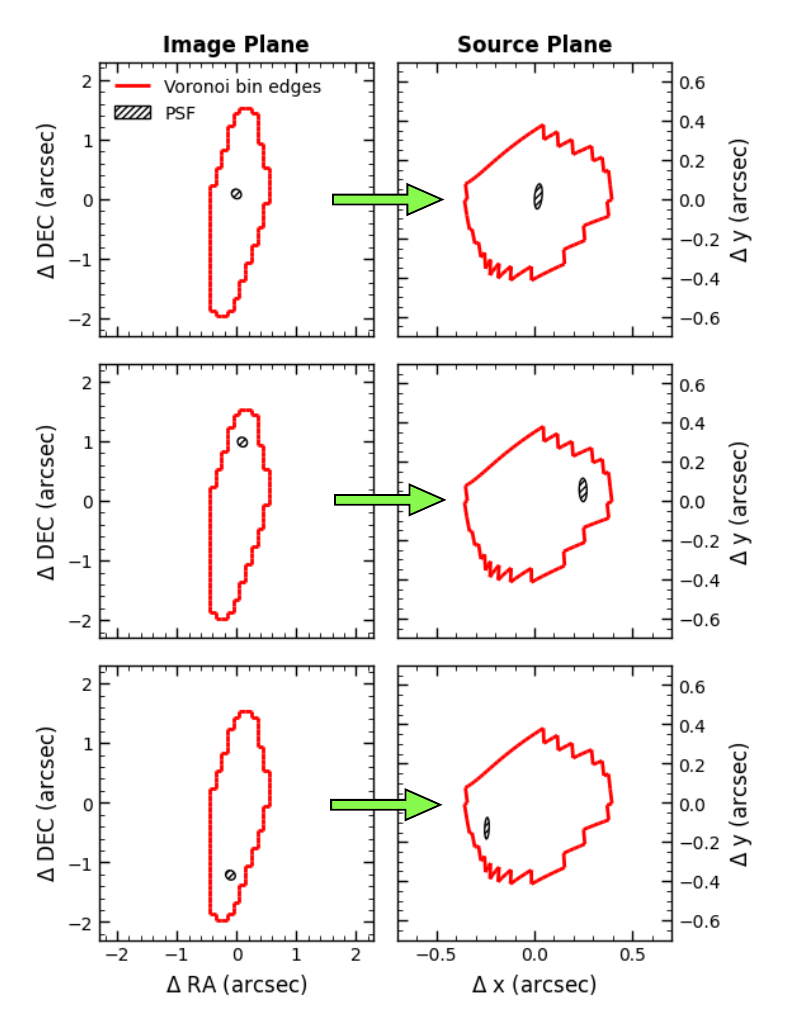}
    \caption{NIRSpec/IFU PSF positioned in three different points of the galaxy disk (left column) and projected onto the source plane using the lens model (right column). The red contours mark the external perimeter of the analyzed Voronoi bins, both in the image plane and in the source plane.}
    \label{fig:PSF}
\end{figure}

We verify that the magnification does not vary greatly across the analyzed spaxels (as visible in Fig.\ref{fig:magnification_map}), and thus we should be able to apply the ``centrally de-lensed'' PSF across the entire cube. However, spaxels belonging to the bin that is closer to the critical line are experiencing higher levels of magnification than the rest of the galaxy ($\mu \sim 12$ instead of $\mu \sim 4$): applying the same "centrally de-lensed" PSF to these spaxels may not be accurate, and may affect the resulting 3D reconstruction of the outflow by the fitting code. To verify that this is not the case, we show in Fig. \ref{fig:PSF} the shape of NIRSpec/IFU's PSF overlapped onto the galaxy's centroid in the image plane (upper left) and how that gets distorted when projecting it onto the source plane (upper right). In the central row of panels we verify that changing the PSF position away from the galaxy centroid in the image plane (left) results in a source-plane PSF that does not differ significantly from the centrally de-lensed one. Fitting both source-plane PSFs with an anisotropic 2D Gaussian model, we find only a small variation of the FWHM of the dislocated PSF's of 8\% along the minor axis and 0.2\% along the major axis with respect to the centrally-located one. However, when positioning the PSF onto spaxels nearing the critical line in the image plane (last row), the magnification jumps to 3x the median value elsewhere and the PSF gets highly distorted along a preferential axis, with a variation in FWHM of 60\% along the minor axis and 7\% along the major one with respect to the center. Nonetheless, spaxels affected by the higher magnification form a small minority of all the spaxels analyzed, on top of being lower in signal-to-noise with respect to the central parts of the galaxy (see Fig.\ref{fig:general image}). For this reason, the spatial bins are large near the critical line, and convolution by the PSF has only a very small effect. Thus, we decide to apply the centrally-projected de-lensed PSF to the entire resampled and de-lensed cubes to be fitted with MOKA3D, neglecting any change in magnification levels. The fact that we get consistent results from the MOKA3D fitting of the \ha emission -- detected in all the analyzed bins  -- and the [OIII] line -- which is not detected in the external parts of the galaxy, thus the fit does not include any of the greatly magnified spaxels -- is evidence that our central PSF approximation is not significantly affecting the 3D kinematics fitting results.  

\begin{figure}
    \centering
    \includegraphics[width=\linewidth]{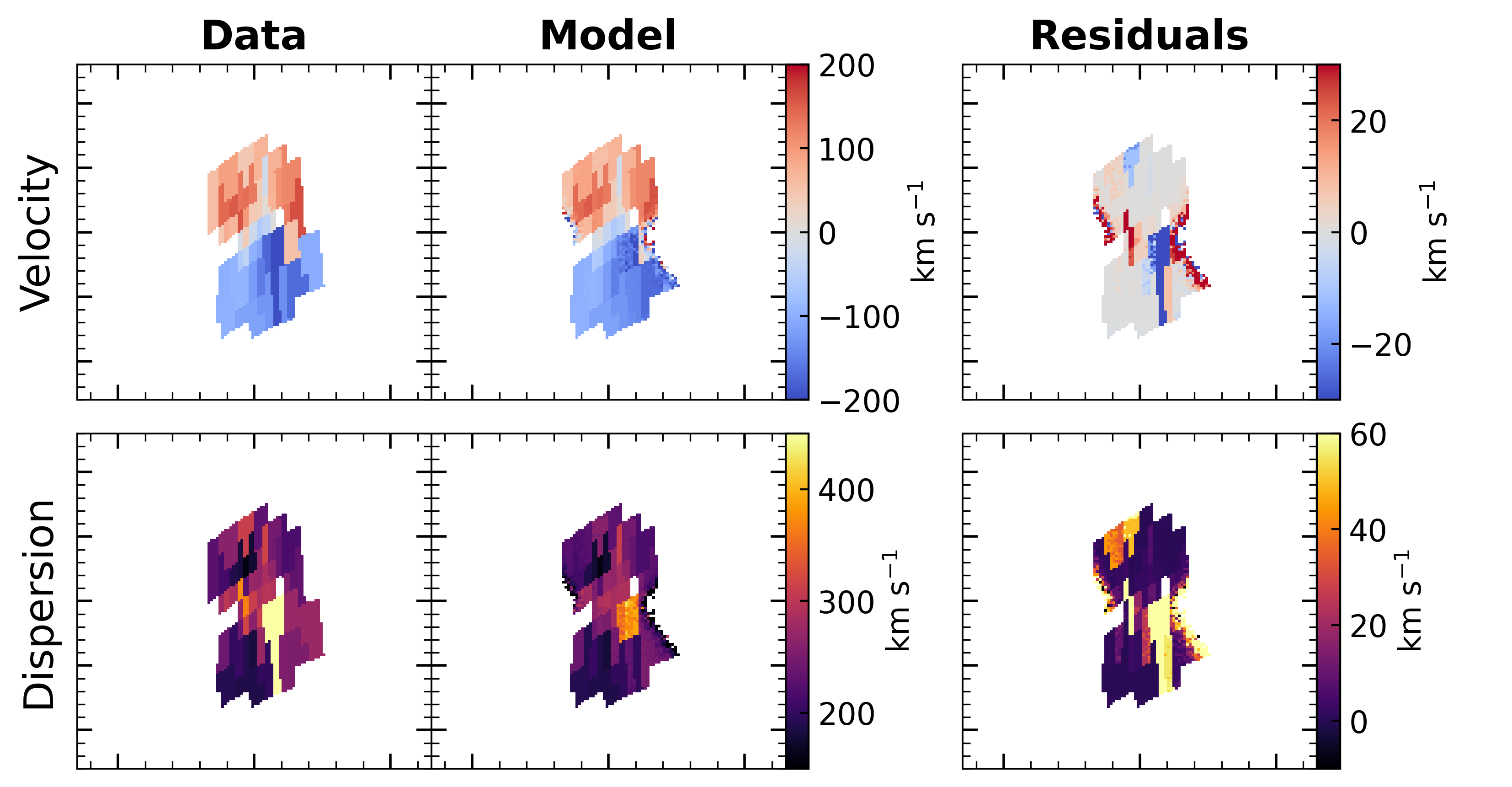}
    \caption{Comparison between the observed and model gas velocity (moment 1) and dispersion (moment 2) maps from the [OIII] line profiles.}
    \label{fig:moka3d_results_o3}
\end{figure}

\begin{figure}
    \centering
    \includegraphics[width=\linewidth]{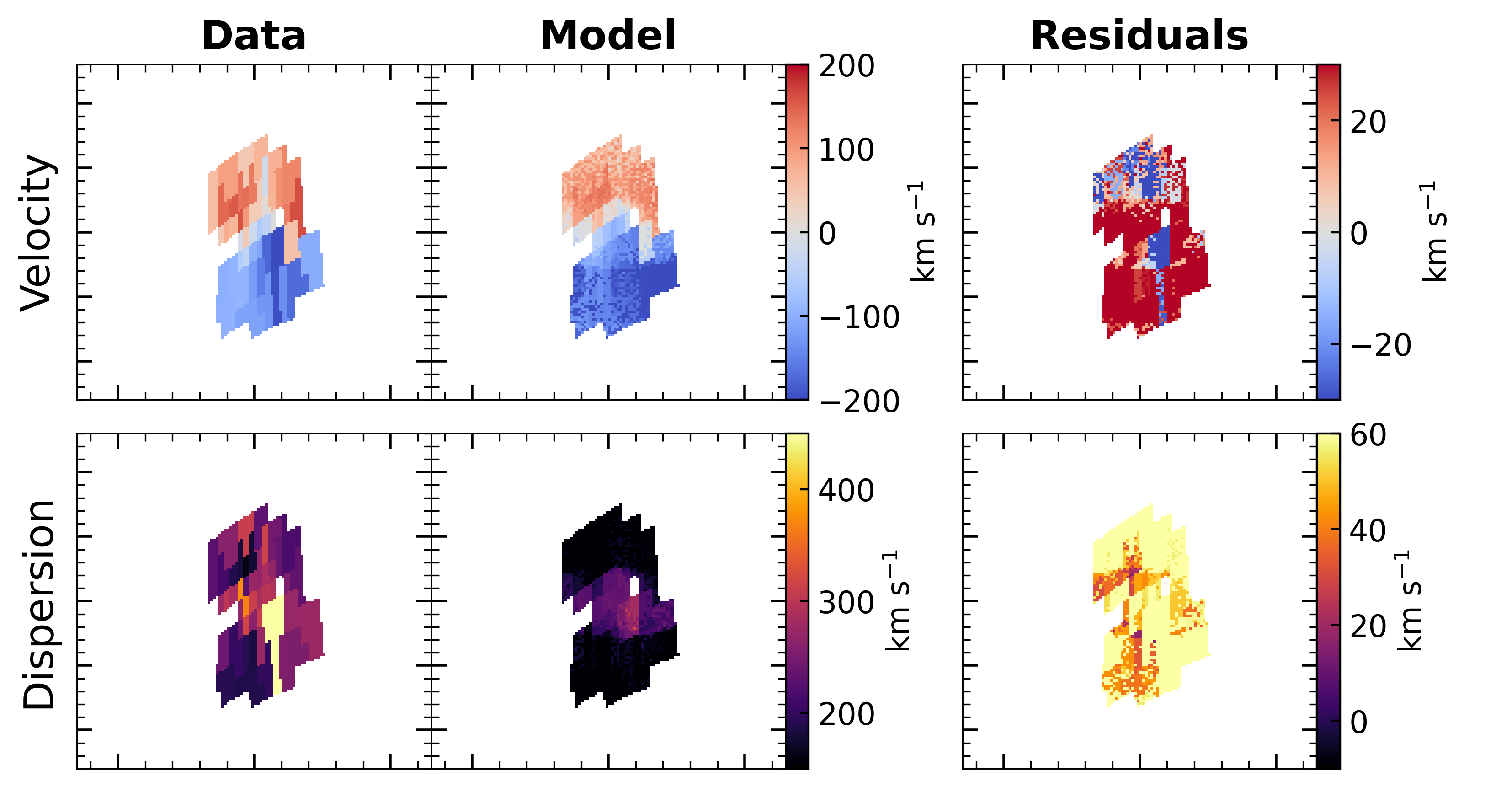}
    \caption{Comparison between the observed and model gas velocity (moment 1) and dispersion (moment 2) maps from the [OIII] line profiles, assuming a fixed inclination of $30^\circ$.}
    \label{fig:moka3d_results_forcedinclination}
\end{figure}

The results of the fit using the [OIII] emission line is visible in Fig.\ref{fig:moka3d_results_o3}. We also show, in Fig.~\ref{fig:moka3d_results_forcedinclination}, the results of the [OIII] kinematics fit using the model with forced inclination of $30^\circ$ for the bi-cone discussed in Sect.\ref{sec:ionizeedgasoutflow}: clearly, residuals are much higher than what is found in the inclination-free case and the model does not perform as well in reproducing the data.

\bibliographystyle{aa}
\bibliography{articolo_m0150}{}

@ARTICLE{skarbinkimaya26,
       author = {{Skarbinski}, Maya and {Rowlands}, Kate and {Alatalo}, Katherine and {Wild}, Vivienne and {Carnall}, Adam C. and {Almaini}, Omar and {Maltby}, David and {de Lisle}, Thomas and {Heckman}, Timothy and {Begley}, Ryan and {Cullen}, Fergus and {Dunlop}, James S. and {Hewitt}, Guillaume and {Leung}, Ho-Hin and {McLeod}, Derek and {McLure}, Ross and {Otter}, Justin Atsushi and {Patil}, Pallavi and {Petric}, Andreea and {Shapley}, Alice E. and {Stevenson}, Struan and {Taylor}, Elizabeth},
        title = "{The JWST EXCELS Survey: Insights into the Nature of Quenching at Cosmic Noon}",
      journal = {\apj},
         year = 2026,
        month = apr,
       volume = {1000},
       number = {2},
          eid = {191},
        pages = {191},
          doi = {10.3847/1538-4357/ae459d},
archivePrefix = {arXiv},
       eprint = {2509.18278},
 primaryClass = {astro-ph.GA},
       adsurl = {https://ui.adsabs.harvard.edu/abs/2026ApJ..1000..191S}
}

@ARTICLE{kim_smaug_2020,
       author = {{Kim}, Chang-Goo and {Ostriker}, Eve C. and {Somerville}, Rachel S. and {Bryan}, Greg L. and {Fielding}, Drummond B. and {Forbes}, John C. and {Hayward}, Christopher C. and {Hernquist}, Lars and {Pandya}, Viraj},
        title = "{First Results from SMAUG: Characterization of Multiphase Galactic Outflows from a Suite of Local Star-forming Galactic Disk Simulations}",
      journal = {\apj},
         year = 2020,
        month = sep,
       volume = {900},
       number = {1},
          eid = {61},
        pages = {61},
          doi = {10.3847/1538-4357/aba962},
archivePrefix = {arXiv},
       eprint = {2006.16315},
 primaryClass = {astro-ph.GA},
       adsurl = {https://ui.adsabs.harvard.edu/abs/2020ApJ...900...61K}
}

@ARTICLE{mypaper25,
       author = {{Bugiani}, Letizia and {Belli}, Sirio and {Park}, Minjung and {Davies}, Rebecca L. and {Mendel}, J. Trevor and {Johnson}, Benjamin D. and {Khoram}, Amir H. and {Benton}, Chlo{\"e} and {Cimatti}, Andrea and {Conroy}, Charlie and {Emami}, Razieh and {Leja}, Joel and {Li}, Yijia and {Maheson}, Gabriel and {Mathews}, Elijah P. and {Naidu}, Rohan P. and {Nelson}, Erica J. and {Tacchella}, Sandro and {Terrazas}, Bryan A. and {Weinberger}, Rainer},
        title = "{Active Galactic Nucleus Feedback in Quiescent Galaxies at Cosmic Noon Traced by Ionized Gas Emission}",
      journal = {\apj},
         year = 2025,
        month = mar,
       volume = {981},
       number = {1},
          eid = {25},
        pages = {25},
          doi = {10.3847/1538-4357/adaeaf},
archivePrefix = {arXiv},
       eprint = {2406.08547},
 primaryClass = {astro-ph.GA},
       adsurl = {https://ui.adsabs.harvard.edu/abs/2025ApJ...981...25B}
}

@ARTICLE{taylor26,
       author = {{Taylor}, Elizabeth and {Carnall}, Adam C. and {Maltby}, David and {Almaini}, Omar and {Leung}, Ho-Hin and {Stevenson}, Struan D. and {Negri}, Andrea and {Cullen}, Fergus and {Wild}, Vivienne and {McLure}, Ross J. and {Shapley}, Alice E. and {Arellano-C{\'o}rdova}, Karla Z. and {Begley}, Ryan and {Bondestam}, Cecilia and {de Lisle}, Thomas and {Donnan}, Callum T. and {Dunlop}, James S. and {Ellis}, Richard and {Hewitt}, Guillaume and {Koekemoer}, Anton M. and {Liu}, Feng-Yuan Frey and {McLeod}, Derek J. and {Rowlands}, Kate and {Sanders}, Ryan L. and {Scholte}, Dirk and {Skarbinski}, Maya and {Stanton}, Thomas M.},
        title = "{The JWST EXCELS survey: outflows in 1.5 < z < 5 quiescent and recently quenched galaxies are likely relics from episodic AGN activity}",
      journal = {\mnras},
         year = 2026,
        month = jul,
       volume = {549},
       number = {3},
          eid = {stag918},
        pages = {stag918},
          doi = {10.1093/mnras/stag918},
archivePrefix = {arXiv},
       eprint = {2601.02269},
 primaryClass = {astro-ph.GA},
       adsurl = {https://ui.adsabs.harvard.edu/abs/2026MNRAS.549ag918T}
}

@ARTICLE{heckman_growth_2004,
       author = {{Heckman}, Timothy M. and {Kauffmann}, Guinevere and {Brinchmann}, Jarle and {Charlot}, St{\'e}phane and {Tremonti}, Christy and {White}, Simon D.~M.},
        title = "{Present-Day Growth of Black Holes and Bulges: The Sloan Digital Sky Survey Perspective}",
      journal = {\apj},
         year = 2004,
        month = sep,
       volume = {613},
       number = {1},
        pages = {109-118},
          doi = {10.1086/422872},
archivePrefix = {arXiv},
       eprint = {astro-ph/0406218},
 primaryClass = {astro-ph},
       adsurl = {https://ui.adsabs.harvard.edu/abs/2004ApJ...613..109H}
}

@article{ho_nuclear_2008,
	title = {Nuclear {Activity} in {Nearby} {Galaxies}},
	volume = {46},
	issn = {0066-4146},
	url = {https://www.annualreviews.org/content/journals/10.1146/annurev.astro.45.051806.110546},
	doi = {10.1146/annurev.astro.45.051806.110546},
	number = {Volume 46, 2008},
	journal = {Annual Review of Astronomy and Astrophysics},
	author = {Ho, L. C.},
	month = sep,
	year = {2008},
	pages = {475--539},
}

@ARTICLE{mcconnell_blackholes_2011,
       author = {{McConnell}, Nicholas J. and {Ma}, Chung-Pei and {Gebhardt}, Karl and {Wright}, Shelley A. and {Murphy}, Jeremy D. and {Lauer}, Tod R. and {Graham}, James R. and {Richstone}, Douglas O.},
        title = "{Two ten-billion-solar-mass black holes at the centres of giant elliptical galaxies}",
      journal = {\nat},
         year = 2011,
        month = dec,
       volume = {480},
       number = {7376},
        pages = {215-218},
          doi = {10.1038/nature10636},
archivePrefix = {arXiv},
       eprint = {1112.1078},
 primaryClass = {astro-ph.CO},
       adsurl = {https://ui.adsabs.harvard.edu/abs/2011Natur.480..215M}
}

@ARTICLE{dimatteo_2005,
       author = {{Di Matteo}, Tiziana and {Springel}, Volker and {Hernquist}, Lars},
        title = "{Energy input from quasars regulates the growth and activity of black holes and their host galaxies}",
      journal = {\nat},
         year = 2005,
        month = feb,
       volume = {433},
       number = {7026},
        pages = {604-607},
          doi = {10.1038/nature03335},
archivePrefix = {arXiv},
       eprint = {astro-ph/0502199},
 primaryClass = {astro-ph},
       adsurl = {https://ui.adsabs.harvard.edu/abs/2005Natur.433..604D}
}

@ARTICLE{muzzin_2013-cosmosultravista,
       author = {{Muzzin}, Adam and {Marchesini}, Danilo and {Stefanon}, Mauro and {Franx}, Marijn and {McCracken}, Henry J. and {Milvang-Jensen}, Bo and {Dunlop}, James S. and {Fynbo}, J.~P.~U. and {Brammer}, Gabriel and {Labb{\'e}}, Ivo and {van Dokkum}, Pieter G.},
        title = "{The Evolution of the Stellar Mass Functions of Star-forming and Quiescent Galaxies to z = 4 from the COSMOS/UltraVISTA Survey}",
      journal = {\apj},
         year = 2013,
        month = nov,
       volume = {777},
       number = {1},
          eid = {18},
        pages = {18},
          doi = {10.1088/0004-637X/777/1/18},
archivePrefix = {arXiv},
       eprint = {1303.4409},
 primaryClass = {astro-ph.CO},
       adsurl = {https://ui.adsabs.harvard.edu/abs/2013ApJ...777...18M}
}

@ARTICLE{weiner_2009,
       author = {{Weiner}, Benjamin J. and {Coil}, Alison L. and {Prochaska}, Jason X. and {Newman}, Jeffrey A. and {Cooper}, Michael C. and {Bundy}, Kevin and {Conselice}, Christopher J. and {Dutton}, Aaron A. and {Faber}, S.~M. and {Koo}, David C. and {Lotz}, Jennifer M. and {Rieke}, G.~H. and {Rubin}, K.~H.~R.},
        title = "{Ubiquitous Outflows in DEEP2 Spectra of Star-Forming Galaxies at z = 1.4}",
      journal = {\apj},
         year = 2009,
        month = feb,
       volume = {692},
       number = {1},
        pages = {187-211},
          doi = {10.1088/0004-637X/692/1/187},
archivePrefix = {arXiv},
       eprint = {0804.4686},
 primaryClass = {astro-ph},
       adsurl = {https://ui.adsabs.harvard.edu/abs/2009ApJ...692..187W}
}

@ARTICLE{wupofeng_2025,
       author = {{Wu}, Po-Feng},
        title = "{Ejective Feedback as a Quenching Mechanism in the First 1.5 Billion Years of the Universe: Detection of Neutral Gas Outflow in a z = 4 Recently Quenched Galaxy}",
      journal = {\apj},
         year = 2025,
        month = jan,
       volume = {978},
       number = {2},
          eid = {131},
        pages = {131},
          doi = {10.3847/1538-4357/ad98ef},
archivePrefix = {arXiv},
       eprint = {2409.00471},
 primaryClass = {astro-ph.GA},
       adsurl = {https://ui.adsabs.harvard.edu/abs/2025ApJ...978..131W}
}

@ARTICLE{fabian2012,
       author = {{Fabian}, A.~C.},
        title = "{Observational Evidence of Active Galactic Nuclei Feedback}",
      journal = {\araa},
         year = 2012,
        month = sep,
       volume = {50},
        pages = {455-489},
          doi = {10.1146/annurev-astro-081811-125521},
archivePrefix = {arXiv},
       eprint = {1204.4114},
 primaryClass = {astro-ph.CO},
       adsurl = {https://ui.adsabs.harvard.edu/abs/2012ARA&A..50..455F}
}

@ARTICLE{hopkins_2006,
       author = {{Hopkins}, Philip F. and {Hernquist}, Lars and {Cox}, Thomas J. and {Di Matteo}, Tiziana and {Robertson}, Brant and {Springel}, Volker},
        title = "{A Unified, Merger-driven Model of the Origin of Starbursts, Quasars, the Cosmic X-Ray Background, Supermassive Black Holes, and Galaxy Spheroids}",
      journal = {\apjs},
         year = 2006,
        month = mar,
       volume = {163},
       number = {1},
        pages = {1-49},
          doi = {10.1086/499298},
archivePrefix = {arXiv},
       eprint = {astro-ph/0506398},
 primaryClass = {astro-ph},
       adsurl = {https://ui.adsabs.harvard.edu/abs/2006ApJS..163....1H}
}

@ARTICLE{bower_2006,
       author = {{Bower}, R.~G. and {Benson}, A.~J. and {Malbon}, R. and {Helly}, J.~C. and {Frenk}, C.~S. and {Baugh}, C.~M. and {Cole}, S. and {Lacey}, C.~G.},
        title = "{Breaking the hierarchy of galaxy formation}",
      journal = {\mnras},
         year = 2006,
        month = aug,
       volume = {370},
       number = {2},
        pages = {645-655},
          doi = {10.1111/j.1365-2966.2006.10519.x},
archivePrefix = {arXiv},
       eprint = {astro-ph/0511338},
 primaryClass = {astro-ph},
       adsurl = {https://ui.adsabs.harvard.edu/abs/2006MNRAS.370..645B}
}

@ARTICLE{somerville_2008,
       author = {{Somerville}, Rachel S. and {Hopkins}, Philip F. and {Cox}, Thomas J. and {Robertson}, Brant E. and {Hernquist}, Lars},
        title = "{A semi-analytic model for the co-evolution of galaxies, black holes and active galactic nuclei}",
      journal = {\mnras},
         year = 2008,
        month = dec,
       volume = {391},
       number = {2},
        pages = {481-506},
          doi = {10.1111/j.1365-2966.2008.13805.x},
archivePrefix = {arXiv},
       eprint = {0808.1227},
 primaryClass = {astro-ph},
       adsurl = {https://ui.adsabs.harvard.edu/abs/2008MNRAS.391..481S}
}

@ARTICLE{dave_2019,
       author = {{Dav{\'e}}, Romeel and {Angl{\'e}s-Alc{\'a}zar}, Daniel and {Narayanan}, Desika and {Li}, Qi and {Rafieferantsoa}, Mika H. and {Appleby}, Sarah},
        title = "{SIMBA: Cosmological simulations with black hole growth and feedback}",
      journal = {\mnras},
         year = 2019,
        month = jun,
       volume = {486},
       number = {2},
        pages = {2827-2849},
          doi = {10.1093/mnras/stz937},
archivePrefix = {arXiv},
       eprint = {1901.10203},
 primaryClass = {astro-ph.GA},
       adsurl = {https://ui.adsabs.harvard.edu/abs/2019MNRAS.486.2827D}
}

@ARTICLE{fachur_gigeurts_2012,
       author = {{Faucher-Gigu{\`e}re}, Claude-Andr{\'e} and {Quataert}, Eliot},
        title = "{The physics of galactic winds driven by active galactic nuclei}",
      journal = {\mnras},
         year = 2012,
        month = sep,
       volume = {425},
       number = {1},
        pages = {605-622},
          doi = {10.1111/j.1365-2966.2012.21512.x},
archivePrefix = {arXiv},
       eprint = {1204.2547},
 primaryClass = {astro-ph.CO},
       adsurl = {https://ui.adsabs.harvard.edu/abs/2012MNRAS.425..605F}
}

@ARTICLE{king_pounds_2015,
       author = {{King}, Andrew and {Pounds}, Ken},
        title = "{Powerful Outflows and Feedback from Active Galactic Nuclei}",
      journal = {\araa},
         year = 2015,
        month = aug,
       volume = {53},
        pages = {115-154},
          doi = {10.1146/annurev-astro-082214-122316},
archivePrefix = {arXiv},
       eprint = {1503.05206},
 primaryClass = {astro-ph.GA},
       adsurl = {https://ui.adsabs.harvard.edu/abs/2015ARA&A..53..115K}
}

@ARTICLE{hewitt_monsters_2026,
       author = {{Hewitt}, Guillaume and {Almaini}, Omar and {Maltby}, David and {Chapman}, Emma and {de Lisle}, Thomas and {Patil}, Pallavi and {Rowlands}, Kate and {Skarbinski}, Maya and {Taylor}, Elizabeth and {Wild}, Vivienne and {Carnall}, Adam C. and {Dunlop}, James S. and {Grogin}, Norman and {Koekemoer}, Anton M. and {McLeod}, Derek J. and {P{\'e}rez-Gonz{\'a}lez}, Pablo G.},
        title = "{No Hidden Monsters: Probing recently quenched galaxies for obscured AGN with JWST-PRIMER MIRI and NIRCam}",
      journal = {\mnras},
         year = 2026,
        month = aug,
       volume = {550},
       number = {2},
          eid = {stag1180},
        pages = {stag1180},
          doi = {10.1093/mnras/stag1180},
archivePrefix = {arXiv},
       eprint = {2606.20800},
 primaryClass = {astro-ph.GA},
       adsurl = {https://ui.adsabs.harvard.edu/abs/2026MNRAS.550g1180H}
}

@ARTICLE{cidfernandes_2010,
       author = {{Cid Fernandes}, R. and {Stasi{\'n}ska}, G. and {Schlickmann}, M.~S. and {Mateus}, A. and {Vale Asari}, N. and {Schoenell}, W. and {Sodr{\'e}}, L.},
        title = "{Alternative diagnostic diagrams and the `forgotten' population of weak line galaxies in the SDSS}",
      journal = {\mnras},
         year = 2010,
        month = apr,
       volume = {403},
       number = {2},
        pages = {1036-1053},
          doi = {10.1111/j.1365-2966.2009.16185.x},
archivePrefix = {arXiv},
       eprint = {0912.1643},
 primaryClass = {astro-ph.CO},
       adsurl = {https://ui.adsabs.harvard.edu/abs/2010MNRAS.403.1036C}
}

@ARTICLE{dopita_2015b,
       author = {{Dopita}, Michael A. and {Ho}, I.-Ting and {Dressel}, Linda L. and {Sutherland}, Ralph and {Kewley}, Lisa and {Davies}, Rebecca and {Hampton}, Elise and {Shastri}, Prajval and {Kharb}, Preeti and {Jose}, Jessy and {Bhatt}, Harish and {Ramya}, S. and {Scharw{\"a}chter}, Julia and {Jin}, Chichuan and {Banfield}, Julie and {Zaw}, Ingyin and {James}, Bethan and {Juneau}, St{\'e}phanie and {Srivastava}, Shweta},
        title = "{Probing the Physics of Narrow-line Regions in Active Galaxies. III. Accretion and Cocoon Shocks in the LINER NGC 1052}",
      journal = {\apj},
         year = 2015,
        month = mar,
       volume = {801},
       number = {1},
          eid = {42},
        pages = {42},
          doi = {10.1088/0004-637X/801/1/42},
archivePrefix = {arXiv},
       eprint = {1501.02507},
 primaryClass = {astro-ph.GA},
       adsurl = {https://ui.adsabs.harvard.edu/abs/2015ApJ...801...42D}
}

@ARTICLE{belfiore_2016,
       author = {{Belfiore}, Francesco and {Maiolino}, Roberto and {Maraston}, Claudia and {Emsellem}, Eric and {Bershady}, Matthew A. and {Masters}, Karen L. and {Yan}, Renbin and {Bizyaev}, Dmitry and {Boquien}, M{\'e}d{\'e}ric and {Brownstein}, Joel R. and {Bundy}, Kevin and {Drory}, Niv and {Heckman}, Timothy M. and {Law}, David R. and {Roman-Lopes}, Alexandre and {Pan}, Kaike and {Stanghellini}, Letizia and {Thomas}, Daniel and {Weijmans}, Anne-Marie and {Westfall}, Kyle B.},
        title = "{SDSS IV MaNGA - spatially resolved diagnostic diagrams: a proof that many galaxies are LIERs}",
      journal = {\mnras},
         year = 2016,
        month = sep,
       volume = {461},
       number = {3},
        pages = {3111-3134},
          doi = {10.1093/mnras/stw1234},
archivePrefix = {arXiv},
       eprint = {1605.07189},
 primaryClass = {astro-ph.GA},
       adsurl = {https://ui.adsabs.harvard.edu/abs/2016MNRAS.461.3111B}
}

@ARTICLE{miles_lib,
       author = {{Falc{\'o}n-Barroso}, J. and {S{\'a}nchez-Bl{\'a}zquez}, P. and {Vazdekis}, A. and {Ricciardelli}, E. and {Cardiel}, N. and {Cenarro}, A.~J. and {Gorgas}, J. and {Peletier}, R.~F.},
        title = "{An updated MILES stellar library and stellar population models}",
      journal = {\aap},
         year = 2011,
        month = aug,
       volume = {532},
          eid = {A95},
        pages = {A95},
          doi = {10.1051/0004-6361/201116842},
archivePrefix = {arXiv},
       eprint = {1107.2303},
 primaryClass = {astro-ph.CO},
       adsurl = {https://ui.adsabs.harvard.edu/abs/2011A&A...532A..95F}
}

@ARTICLE{mist_isochrones,
       author = {{Dotter}, Aaron},
        title = "{MESA Isochrones and Stellar Tracks (MIST) 0: Methods for the Construction of Stellar Isochrones}",
      journal = {\apjs},
         year = 2016,
        month = jan,
       volume = {222},
       number = {1},
          eid = {8},
        pages = {8},
          doi = {10.3847/0067-0049/222/1/8},
archivePrefix = {arXiv},
       eprint = {1601.05144},
 primaryClass = {astro-ph.SR},
       adsurl = {https://ui.adsabs.harvard.edu/abs/2016ApJS..222....8D}
}

@ARTICLE{almaini2025,
       author = {{Almaini}, Omar and {Wild}, Vivienne and {Maltby}, David and {Taylor}, Elizabeth and {Rowlands}, Kate and {de Lisle}, Thomas and {Alatalo}, Katherine and {Harrold}, Jimi and {Hewitt}, Guillaume and {Patil}, Pallavi and {Skarbinski}, Maya},
        title = "{No evidence for excess AGN activity in recently quenched massive galaxies at cosmic noon}",
      journal = {\mnras},
         year = 2025,
        month = jun,
       volume = {539},
       number = {4},
        pages = {3568-3581},
          doi = {10.1093/mnras/staf659},
archivePrefix = {arXiv},
       eprint = {2504.15342},
 primaryClass = {astro-ph.GA},
       adsurl = {https://ui.adsabs.harvard.edu/abs/2025MNRAS.539.3568A}
}

@article{cid_fernandes_comprehensive_2011,
	title = {A comprehensive classification of galaxies in the {Sloan} {Digital} {Sky} {Survey}: how to tell true from fake {AGN}?},
	volume = {413},
	doi = {10.1111/j.1365-2966.2011.18244.x},
	number = {3},
	journal = {{\textbackslash}mnras},
	author = {Cid Fernandes, R. and Stasińska, G. and Mateus, A. and Vale Asari, N.},
	month = may,
	year = {2011},
	note = {\_eprint: 1012.4426},
	pages = {1687--1699},
}

@article{baldwin_classification_1981,
	title = {Classification parameters for the emission-line spectra of extragalactic objects.},
	volume = {93},
	doi = {10.1086/130766},
	journal = {{\textbackslash}pasp},
	author = {Baldwin, J. A. and Phillips, M. M. and Terlevich, R.},
	month = feb,
	year = {1981},
	pages = {5--19},
}

@article{roy_evidence_2021,
	title = {Evidence of {Wind} {Signatures} in the {Gas} {Velocity} {Profiles} of {Red} {Geysers}},
	volume = {913},
	issn = {0004-637X},
	url = {https://ui.adsabs.harvard.edu/abs/2021ApJ...913...33R},
	doi = {10.3847/1538-4357/abf1e6},
	journal = {The Astrophysical Journal},
	publisher = {IOP},
	author = {Roy, Namrata and Bundy, Kevin and Nevin, Rebecca and Belfiore, Francesco and Yan, Renbin and Campbell, Stephanie and Riffel, Rogemar A. and Riffel, Rogerio and Bershady, Matthew and Westfall, Kyle and Drory, Niv and Zhang, Kai},
	month = may,
	year = {2021},
	pages = {33},
}

@ARTICLE{withaker_quenchingreview_2026,
       author = {{Whitaker}, Katherine E. and {Bezanson}, Rachel},
        title = "{Quenching of Star Formation in Massive Galaxies}",
      journal = {arXiv e-prints},
         year = 2026,
        month = jun,
          eid = {arXiv:2606.12156},
        pages = {arXiv:2606.12156},
          doi = {10.48550/arXiv.2606.12156},
archivePrefix = {arXiv},
       eprint = {2606.12156},
 primaryClass = {astro-ph.GA},
       adsurl = {https://ui.adsabs.harvard.edu/abs/2026arXiv260612156W}
}

@ARTICLE{lombardi_gravity_2024,
       author = {{Lombardi}, Marco},
        title = "{Gravity.jl: Fast and accurate gravitational lens modeling in Julia: I. Point-like and linearized extended sources}",
      journal = {\aap},
         year = 2024,
        month = oct,
       volume = {690},
          eid = {A346},
        pages = {A346},
          doi = {10.1051/0004-6361/202451214},
archivePrefix = {arXiv},
       eprint = {2406.15280},
 primaryClass = {astro-ph.IM},
       adsurl = {https://ui.adsabs.harvard.edu/abs/2024A&A...690A.346L}
}

@ARTICLE{roberts_multiphase_2020,
       author = {{Roberts-Borsani}, G.~W.},
        title = "{Observational constraints on the multiphase nature of outflows using large spectroscopic surveys at z {\ensuremath{\sim}} 0}",
      journal = {\mnras},
         year = 2020,
        month = may,
       volume = {494},
       number = {3},
        pages = {4266-4278},
          doi = {10.1093/mnras/staa1006},
archivePrefix = {arXiv},
       eprint = {2004.06116},
 primaryClass = {astro-ph.GA},
       adsurl = {https://ui.adsabs.harvard.edu/abs/2020MNRAS.494.4266R}
}

@ARTICLE{venturi_ganifs_2025,
       author = {{Venturi}, Giacomo and {Carniani}, Stefano and {Bertola}, Elena and {Circosta}, Chiara and {Parlanti}, Eleonora and {Perna}, Michele and {Arribas}, Santiago and {B{\"o}ker}, Torsten and {Bunker}, Andrew and {Charlot}, St{\'e}phane and {D'Eugenio}, Francesco and {Maiolino}, Roberto and {Rodr{\'\i}guez del Pino}, Bruno and {{\"U}bler}, Hannah and {Cresci}, Giovanni and {Jones}, Gareth C. and {Kumari}, Nimisha and {Lamperti}, Isabella and {Marshall}, Madeline A. and {Scholtz}, Jan and {Zamora}, Sandra},
        title = "{GA-NIFS: Powerful and frequent outflows in moderate-luminosity active galactic nuclei at $z\sim3-6$}",
      journal = {arXiv e-prints},
         year = 2025,
        month = dec,
          eid = {arXiv:2512.09996},
        pages = {arXiv:2512.09996},
          doi = {10.48550/arXiv.2512.09996},
archivePrefix = {arXiv},
       eprint = {2512.09996},
 primaryClass = {astro-ph.GA},
       adsurl = {https://ui.adsabs.harvard.edu/abs/2025arXiv251209996V}
}

@ARTICLE{bertola_mapping_outflows_2025,
       author = {{Bertola}, E. and {Cresci}, G. and {Venturi}, G. and {Perna}, M. and {Circosta}, C. and {Tozzi}, G. and {Lamperti}, I. and {Vignali}, C. and {Arribas}, S. and {Bunker}, A.~J. and {Charlot}, S. and {Carniani}, S. and {Maiolino}, R. and {Rodr{\'\i}guez Del Pino}, B. and {{\"U}bler}, H. and {Willott}, C.~J. and {B{\"o}ker}, T. and {Marshall}, M.~A. and {Parlanti}, E. and {Scholtz}, J.},
        title = "{GA-NIFS: Mapping z ≃ 3.5 AGN-driven ionized outflows in the COSMOS field}",
      journal = {\aap},
         year = 2025,
        month = jul,
       volume = {699},
          eid = {A220},
        pages = {A220},
          doi = {10.1051/0004-6361/202554281},
archivePrefix = {arXiv},
       eprint = {2505.08867},
 primaryClass = {astro-ph.GA},
       adsurl = {https://ui.adsabs.harvard.edu/abs/2025A&A...699A.220B}
}

@ARTICLE{fiore_agnwinds_2017,
       author = {{Fiore}, F. and {Feruglio}, C. and {Shankar}, F. and {Bischetti}, M. and {Bongiorno}, A. and {Brusa}, M. and {Carniani}, S. and {Cicone}, C. and {Duras}, F. and {Lamastra}, A. and {Mainieri}, V. and {Marconi}, A. and {Menci}, N. and {Maiolino}, R. and {Piconcelli}, E. and {Vietri}, G. and {Zappacosta}, L.},
        title = "{AGN wind scaling relations and the co-evolution of black holes and galaxies}",
      journal = {\aap},
         year = 2017,
        month = may,
       volume = {601},
          eid = {A143},
        pages = {A143},
          doi = {10.1051/0004-6361/201629478},
archivePrefix = {arXiv},
       eprint = {1702.04507},
 primaryClass = {astro-ph.GA},
       adsurl = {https://ui.adsabs.harvard.edu/abs/2017A&A...601A.143F}
}

@ARTICLE{brusa_xshooter_2015,
       author = {{Brusa}, M. and {Bongiorno}, A. and {Cresci}, G. and {Perna}, M. and {Marconi}, A. and {Mainieri}, V. and {Maiolino}, R. and {Salvato}, M. and {Lusso}, E. and {Santini}, P. and {Comastri}, A. and {Fiore}, F. and {Gilli}, R. and {La Franca}, F. and {Lanzuisi}, G. and {Lutz}, D. and {Merloni}, A. and {Mignoli}, M. and {Onori}, F. and {Piconcelli}, E. and {Rosario}, D. and {Vignali}, C. and {Zamorani}, G.},
        title = "{X-shooter reveals powerful outflows in z {\ensuremath{\sim}} 1.5 X-ray selected obscured quasi-stellar objects}",
      journal = {\mnras},
         year = 2015,
        month = jan,
       volume = {446},
       number = {3},
        pages = {2394-2417},
          doi = {10.1093/mnras/stu2117},
archivePrefix = {arXiv},
       eprint = {1409.1615},
 primaryClass = {astro-ph.GA},
       adsurl = {https://ui.adsabs.harvard.edu/abs/2015MNRAS.446.2394B}
}

@ARTICLE{cresci_blowing_2015,
       author = {{Cresci}, G. and {Mainieri}, V. and {Brusa}, M. and {Marconi}, A. and {Perna}, M. and {Mannucci}, F. and {Piconcelli}, E. and {Maiolino}, R. and {Feruglio}, C. and {Fiore}, F. and {Bongiorno}, A. and {Lanzuisi}, G. and {Merloni}, A. and {Schramm}, M. and {Silverman}, J.~D. and {Civano}, F.},
        title = "{Blowin' in the Wind: Both ``Negative'' and ``Positive'' Feedback in an Obscured High-z Quasar}",
      journal = {\apj},
         year = 2015,
        month = jan,
       volume = {799},
       number = {1},
          eid = {82},
        pages = {82},
          doi = {10.1088/0004-637X/799/1/82},
archivePrefix = {arXiv},
       eprint = {1411.4208},
 primaryClass = {astro-ph.GA},
       adsurl = {https://ui.adsabs.harvard.edu/abs/2015ApJ...799...82C}
}

@ARTICLE{veilleux2018,
       author = {{Veilleux}, Sylvain and {Maiolino}, Roberto and {Bolatto}, Alberto D. and {Aalto}, Susanne},
        title = "{Cool outflows in galaxies and their implications}",
      journal = {\aapr},
         year = 2020,
        month = apr,
       volume = {28},
       number = {1},
          eid = {2},
        pages = {2},
          doi = {10.1007/s00159-019-0121-9},
archivePrefix = {arXiv},
       eprint = {2002.07765},
 primaryClass = {astro-ph.GA},
       adsurl = {https://ui.adsabs.harvard.edu/abs/2020A&ARv..28....2V}
}

@ARTICLE{daviesbaron2020,
       author = {{Davies}, R. and {Baron}, D. and {Shimizu}, T. and {Netzer}, H. and {Burtscher}, L. and {de Zeeuw}, P.~T. and {Genzel}, R. and {Hicks}, E.~K.~S. and {Koss}, M. and {Lin}, M.-Y. and {Lutz}, D. and {Maciejewski}, W. and {M{\"u}ller-S{\'a}nchez}, F. and {Orban de Xivry}, G. and {Ricci}, C. and {Riffel}, R. and {Riffel}, R.~A. and {Rosario}, D. and {Schartmann}, M. and {Schnorr-M{\"u}ller}, A. and {Shangguan}, J. and {Sternberg}, A. and {Sturm}, E. and {Storchi-Bergmann}, T. and {Tacconi}, L. and {Veilleux}, S.},
        title = "{Ionized outflows in local luminous AGN: what are the real densities and outflow rates?}",
      journal = {\mnras},
         year = 2020,
        month = nov,
       volume = {498},
       number = {3},
        pages = {4150-4177},
          doi = {10.1093/mnras/staa2413},
archivePrefix = {arXiv},
       eprint = {2003.06153},
 primaryClass = {astro-ph.GA},
       adsurl = {https://ui.adsabs.harvard.edu/abs/2020MNRAS.498.4150D}
}

@article{Sun_2026_neutral,
   title={Extreme Neutral Outflow in a Non-active Galactic Nucleus Quiescent Galaxy at
                    <i>z</i>
                    ∼ 1.3},
   volume={997},
   ISSN={1538-4357},
   url={http://dx.doi.org/10.3847/1538-4357/ae2c83},
   DOI={10.3847/1538-4357/ae2c83},
   number={2},
   journal={The Astrophysical Journal},
   publisher={American Astronomical Society},
   author={Sun, Yang and Ji, Zhiyuan and Rieke, George H. and D’Eugenio, Francesco and Zhu, Yongda and Sun, Fengwu and Lin, Xiaojing and Bunker, Andrew J. and Lyu 吕, Jianwei 建伟 and Rinaldi, Pierluigi and Willmer, Christopher N. A.},
   year={2026},
   month=Jan, pages={140} }

@ARTICLE{newman_resolving2_2018,
       author = {{Newman}, Andrew B. and {Belli}, Sirio and {Ellis}, Richard S. and {Patel}, Shannon G.},
        title = "{Resolving Quiescent Galaxies at z {\ensuremath{\gtrsim}} 2. II. Direct Measures of Rotational Support}",
      journal = {\apj},
         year = 2018,
        month = aug,
       volume = {862},
       number = {2},
          eid = {126},
        pages = {126},
          doi = {10.3847/1538-4357/aacd4f},
archivePrefix = {arXiv},
       eprint = {1806.06815},
 primaryClass = {astro-ph.GA},
       adsurl = {https://ui.adsabs.harvard.edu/abs/2018ApJ...862..126N}
}

@ARTICLE{emiles_library,
       author = {{Vazdekis}, A. and {Koleva}, M. and {Ricciardelli}, E. and {R{\"o}ck}, B. and {Falc{\'o}n-Barroso}, J.},
        title = "{UV-extended E-MILES stellar population models: young components in massive early-type galaxies}",
      journal = {\mnras},
         year = 2016,
        month = dec,
       volume = {463},
       number = {4},
        pages = {3409-3436},
          doi = {10.1093/mnras/stw2231},
archivePrefix = {arXiv},
       eprint = {1612.01187},
 primaryClass = {astro-ph.GA},
       adsurl = {https://ui.adsabs.harvard.edu/abs/2016MNRAS.463.3409V}
}

@ARTICLE{Steidel2010,
       author = {{Steidel}, Charles C. and {Erb}, Dawn K. and {Shapley}, Alice E. and {Pettini}, Max and {Reddy}, Naveen and {Bogosavljevi{\'c}}, Milan and {Rudie}, Gwen C. and {Rakic}, Olivera},
        title = "{The Structure and Kinematics of the Circumgalactic Medium from Far-ultraviolet Spectra of z \raisebox{-0.5ex}\textasciitilde= 2-3 Galaxies}",
      journal = {\apj},
         year = 2010,
        month = jul,
       volume = {717},
       number = {1},
        pages = {289-322},
          doi = {10.1088/0004-637X/717/1/289},
archivePrefix = {arXiv},
       eprint = {1003.0679},
 primaryClass = {astro-ph.CO},
       adsurl = {https://ui.adsabs.harvard.edu/abs/2010ApJ...717..289S}
}

@ARTICLE{Byler2019,
       author = {{Byler}, Nell and {Dalcanton}, Julianne J. and {Conroy}, Charlie and {Johnson}, Benjamin D. and {Choi}, Jieun and {Dotter}, Aaron and {Rosenfield}, Philip},
        title = "{Self-consistent Predictions for LIER-like Emission Lines from Post-AGB Stars}",
      journal = {\aj},
         year = 2019,
        month = jul,
       volume = {158},
       number = {1},
          eid = {2},
        pages = {2},
          doi = {10.3847/1538-3881/ab1b70},
archivePrefix = {arXiv},
       eprint = {1904.10978},
 primaryClass = {astro-ph.GA},
       adsurl = {https://ui.adsabs.harvard.edu/abs/2019AJ....158....2B}
}

@ARTICLE{Bergamini2019,
       author = {{Bergamini}, P. and {Rosati}, P. and {Mercurio}, A. and {Grillo}, C. and {Caminha}, G.~B. and {Meneghetti}, M. and {Agnello}, A. and {Biviano}, A. and {Calura}, F. and {Giocoli}, C. and {Lombardi}, M. and {Rodighiero}, G. and {Vanzella}, E.},
        title = "{Enhanced cluster lensing models with measured galaxy kinematics}",
      journal = {\aap},
         year = 2019,
        month = nov,
       volume = {631},
          eid = {A130},
        pages = {A130},
          doi = {10.1051/0004-6361/201935974},
archivePrefix = {arXiv},
       eprint = {1905.13236},
 primaryClass = {astro-ph.GA},
       adsurl = {https://ui.adsabs.harvard.edu/abs/2019A&A...631A.130B}
}

@ARTICLE{Oguri2010,
       author = {{Oguri}, Masamune},
        title = "{The Mass Distribution of SDSS J1004+4112 Revisited}",
      journal = {\pasj},
         year = 2010,
        month = aug,
       volume = {62},
        pages = {1017},
          doi = {10.1093/pasj/62.4.1017},
archivePrefix = {arXiv},
       eprint = {1005.3103},
 primaryClass = {astro-ph.CO},
       adsurl = {https://ui.adsabs.harvard.edu/abs/2010PASJ...62.1017O}
}

@ARTICLE{Zubovas2023,
       author = {{Zubovas}, Kastytis and {Maskeli{\={u}}nas}, Gediminas},
        title = "{Life after AGN switch off: evolution and properties of fossil galactic outflows}",
      journal = {\mnras},
         year = 2023,
        month = oct,
       volume = {524},
       number = {4},
        pages = {4819-4840},
          doi = {10.1093/mnras/stad1661},
archivePrefix = {arXiv},
       eprint = {2306.00518},
 primaryClass = {astro-ph.GA},
       adsurl = {https://ui.adsabs.harvard.edu/abs/2023MNRAS.524.4819Z}
}

@ARTICLE{Valentino2026,
       author = {{Valentino}, F. and {Ito}, K. and {Farcy}, M. and {Fontanot}, F. and {Lagos}, C. and {De Lucia}, G. and {Hirschmann}, M. and {Brammer}, G. and {Kokorev}, V. and {Hamadouche}, M. and {Zhu}, P. and {Scarpe}, G. and {Pensabene}, A. and {Whitaker}, K.~E. and {Baker}, W.~M. and {Araya-Araya}, P. and {Antwi-Danso}, J. and {Ceverino}, D. and {Faisst}, A.~L. and {Fujimoto}, S. and {Gillman}, S. and {Ilbert}, O. and {Jespersen}, C.~K. and {Kakimoto}, T. and {Kubo}, M. and {Man}, A.~W.~S. and {Magdis}, G. and {Onodera}, M. and {Shimakawa}, R. and {Tanaka}, M. and {Toft}, S. and {Xie}, L. and {Weaver}, J.~R. and {Wu}, P.-F.},
        title = "{High-ionization coronal lines trace quasar-like activity in recently quenched galaxies at high redshift}",
      journal = {arXiv e-prints},
         year = 2026,
        month = may,
          eid = {arXiv:2605.30424},
        pages = {arXiv:2605.30424},
          doi = {10.48550/arXiv.2605.30424},
archivePrefix = {arXiv},
       eprint = {2605.30424},
 primaryClass = {astro-ph.GA},
       adsurl = {https://ui.adsabs.harvard.edu/abs/2026arXiv260530424V}
}

@ARTICLE{Johnson_prospector_2021,
       author = {{Johnson}, Benjamin D. and {Leja}, Joel and {Conroy}, Charlie and {Speagle}, Joshua S.},
        title = "{Stellar Population Inference with Prospector}",
      journal = {\apjs},
         year = 2021,
        month = jun,
       volume = {254},
       number = {2},
          eid = {22},
        pages = {22},
          doi = {10.3847/1538-4365/abef67},
archivePrefix = {arXiv},
       eprint = {2012.01426},
 primaryClass = {astro-ph.GA},
       adsurl = {https://ui.adsabs.harvard.edu/abs/2021ApJS..254...22J}
}

@article{Straatman2014,
	title = {A {Substantial} {Population} of {Massive} {Quiescent} {Galaxies} at z $\sim$ 4 from {ZFOURGE}},
	volume = {783},
	doi = {10.1088/2041-8205/783/1/L14},
	number = {1},
	journal = {\apjl},
	author = {Straatman, Caroline M. S. and Labb{\'{e}}, Ivo and Spitler, Lee R. and Allen, Rebecca and Altieri, Bruno and Brammer, Gabriel B. and Dickinson, Mark and van Dokkum, Pieter and Inami, Hanae and Glazebrook, Karl and Kacprzak, Glenn G. and Kawinwanichakij, Lalit and Kelson, Daniel D. and McCarthy, Patrick J. and Mehrtens, Nicola and Monson, Andy and Murphy, David and Papovich, Casey and Persson, S. Eric and Quadri, Ryan and Rees, Glen and Tomczak, Adam and Tran, Kim-Vy H. and Tilvi, Vithal},
	month = mar,
	year = {2014},
	pages = {L14},
}

@article{hinshaw13,
  author        = {{Hinshaw}, G. and {Larson}, D. and {Komatsu}, E. and {Spergel}, D.~N. and {Bennett}, C.~L. and {Dunkley}, J. and {Nolta}, M.~R. and {Halpern}, M. and {Hill}, R.~S. and {Odegard}, N. and {Page}, L. and {Smith}, K.~M. and {Weiland}, J.~L. and {Gold}, B. and {Jarosik}, N. and {Kogut}, A. and {Limon}, M. and {Meyer}, S.~S. and {Tucker}, G.~S. and {Wollack}, E. and {Wright}, E.~L.},
  title         = {{Nine-year Wilkinson Microwave Anisotropy Probe (WMAP) Observations: Cosmological Parameter Results}},
  journal       = {\apjs},
  year          = 2013,
  month         = oct,
  volume        = {208},
  number        = {2},
  eid           = {19},
  pages         = {19},
  doi           = {10.1088/0067-0049/208/2/19},
  archiveprefix = {arXiv},
  eprint        = {1212.5226},
  primaryclass  = {astro-ph.CO},
  adsurl        = {https://ui.adsabs.harvard.edu/abs/2013ApJS..208...19H}
}

@article{CHABRIER_IMF,
  author        = {{Chabrier}, Gilles},
  title         = {{Galactic Stellar and Substellar Initial Mass Function}},
  journal       = {PASP},
  year          = 2003,
  month         = jul,
  volume        = {115},
  number        = {809},
  pages         = {763-795},
  doi           = {10.1086/376392},
  archiveprefix = {arXiv},
  eprint        = {astro-ph/0304382},
  primaryclass  = {astro-ph},
  adsurl        = {https://ui.adsabs.harvard.edu/abs/2003PASP..115..763C}
}

@ARTICLE{Davies2019,
       author = {{Davies}, R.~L. and {F{\"o}rster Schreiber}, N.~M. and {{\"U}bler}, H. and {Genzel}, R. and {Lutz}, D. and {Renzini}, A. and {Tacchella}, S. and {Tacconi}, L.~J. and {Belli}, S. and {Burkert}, A. and {Carollo}, C.~M. and {Davies}, R.~I. and {Herrera-Camus}, R. and {Lilly}, S.~J. and {Mancini}, C. and {Naab}, T. and {Nelson}, E.~J. and {Price}, S.~H. and {Shimizu}, T.~T. and {Sternberg}, A. and {Wisnioski}, E. and {Wuyts}, S.},
        title = "{Kiloparsec Scale Properties of Star Formation Driven Outflows at z {\ensuremath{\sim}} 2.3 in the SINS/zC-SINF AO Survey}",
      journal = {\apj},
         year = 2019,
        month = mar,
       volume = {873},
       number = {2},
          eid = {122},
        pages = {122},
          doi = {10.3847/1538-4357/ab06f1},
archivePrefix = {arXiv},
       eprint = {1808.10700},
 primaryClass = {astro-ph.GA},
       adsurl = {https://ui.adsabs.harvard.edu/abs/2019ApJ...873..122D}
}

@ARTICLE{Donevski2026,
       author = {{Donevski}, D. and {Nanni}, A. and {Whitaker}, K.~E. and {Man}, A.~W.~S. and {Faisst}, A. and {Lapi}, A. and {Garc{\'\i}a-Bernete}, I. and {Romano}, M. and {Petrushevska}, T. and {Gururajan}, G. and {Lorenzon}, G. and {Narayanan}, D.},
        title = "{Dust and PAHs in late-stage galaxy evolution: Imprints of TP-AGB dust injection, grain growth and AGN feedback in high-z quiescent galaxies with JWST and ALMA}",
      journal = {arXiv e-prints},
         year = 2026,
        month = aug,
          eid = {arXiv:2608.27571},
        pages = {arXiv:2608.27571},
          doi = {10.48550/arXiv.2608.27571},
archivePrefix = {arXiv},
       eprint = {2608.27571},
 primaryClass = {astro-ph.GA},
       adsurl = {https://ui.adsabs.harvard.edu/abs/2026arXiv260827571D}
}

@article{Weaver2023,
	title = {{COSMOS2020}: {The} galaxy stellar mass function. {The} assembly and star formation cessation of galaxies at 0.2{\textless} z $\leq$ 7.5},
	volume = {677},
	doi = {10.1051/0004-6361/202245581},
	journal = {\aap},
	author = {Weaver, J. R. and Davidzon, I. and Toft, S. and Ilbert, O. and McCracken, H. J. and Gould, K. M. L. and Jespersen, C. K. and Steinhardt, C. and Lagos, C. D. P. and Capak, P. L. and Casey, C. M. and Chartab, N. and Faisst, A. L. and Hayward, C. C. and Kartaltepe, J. S. and Kauffmann, O. B. and Koekemoer, A. M. and Kokorev, V. and Laigle, C. and Liu, D. and Long, A. and Magdis, G. E. and McPartland, C. J. R. and Milvang-Jensen, B. and Mobasher, B. and Moneti, A. and Peng, Y. and Sanders, D. B. and Shuntov, M. and Sneppen, A. and Valentino, F. and Zalesky, L. and Zamorani, G.},
	month = sep,
	year = {2023},
	pages = {A184},
}

@article{Tomczak2014,
	title = {Galaxy {Stellar} {Mass} {Functions} from {ZFOURGE}/{CANDELS}: {An} {Excess} of {Low}-mass {Galaxies} since z = 2 and the {Rapid} {Buildup} of {Quiescent} {Galaxies}},
	volume = {783},
	doi = {10.1088/0004-637X/783/2/85},
	number = {2},
	journal = {\apj},
	author = {Tomczak, Adam R. and Quadri, Ryan F. and Tran, Kim-Vy H. and Labb{\'{e}}, Ivo and Straatman, Caroline M. S. and Papovich, Casey and Glazebrook, Karl and Allen, Rebecca and Brammer, Gabriel B. and Kacprzak, Glenn G. and Kawinwanichakij, Lalitwadee and Kelson, Daniel D. and McCarthy, Patrick J. and Mehrtens, Nicola and Monson, Andrew J. and Persson, S. Eric and Spitler, Lee R. and Tilvi, Vithal and van Dokkum, Pieter},
	month = mar,
	year = {2014},
	pages = {85},
}

@article{Leja19_SFH,
  author        = {{Leja}, Joel and {Carnall}, Adam C. and {Johnson}, Benjamin D. and {Conroy}, Charlie and {Speagle}, Joshua S.},
  title         = {{How to Measure Galaxy Star Formation Histories. II. Nonparametric Models}},
  journal       = {ApJ},
  year          = 2019,
  month         = may,
  volume        = {876},
  number        = {1},
  eid           = {3},
  pages         = {3},
  doi           = {10.3847/1538-4357/ab133c},
  archiveprefix = {arXiv},
  eprint        = {1811.03637},
  primaryclass  = {astro-ph.GA},
  adsurl        = {https://ui.adsabs.harvard.edu/abs/2019ApJ...876....3L}
}

@ARTICLE{zubovas_outflows_2022,
       author = {{Zubovas}, Kastytis and {Bialopetravi{\v{c}}ius}, Jonas and {Kazlauskait{\.{e}}}, Monika},
        title = "{Determining active galactic nucleus luminosity histories using present-day outflow properties: a neural network-based approach}",
      journal = {\mnras},
         year = 2022,
        month = sep,
       volume = {515},
       number = {2},
        pages = {1705-1722},
          doi = {10.1093/mnras/stac1887},
archivePrefix = {arXiv},
       eprint = {2207.01959},
 primaryClass = {astro-ph.GA},
       adsurl = {https://ui.adsabs.harvard.edu/abs/2022MNRAS.515.1705Z}
}

@ARTICLE{emcee,
       author = {{Foreman-Mackey}, Daniel and {Hogg}, David W. and {Lang}, Dustin and {Goodman}, Jonathan},
        title = "{emcee: The MCMC Hammer}",
      journal = {\pasp},
         year = 2013,
        month = mar,
       volume = {125},
       number = {925},
        pages = {306},
          doi = {10.1086/670067},
archivePrefix = {arXiv},
       eprint = {1202.3665},
 primaryClass = {astro-ph.IM},
       adsurl = {https://ui.adsabs.harvard.edu/abs/2013PASP..125..306F}
}

@ARTICLE{matteos_article_sodium_2026,
       author = {{Sapori}, Matteo and {Belli}, Sirio and {Bugiani}, Letizia and {Khoram}, Amir H. and {Liboni}, Caterina and {Maheson}, Gabriel and {Brammer}, Gabriel and {Davies}, Rebecca and {Dayal}, Pratika and {de Graaff}, Anna and {Leja}, Joel and {Naidu}, Rohan and {Oesch}, Pascal and {Sotira}, Stefano and {Tacchella}, Sandro and {Wang}, Bingjie and {Whitaker}, Katherine E.},
        title = "{Outflows in the Early Universe: Neutral gas Absorption in Galaxies at z > 3 from low-resolution JWST Spectroscopy}",
      journal = {arXiv e-prints},
         year = 2026,
        month = jul,
          eid = {arXiv:2607.20606},
        pages = {arXiv:2607.20606},
          doi = {10.48550/arXiv.2607.20606},
archivePrefix = {arXiv},
       eprint = {2607.20606},
 primaryClass = {astro-ph.GA},
       adsurl = {https://ui.adsabs.harvard.edu/abs/2026arXiv260720606S}
}

@ARTICLE{swinbank_energetics_2019,
       author = {{Swinbank}, A.~M. and {Harrison}, C.~M. and {Tiley}, A.~L. and {Johnson}, H.~L. and {Smail}, Ian and {Stott}, J.~P. and {Best}, P.~N. and {Bower}, R.~G. and {Bureau}, M. and {Bunker}, A. and {Cirasuolo}, M. and {Jarvis}, M. and {Magdis}, G.~E. and {Sharples}, R.~M. and {Sobral}, D.},
        title = "{The energetics of starburst-driven outflows at z {\ensuremath{\sim}} 1 from KMOS}",
      journal = {\mnras},
         year = 2019,
        month = jul,
       volume = {487},
       number = {1},
        pages = {381-393},
          doi = {10.1093/mnras/stz1275},
       adsurl = {https://ui.adsabs.harvard.edu/abs/2019MNRAS.487..381S}
}

@ARTICLE{canodiaz_2012,
       author = {{Cano-D{\'\i}az}, M. and {Maiolino}, R. and {Marconi}, A. and {Netzer}, H. and {Shemmer}, O. and {Cresci}, G.},
        title = "{Observational evidence of quasar feedback quenching star formation at high redshift}",
      journal = {\aap},
         year = 2012,
        month = jan,
       volume = {537},
          eid = {L8},
        pages = {L8},
          doi = {10.1051/0004-6361/201118358},
archivePrefix = {arXiv},
       eprint = {1112.3071},
 primaryClass = {astro-ph.CO},
       adsurl = {https://ui.adsabs.harvard.edu/abs/2012A&A...537L...8C}
}

@ARTICLE{ho_2009_bolcorr,
       author = {{Ho}, Luis C.},
        title = "{Radiatively Inefficient Accretion in Nearby Galaxies}",
      journal = {\apj},
         year = 2009,
        month = jul,
       volume = {699},
       number = {1},
        pages = {626-637},
          doi = {10.1088/0004-637X/699/1/626},
archivePrefix = {arXiv},
       eprint = {0906.4104},
 primaryClass = {astro-ph.GA},
       adsurl = {https://ui.adsabs.harvard.edu/abs/2009ApJ...699..626H}
}

@ARTICLE{newman_discovery_2015,
       author = {{Newman}, Andrew B. and {Belli}, Sirio and {Ellis}, Richard S.},
        title = "{Discovery of a Strongly Lensed Massive Quiescent Galaxy at z = 2.636: Spatially Resolved Spectroscopy and Indications of Rotation}",
      journal = {\apjl},
         year = 2015,
        month = nov,
       volume = {813},
       number = {1},
          eid = {L7},
        pages = {L7},
          doi = {10.1088/2041-8205/813/1/L7},
archivePrefix = {arXiv},
       eprint = {1509.04345},
 primaryClass = {astro-ph.GA},
       adsurl = {https://ui.adsabs.harvard.edu/abs/2015ApJ...813L...7N}
}

@ARTICLE{cappellari_vorbin_2003,
       author = {{Cappellari}, Michele and {Copin}, Yannick},
        title = "{Adaptive spatial binning of integral-field spectroscopic data using Voronoi tessellations}",
      journal = {\mnras},
         year = 2003,
        month = jun,
       volume = {342},
       number = {2},
        pages = {345-354},
          doi = {10.1046/j.1365-8711.2003.06541.x},
archivePrefix = {arXiv},
       eprint = {astro-ph/0302262},
 primaryClass = {astro-ph},
       adsurl = {https://ui.adsabs.harvard.edu/abs/2003MNRAS.342..345C}
}

@article{forster_schreiber_kmos3d_2019,
	title = {The {KMOS3D} {Survey}: {Demographics} and {Properties} of {Galactic} {Outflows} at z = 0.6-2.7},
	volume = {875},
	issn = {0004-637X},
	shorttitle = {The {KMOS3D} {Survey}},
	url = {https://ui.adsabs.harvard.edu/abs/2019ApJ...875...21F},
	doi = {10.3847/1538-4357/ab0ca2},
	urldate = {2026-08-19},
	journal = {The Astrophysical Journal},
	publisher = {IOP},
	author = {Förster Schreiber, N. M. and Übler, H. and Davies, R. L. and Genzel, R. and Wisnioski, E. and Belli, S. and Shimizu, T. and Lutz, D. and Fossati, M. and Herrera-Camus, R. and Mendel, J. T. and Tacconi, L. J. and Wilman, D. and Beifiori, A. and Brammer, G. B. and Burkert, A. and Carollo, C. M. and Davies, R. I. and Eisenhauer, F. and Fabricius, M. and Lilly, S. J. and Momcheva, I. and Naab, T. and Nelson, E. J. and Price, S. H. and Renzini, A. and Saglia, R. and Sternberg, A. and van Dokkum, P. and Wuyts, S.},
	month = apr,
	year = {2019},
	note = {ADS Bibcode: 2019ApJ...875...21F},
	pages = {21},
}

@article{cresci_blowin_2015,
	title = {Blowin' in the {Wind}: {Both} "{Negative}" and "{Positive}" {Feedback} in an {Obscured} {High}-z {Quasar}},
	volume = {799},
	issn = {0004-637X},
	shorttitle = {Blowin' in the {Wind}},
	url = {https://ui.adsabs.harvard.edu/abs/2015ApJ...799...82C},
	doi = {10.1088/0004-637X/799/1/82},
	urldate = {2026-08-19},
	journal = {The Astrophysical Journal},
	publisher = {IOP},
	author = {Cresci, G. and Mainieri, V. and Brusa, M. and Marconi, A. and Perna, M. and Mannucci, F. and Piconcelli, E. and Maiolino, R. and Feruglio, C. and Fiore, F. and Bongiorno, A. and Lanzuisi, G. and Merloni, A. and Schramm, M. and Silverman, J. D. and Civano, F.},
	month = jan,
	year = {2015},
	note = {ADS Bibcode: 2015ApJ...799...82C},
	pages = {82},
}

@article{kakkad_tracing_2016,
	title = {Tracing outflows in the {AGN} forbidden region with {SINFONI}},
	volume = {592},
	issn = {0004-6361},
	url = {https://ui.adsabs.harvard.edu/abs/2016A&A...592A.148K},
	doi = {10.1051/0004-6361/201527968},
	urldate = {2026-08-19},
	journal = {Astronomy and Astrophysics},
	publisher = {EDP},
	author = {Kakkad, D. and Mainieri, V. and Padovani, P. and Cresci, G. and Husemann, B. and Carniani, S. and Brusa, M. and Lamastra, A. and Lanzuisi, G. and Piconcelli, E. and Schramm, M.},
	month = aug,
	year = {2016},
	note = {ADS Bibcode: 2016A\&A...592A.148K},
	pages = {A148},
}

@article{davies_nuclear_2020,
	title = {From {Nuclear} to {Circumgalactic}: {Zooming} in on {AGN}-driven {Outflows} at z ∼ 2.2 with {SINFONI}},
	volume = {894},
	issn = {0004-637X},
	shorttitle = {From {Nuclear} to {Circumgalactic}},
	url = {https://ui.adsabs.harvard.edu/abs/2020ApJ...894...28D},
	doi = {10.3847/1538-4357/ab86ad},
	urldate = {2026-08-19},
	journal = {The Astrophysical Journal},
	publisher = {IOP},
	author = {Davies, Rebecca L. and Förster Schreiber, N. M. and Lutz, D. and Genzel, R. and Belli, S. and Shimizu, T. T. and Contursi, A. and Davies, R. I. and Herrera-Camus, R. and Lee, M. M. and Naab, T. and Price, S. H. and Renzini, A. and Schruba, A. and Sternberg, A. and Tacconi, L. J. and Übler, H. and Wisnioski, E. and Wuyts, S.},
	month = may,
	year = {2020},
	note = {ADS Bibcode: 2020ApJ...894...28D},
	pages = {28},
}

@article{carnall_surprising_2023,
	title = {A surprising abundance of massive quiescent galaxies at 3 {\textless} z {\textless} 5 in the first data from {JWST} {CEERS}},
	volume = {520},
	issn = {0035-8711},
	url = {https://ui.adsabs.harvard.edu/abs/2023MNRAS.520.3974C},
	doi = {10.1093/mnras/stad369},
	urldate = {2026-08-19},
	journal = {Monthly Notices of the Royal Astronomical Society},
	publisher = {OUP},
	author = {Carnall, A. C. and McLeod, D. J. and McLure, R. J. and Dunlop, J. S. and Begley, R. and Cullen, F. and Donnan, C. T. and Hamadouche, M. L. and Jewell, S. M. and Jones, E. W. and Pollock, C. L. and Wild, V.},
	month = apr,
	year = {2023},
	note = {ADS Bibcode: 2023MNRAS.520.3974C},
	pages = {3974--3985},
}

@article{nanayakkara_population_2024,
	title = {A population of faint, old, and massive quiescent galaxies at 3},
	volume = {14},
	issn = {2045-2322},
	url = {https://ui.adsabs.harvard.edu/abs/2024NatSR..14.3724N},
	doi = {10.1038/s41598-024-52585-4},
	urldate = {2026-08-19},
	journal = {Scientific Reports},
	publisher = {Springer},
	author = {Nanayakkara, Themiya and Glazebrook, Karl and Jacobs, Colin and Kawinwanichakij, Lalitwadee and Schreiber, Corentin and Brammer, Gabriel and Esdaile, James and Kacprzak, Glenn G. and Labbe, Ivo and Lagos, Claudia and Marchesini, Danilo and Marsan, Z. Cemile and Oesch, Pascal A. and Papovich, Casey and Remus, Rhea-Silvia and Tran, Kim-Vy H.},
	month = feb,
	year = {2024},
	note = {ADS Bibcode: 2024NatSR..14.3724N},
	pages = {3724},
}

@article{de_graaff_efficient_2025,
	title = {Efficient formation of a massive quiescent galaxy at redshift 4.9},
	volume = {9},
	issn = {2397-3366},
	url = {https://ui.adsabs.harvard.edu/abs/2025NatAs...9..280D},
	doi = {10.1038/s41550-024-02424-3},
	urldate = {2026-08-19},
	journal = {Nature Astronomy},
	author = {de Graaff, Anna and Setton, David J. and Brammer, Gabriel and Cutler, Sam and Suess, Katherine A. and Labbé, Ivo and Leja, Joel and Weibel, Andrea and Maseda, Michael V. and Whitaker, Katherine E. and Bezanson, Rachel and Boogaard, Leindert A. and Cleri, Nikko J. and De Lucia, Gabriella and Franx, Marijn and Greene, Jenny E. and Hirschmann, Michaela and Matthee, Jorryt and McConachie, Ian and Naidu, Rohan P. and Oesch, Pascal A. and Price, Sedona H. and Rix, Hans-Walter and Valentino, Francesco and Wang, Bingjie and Williams, Christina C.},
	month = feb,
	year = {2025},
	note = {ADS Bibcode: 2025NatAs...9..280D},
	pages = {280--292},
}

@article{weibel_rubies_2025,
	title = {{RUBIES} {Reveals} a {Massive} {Quiescent} {Galaxy} at z = 7.3},
	volume = {983},
	issn = {0004-637X},
	url = {https://ui.adsabs.harvard.edu/abs/2025ApJ...983...11W},
	doi = {10.3847/1538-4357/adab7a},
	urldate = {2026-08-19},
	journal = {The Astrophysical Journal},
	publisher = {IOP},
	author = {Weibel, Andrea and de Graaff, Anna and Setton, David J. and Miller, Tim B. and Oesch, Pascal A. and Brammer, Gabriel and Lagos, Claudia D. P. and Whitaker, Katherine E. and Williams, Christina C. and Baggen, Josephine F. W. and Bezanson, Rachel and Boogaard, Leindert A. and Cleri, Nikko J. and Greene, Jenny E. and Hirschmann, Michaela and Hviding, Raphael E. and Kuruvanthodi, Adarsh and Labbé, Ivo and Leja, Joel and Maseda, Michael V. and Matthee, Jorryt and McConachie, Ian and Naidu, Rohan P. and Roberts-Borsani, Guido and Schaerer, Daniel and Suess, Katherine A. and Valentino, Francesco and van Dokkum, Pieter and Wang, Bingjie},
	month = apr,
	year = {2025},
	note = {ADS Bibcode: 2025ApJ...983...11W},
	pages = {11},
}

@software{cappellari_vorbin_2012,
       author = {{Cappellari}, Michele and {Copin}, Yannick},
        title = "{VorBin: Voronoi binning method}",
 howpublished = {Astrophysics Source Code Library, record ascl:1211.006},
         year = 2012,
        month = nov,
          eid = {ascl:1211.006},
archivePrefix = {ascl},
       eprint = {1211.006},
       adsurl = {https://ui.adsabs.harvard.edu/abs/2012ascl.soft11006C}
}

@ARTICLE{veilleux_galactic_winds_2005,
       author = {{Veilleux}, Sylvain and {Cecil}, Gerald and {Bland-Hawthorn}, Joss},
        title = "{Galactic Winds}",
      journal = {\araa},
         year = 2005,
        month = sep,
       volume = {43},
       number = {1},
        pages = {769-826},
          doi = {10.1146/annurev.astro.43.072103.150610},
archivePrefix = {arXiv},
       eprint = {astro-ph/0504435},
 primaryClass = {astro-ph},
       adsurl = {https://ui.adsabs.harvard.edu/abs/2005ARA&A..43..769V}
}

@ARTICLE{valentino_recently_quenched,
       author = {{Valentino}, F. and {Heintz}, K.~E. and {Brammer}, G. and {Ito}, K. and {Kokorev}, V. and {Whitaker}, K.~E. and {Gallazzi}, A. and {de Graaff}, A. and {Weibel}, A. and {Frye}, B.~L. and {Kamieneski}, P.~S. and {Jin}, S. and {Ceverino}, D. and {Faisst}, A. and {Farcy}, M. and {Fujimoto}, S. and {Gillman}, S. and {Gottumukkala}, R. and {Hamadouche}, M. and {Harrington}, K.~C. and {Hirschmann}, M. and {Jespersen}, C.~K. and {Kakimoto}, T. and {Kubo}, M. and {Lagos}, C. d. P. and {Lee}, M. and {Magdis}, G.~E. and {Man}, A.~W.~S. and {Onodera}, M. and {Rizzo}, F. and {Shimakawa}, R. and {Setton}, D.~J. and {Tanaka}, M. and {Toft}, S. and {Wu}, P.-F. and {Zhu}, P.},
        title = "{Gas outflows in two recently quenched galaxies at z = 4 and 7}",
      journal = {\aap},
         year = 2025,
        month = jul,
       volume = {699},
          eid = {A358},
        pages = {A358},
          doi = {10.1051/0004-6361/202553908},
archivePrefix = {arXiv},
       eprint = {2503.01990},
 primaryClass = {astro-ph.GA},
       adsurl = {https://ui.adsabs.harvard.edu/abs/2025A&A...699A.358V}
}

@ARTICLE{carniani_outflows_2015,
       author = {{Carniani}, S. and {Marconi}, A. and {Maiolino}, R. and {Balmaverde}, B. and {Brusa}, M. and {Cano-D{\'\i}az}, M. and {Cicone}, C. and {Comastri}, A. and {Cresci}, G. and {Fiore}, F. and {Feruglio}, C. and {La Franca}, F. and {Mainieri}, V. and {Mannucci}, F. and {Nagao}, T. and {Netzer}, H. and {Piconcelli}, E. and {Risaliti}, G. and {Schneider}, R. and {Shemmer}, O.},
        title = "{Ionised outflows in z \raisebox{-0.5ex}\textasciitilde 2.4 quasar host galaxies}",
      journal = {\aap},
         year = 2015,
        month = aug,
       volume = {580},
          eid = {A102},
        pages = {A102},
          doi = {10.1051/0004-6361/201526557},
archivePrefix = {arXiv},
       eprint = {1506.03096},
 primaryClass = {astro-ph.GA},
       adsurl = {https://ui.adsabs.harvard.edu/abs/2015A&A...580A.102C}
}

@ARTICLE{heavymetalsurvey,
       author = {{Kriek}, Mariska and {Beverage}, Aliza G. and {Price}, Sedona H. and {Suess}, Katherine A. and {Barro}, Guillermo and {Bezanson}, Rachel S. and {Conroy}, Charlie and {Cutler}, Sam E. and {Franx}, Marijn and {Lin}, Jamie and {Lorenz}, Brian and {Ma}, Yilun and {Momcheva}, Ivelina G. and {Mowla}, Lamiya A. and {Pasha}, Imad and {van Dokkum}, Pieter and {Whitaker}, Katherine E.},
        title = "{The Heavy Metal Survey: Star Formation Constraints and Dynamical Masses of 21 Massive Quiescent Galaxies at z = 1.3─2.3}",
      journal = {\apj},
         year = 2024,
        month = may,
       volume = {966},
       number = {1},
          eid = {36},
        pages = {36},
          doi = {10.3847/1538-4357/ad2df9},
archivePrefix = {arXiv},
       eprint = {2311.16232},
 primaryClass = {astro-ph.GA},
       adsurl = {https://ui.adsabs.harvard.edu/abs/2024ApJ...966...36K}
}

@ARTICLE{belli_2017,
       author = {{Belli}, Sirio and {Genzel}, Reinhard and {F{\"o}rster Schreiber}, Natascha M. and {Wisnioski}, Emily and {Wilman}, David J. and {Wuyts}, Stijn and {Mendel}, J. Trevor and {Beifiori}, Alessandra and {Bender}, Ralf and {Brammer}, Gabriel B. and {Burkert}, Andreas and {Chan}, Jeffrey and {Davies}, Rebecca L. and {Davies}, Ric and {Fabricius}, Maximilian and {Fossati}, Matteo and {Galametz}, Audrey and {Lang}, Philipp and {Lutz}, Dieter and {Momcheva}, Ivelina G. and {Nelson}, Erica J. and {Saglia}, Roberto P. and {Tacconi}, Linda J. and {Tadaki}, Ken-ichi and {{\"U}bler}, Hannah and {van Dokkum}, Pieter},
        title = "{KMOS$^{3D}$ Reveals Low-level Star Formation Activity in Massive Quiescent Galaxies at 0.7 < z < 2.7}",
      journal = {\apjl},
         year = 2017,
        month = may,
       volume = {841},
       number = {1},
          eid = {L6},
        pages = {L6},
          doi = {10.3847/2041-8213/aa70e5},
archivePrefix = {arXiv},
       eprint = {1703.07778},
 primaryClass = {astro-ph.GA},
       adsurl = {https://ui.adsabs.harvard.edu/abs/2017ApJ...841L...6B}
}

@ARTICLE{belli2019,
       author = {{Belli}, Sirio and {Newman}, Andrew B. and {Ellis}, Richard S.},
        title = "{MOSFIRE Spectroscopy of Quiescent Galaxies at 1.5 < z < 2.5. II. Star Formation Histories and Galaxy Quenching}",
      journal = {\apj},
         year = 2019,
        month = mar,
       volume = {874},
       number = {1},
          eid = {17},
        pages = {17},
          doi = {10.3847/1538-4357/ab07af},
archivePrefix = {arXiv},
       eprint = {1810.00008},
 primaryClass = {astro-ph.GA},
       adsurl = {https://ui.adsabs.harvard.edu/abs/2019ApJ...874...17B}
}

@ARTICLE{avery_cooloutflows_2022,
       author = {{Avery}, Charlotte R. and {Wuyts}, Stijn and {F{\"o}rster Schreiber}, Natascha M. and {Villforth}, Carolin and {Bertemes}, Caroline and {Hamer}, Stephen L. and {Sharma}, Raman and {Toshikawa}, Jun and {Zhang}, Junkai},
        title = "{Cool outflows in MaNGA: a systematic study and comparison to the warm phase}",
      journal = {\mnras},
         year = 2022,
        month = apr,
       volume = {511},
       number = {3},
        pages = {4223-4237},
          doi = {10.1093/mnras/stac190},
archivePrefix = {arXiv},
       eprint = {2201.08079},
 primaryClass = {astro-ph.GA},
       adsurl = {https://ui.adsabs.harvard.edu/abs/2022MNRAS.511.4223A}
}

@article{cheung_suppressing_2016,
	title = {Suppressing star formation in quiescent galaxies with supermassive black hole winds},
	volume = {533},
	issn = {0028-0836},
	url = {https://ui.adsabs.harvard.edu/abs/2016Natur.533..504C},
	doi = {10.1038/nature18006},
	urldate = {2025-03-09},
	journal = {Nature},
	author = {Cheung, Edmond and Bundy, Kevin and Cappellari, Michele and Peirani, Sébastien and Rujopakarn, Wiphu and Westfall, Kyle and Yan, Renbin and Bershady, Matthew and Greene, Jenny E. and Heckman, Timothy M. and Drory, Niv and Law, David R. and Masters, Karen L. and Thomas, Daniel and Wake, David A. and Weijmans, Anne-Marie and Rubin, Kate and Belfiore, Francesco and Vulcani, Benedetta and Chen, Yan-Mei and Zhang, Kai and Gelfand, Joseph D. and Bizyaev, Dmitry and Roman-Lopes, A. and Schneider, Donald P.},
	month = may,
	year = {2016},
	note = {ADS Bibcode: 2016Natur.533..504C},
	pages = {504--508},
}

@article{deugenio_fast-rotator_2024,
	title = {A fast-rotator post-starburst galaxy quenched by supermassive black-hole feedback at z = 3},
	volume = {8},
	issn = {2397-3366},
	url = {https://ui.adsabs.harvard.edu/abs/2024NatAs...8.1443D},
	doi = {10.1038/s41550-024-02345-1},
	urldate = {2025-03-17},
	journal = {Nature Astronomy},
	author = {D'Eugenio, Francesco and Pérez-González, Pablo G. and Maiolino, Roberto and Scholtz, Jan and Perna, Michele and Circosta, Chiara and Übler, Hannah and Arribas, Santiago and Böker, Torsten and Bunker, Andrew J. and Carniani, Stefano and Charlot, Stephane and Chevallard, Jacopo and Cresci, Giovanni and Curtis-Lake, Emma and Jones, Gareth C. and Kumari, Nimisha and Lamperti, Isabella and Looser, Tobias J. and Parlanti, Eleonora and Rix, Hans-Walter and Robertson, Brant and Rodríguez Del Pino, Bruno and Tacchella, Sandro and Venturi, Giacomo and Willott, Chris J.},
	month = nov,
	year = {2024},
	note = {ADS Bibcode: 2024NatAs...8.1443D},
	pages = {1443--1456},
}

@article{carnall_massive_2023,
	title = {A massive quiescent galaxy at redshift 4.658},
	volume = {619},
	issn = {0028-0836},
	url = {https://ui.adsabs.harvard.edu/abs/2023Natur.619..716C},
	doi = {10.1038/s41586-023-06158-6},
	urldate = {2025-03-17},
	journal = {Nature},
	author = {Carnall, Adam C. and McLure, Ross J. and Dunlop, James S. and McLeod, Derek J. and Wild, Vivienne and Cullen, Fergus and Magee, Dan and Begley, Ryan and Cimatti, Andrea and Donnan, Callum T. and Hamadouche, Massissilia L. and Jewell, Sophie M. and Walker, Sam},
	month = jul,
	year = {2023},
	note = {ADS Bibcode: 2023Natur.619..716C},
	pages = {716--719},
}

@article{marconcini_moka3d_2023,
	title = {{MOKA3D}: {An} innovative approach to {3D} gas kinematic modelling. {I}. {Application} to {AGN} ionised outflows},
	volume = {677},
	issn = {0004-6361},
	shorttitle = {{MOKA3D}},
	url = {https://ui.adsabs.harvard.edu/abs/2023A&A...677A..58M},
	doi = {10.1051/0004-6361/202346821},
	urldate = {2026-03-05},
	journal = {Astronomy and Astrophysics},
	publisher = {EDP},
	author = {Marconcini, C. and Marconi, A. and Cresci, G. and Venturi, G. and Ulivi, L. and Mannucci, F. and Belfiore, F. and Tozzi, G. and Ginolfi, M. and Marasco, A. and Carniani, S. and Amiri, A. and Di Teodoro, E. and Scialpi, M. and Tomicic, N. and Mingozzi, M. and Brazzini, M. and Moreschini, B.},
	month = sep,
	year = {2023},
	note = {ADS Bibcode: 2023A\&A...677A..58M},
	pages = {A58},
}

@BOOK{osterbrockfernland,
       author = {{Osterbrock}, Donald E. and {Ferland}, Gary J.},
        title = "{Astrophysics of gaseous nebulae and active galactic nuclei}",
         year = 2006,
       adsurl = {https://ui.adsabs.harvard.edu/abs/2006agna.book.....O}
}

@ARTICLE{calzetti2000,
       author = {{Calzetti}, Daniela and {Armus}, Lee and {Bohlin}, Ralph C. and {Kinney}, Anne L. and {Koornneef}, Jan and {Storchi-Bergmann}, Thaisa},
        title = "{The Dust Content and Opacity of Actively Star-forming Galaxies}",
      journal = {\apj},
         year = 2000,
        month = apr,
       volume = {533},
       number = {2},
        pages = {682-695},
          doi = {10.1086/308692},
archivePrefix = {arXiv},
       eprint = {astro-ph/9911459},
 primaryClass = {astro-ph},
       adsurl = {https://ui.adsabs.harvard.edu/abs/2000ApJ...533..682C}
}

@ARTICLE{park2024,
       author = {{Park}, Minjung and {Belli}, Sirio and {Conroy}, Charlie and {Johnson}, Benjamin D. and {Davies}, Rebecca L. and {Leja}, Joel and {Tacchella}, Sandro and {Mendel}, J. Trevor and {Benton}, Chlo{\"e} and {Bugiani}, Letizia and {Emami}, Razieh and {Khoram}, Amir H. and {Li}, Yijia and {Maheson}, Gabriel and {Mathews}, Elijah P. and {Naidu}, Rohan P. and {Nelson}, Erica J. and {Terrazas}, Bryan A. and {Weinberger}, Rainer},
        title = "{Widespread Rapid Quenching at Cosmic Noon Revealed by JWST Deep Spectroscopy}",
      journal = {\apj},
         year = 2024,
        month = nov,
       volume = {976},
       number = {1},
          eid = {72},
        pages = {72},
          doi = {10.3847/1538-4357/ad7e15},
archivePrefix = {arXiv},
       eprint = {2404.17945},
 primaryClass = {astro-ph.GA},
       adsurl = {https://ui.adsabs.harvard.edu/abs/2024ApJ...976...72P}
}

@ARTICLE{kennicutt_sf98,
       author = {{Kennicutt}, Jr., Robert C.},
        title = "{Star Formation in Galaxies Along the Hubble Sequence}",
      journal = {\araa},
         year = 1998,
        month = jan,
       volume = {36},
        pages = {189-232},
          doi = {10.1146/annurev.astro.36.1.189},
archivePrefix = {arXiv},
       eprint = {astro-ph/9807187},
 primaryClass = {astro-ph},
       adsurl = {https://ui.adsabs.harvard.edu/abs/1998ARA&A..36..189K}
}

@article{whitaker_quenching_2021,
	title = {Quenching of star formation from a lack of inflowing gas to galaxies},
	volume = {597},
	issn = {0028-0836},
	url = {https://ui.adsabs.harvard.edu/abs/2021Natur.597..485W},
	doi = {10.1038/s41586-021-03806-7},
	urldate = {2026-05-12},
	journal = {Nature},
	author = {Whitaker, Katherine E. and Williams, Christina C. and Mowla, Lamiya and Spilker, Justin S. and Toft, Sune and Narayanan, Desika and Pope, Alexandra and Magdis, Georgios E. and van Dokkum, Pieter G. and Akhshik, Mohammad and Bezanson, Rachel and Brammer, Gabriel B. and Leja, Joel and Man, Allison and Nelson, Erica J. and Richard, Johan and Pacifici, Camilla and Sharon, Keren and Valentino, Francesco},
	month = sep,
	year = {2021},
	note = {ADS Bibcode: 2021Natur.597..485W},
	pages = {485--488},
}

@misc{gangula_resolved_2026,
	title = {Resolved {Maps} of {Gas} and {Dust} in a {Massive} {Quiescent} {Galaxy} at z=2 from {INQUEST}-{JWST}: {Evidence} of {Accretion} and {Rejuvenation}},
	shorttitle = {Resolved {Maps} of {Gas} and {Dust} in a {Massive} {Quiescent} {Galaxy} at z=2 from {INQUEST}-{JWST}},
	url = {https://ui.adsabs.harvard.edu/abs/2026arXiv260426195G},
	doi = {10.48550/arXiv.2604.26195},
	urldate = {2026-05-18},
	publisher = {arXiv},
	author = {Gangula, Sai and Newman, Andrew B. and Gu, Meng and Belli, Sirio and Whitaker, Katherine E. and Barone, Tania M. and Beverage, Aliza and Bolamperti, Andrea and Bugiani, Letizia and Ellis, Richard S. and Kriek, Mariska and Matthews, Allison and Nanayakkara, Themiya},
	month = apr,
	year = {2026},
	note = {ADS Bibcode: 2026arXiv260426195G},
}

@ARTICLE{rebecca_neutral24,
       author = {{Davies}, Rebecca L. and {Belli}, Sirio and {Park}, Minjung and {Mendel}, J. Trevor and {Johnson}, Benjamin D. and {Conroy}, Charlie and {Benton}, Chlo{\"e} and {Bugiani}, Letizia and {Emami}, Razieh and {Leja}, Joel and {Li}, Yijia and {Maheson}, Gabriel and {Mathews}, Elijah P. and {Naidu}, Rohan P. and {Nelson}, Erica J. and {Tacchella}, Sandro and {Terrazas}, Bryan A. and {Weinberger}, Rainer},
        title = "{JWST reveals widespread AGN-driven neutral gas outflows in massive z   2 galaxies}",
      journal = {\mnras},
         year = 2024,
        month = mar,
       volume = {528},
       number = {3},
        pages = {4976-4992},
          doi = {10.1093/mnras/stae327},
archivePrefix = {arXiv},
       eprint = {2310.17939},
 primaryClass = {astro-ph.GA},
       adsurl = {https://ui.adsabs.harvard.edu/abs/2024MNRAS.528.4976D}
}

@ARTICLE{bellinat_2024,
       author = {{Belli}, Sirio and {Park}, Minjung and {Davies}, Rebecca L. and {Mendel}, J. Trevor and {Johnson}, Benjamin D. and {Conroy}, Charlie and {Benton}, Chlo{\"e} and {Bugiani}, Letizia and {Emami}, Razieh and {Leja}, Joel and {Li}, Yijia and {Maheson}, Gabriel and {Mathews}, Elijah P. and {Naidu}, Rohan P. and {Nelson}, Erica J. and {Tacchella}, Sandro and {Terrazas}, Bryan A. and {Weinberger}, Rainer},
        title = "{Star formation shut down by multiphase gas outflow in a galaxy at a redshift of 2.45}",
      journal = {\nat},
         year = 2024,
        month = jun,
       volume = {630},
       number = {8015},
        pages = {54-58},
          doi = {10.1038/s41586-024-07412-1},
archivePrefix = {arXiv},
       eprint = {2308.05795},
 primaryClass = {astro-ph.GA},
       adsurl = {https://ui.adsabs.harvard.edu/abs/2024Natur.630...54B}
}

@ARTICLE{newman_stellar_2025,
       author = {{Newman}, Andrew B. and {Gu}, Meng and {Belli}, Sirio and {Ellis}, Richard S. and {Gangula}, Sai and {Greene}, Jenny E. and {Walsh}, Jonelle L. and {Suyu}, Sherry H. and {Ertl}, Sebastian and {Caminha}, Gabriel and {Granata}, Giovanni and {Grillo}, Claudio and {Schuldt}, Stefan and {Barone}, Tania M. and {Bird}, Simeon and {Glazebrook}, Karl and {Jafariyazani}, Marziye and {Kriek}, Mariska and {Matthews}, Allison and {Morishita}, Takahiro and {Nanayakkara}, Themiya and {Pierel}, Justin D.~R. and {Acebr{\'o}n}, Ana and {Bergamini}, Pietro and {Cha}, Sangjun and {Diego}, Jose M. and {Foo}, Nicholas and {Frye}, Brenda and {Fudamoto}, Yoshinobu and {Jee}, M. James and {Kamieneski}, Patrick S. and {Koekemoer}, Anton M. and {Meena}, Asish K. and {Nishida}, Shun and {Oguri}, Masamune and {Rosati}, Piero and {Zitrin}, Adi},
        title = "{A stellar dynamical mass measurement of an inactive black hole at redshift 2}",
      journal = {Science},
         year = 2026,
        month = jun,
       volume = {392},
       number = {6802},
        pages = {1065-1068},
          doi = {10.1126/science.adx5816},
archivePrefix = {arXiv},
       eprint = {2503.17478},
 primaryClass = {astro-ph.GA},
       adsurl = {https://ui.adsabs.harvard.edu/abs/2026Sci...392.1065N}
}

@ARTICLE{newman_resolving_2018,
       author = {{Newman}, Andrew B. and {Belli}, Sirio and {Ellis}, Richard S. and {Patel}, Shannon G.},
        title = "{Resolving Quiescent Galaxies at z {\ensuremath{\gtrsim}} 2. I. Search for Gravitationally Lensed Sources and Characterization of Their Structure, Stellar Populations, and Line Emission}",
      journal = {\apj},
         year = 2018,
        month = aug,
       volume = {862},
       number = {2},
          eid = {125},
        pages = {125},
          doi = {10.3847/1538-4357/aacd4d},
archivePrefix = {arXiv},
       eprint = {1806.06814},
 primaryClass = {astro-ph.GA},
       adsurl = {https://ui.adsabs.harvard.edu/abs/2018ApJ...862..125N}
}

@ARTICLE{Bolamperti2024,
       author = {{Bolamperti}, A. and {Grillo}, C. and {Caminha}, G.~B. and {Granata}, G. and {Suyu}, S.~H. and {Ca{\~n}ameras}, R. and {Christensen}, L. and {Vernet}, J. and {Zanella}, A.},
        title = "{Cosmography from accurate mass modeling of the lens group SDSS J0100+1818: Five sources at three different redshifts}",
      journal = {\aap},
         year = 2024,
        month = dec,
       volume = {692},
          eid = {A239},
        pages = {A239},
          doi = {10.1051/0004-6361/202451209},
archivePrefix = {arXiv},
       eprint = {2411.07289},
 primaryClass = {astro-ph.CO},
       adsurl = {https://ui.adsabs.harvard.edu/abs/2024A&A...692A.239B}
}



\end{document}